\documentclass[mnsc,nonblindrev]{informs3_hide}

\OneAndAHalfSpacedXI %

\usepackage[english]{babel}
\usepackage[autostyle, english = american]{csquotes}
\MakeOuterQuote{"}

\usepackage{bbm}
\usepackage[hypertexnames=false]{hyperref}
\usepackage{cleveref}
\crefname{subsection}{subsection}{subsections}
\usepackage{xspace}
\usepackage{mathtools}
\usepackage{algorithm,algpseudocode}

\newcommand{\eps}{\varepsilon}
\newcommand{\bI}{\mathbbm{1}}
\newcommand{\bE}{\mathbb{E}}

\newcommand{\ALG}{\mathsf{ALG}}
\newcommand{\OPT}{\mathsf{OPT}}
\newcommand{\OFF}{\mathsf{OFF}}

\newcommand{\Indep}{\textsc{Indep}\xspace}
\newcommand{\LPtrunc}{\mathsf{LP}^{\mathsf{trunc}}}

\newcommand{\Correl}{\textsc{Correl}\xspace}
\newcommand{\Adv}{\textsc{Adv}\xspace}
\newcommand{\Rand}{\textsc{Rand}\xspace}
\newcommand{\Poisson}{\textsc{Poisson}\xspace}

\newcommand{\var}{\mathrm{Var}}

\newcommand{\Pois}{\mathrm{Pois}}
\newcommand{\LP}{\mathsf{LP}}
\newcommand{\LPcond}{\mathsf{LP}^\mathsf{cond}}
\newcommand{\bD}{{\bf D}}
\newcommand{\bd}{{\bf d}}
\newcommand{\bx}{{\bf x}}

\newcommand{\bp}{{\bf p}}

\newcommand{\cI}{\mathcal{I}}
\newcommand{\cL}{\mathcal{L}}
\newcommand{\cP}{\mathcal{P}}

\newcommand{\ypre}{y^\mathsf{pre}}
\newcommand{\ypost}{y^\mathsf{post}}

\newcommand{\bb}{\mathbb}
\newcommand{\ex}[1]{{\mathbb E} \left[ #1 \right]}

\newcommand{\expar}[1]{{\mathbb E} [ #1 ]}
\newcommand{\pr}[1]{{\rm Pr} \left[ #1 \right]}
\newcommand{\prpartwo}[2]{{\rm Pr}_{#1} [ #2 ]}
\newcommand{\prtwo}[2]{{\rm Pr}_{#1} \left[ #2 \right]}
\newcommand{\prpar}[1]{{\rm Pr} [ #1 ]}

\newcommand{\bs}[1]{\boldsymbol{#1}}

\newcommand{\cq}{\mathsf{cq}}
\newcommand{\start}{\mathsf{start}}
\newcommand{\ennd}{\mathsf{end}}
\newcommand{\inv}{\mathsf{inv}}
\newcommand{\prob}{\mathsf{prob}}
\newcommand{\free}{\mathsf{free}}
\newcommand{\Acc}{\mathsf{Acc}}
\newcommand{\MyAtop}[2]{\genfrac{}{}{0pt}{}{#1}{#2}}
\usepackage{pifont}%

\usepackage[dvipsnames]{xcolor}
\usepackage{subcaption}
\usepackage{graphicx}

\usepackage{natbib,bbm,multirow,multicol}
 \bibpunct[, ]{(}{)}{,}{a}{}{,}%
 \def\bibfont{\small}%

\TheoremsNumberedThrough     %
\ECRepeatTheorems

\EquationsNumberedThrough    %

\MANUSCRIPTNO{MS-RMA-2024-05424.R1} %

\begin{document}

\TITLE{
A Nonparametric Framework for Online Stochastic Matching with Correlated Arrivals
}
\ARTICLEAUTHORS{
\AUTHOR{Ali Aouad}
\AFF{Sloan School of Management, Massachusetts Institute of Technology, Cambridge, MA 02139, \EMAIL{maouad@mit.edu}}
\AUTHOR{Will Ma}
\AFF{Graduate School of Business and Data Science Institute, Columbia University, New York, NY 10027, \EMAIL{wm2428@gsb.columbia.edu}}
}

\ABSTRACT{
The design of online algorithms for matching markets and revenue management settings is usually bound by the assumption that the demand process is formed by a fixed-length sequence of queries with unknown types, each drawn independently. This notion of {\em serial independence}  implies that the demand of each type, i.e., the number of queries of a given type, has low variance and is approximately Poisson-distributed. 

This paper
proposes a nonparametric framework for modeling arrival sequences in online stochastic matching that departs from the serial independent assumption.
We propose two models, \Indep and \Correl, that capture different forms of serial correlations by combining a nonparametric distribution for the demand with standard assumptions on the arrival patterns---adversarial or random order. The \Indep model can capture arbitrary serial correlations within each customer type but assumes cross-sectional independence across types, whereas the \Correl model captures common shocks across customer types.  We demonstrate that fluid relaxations, which rely solely on demand expectations, have arbitrarily bad performance guarantees. In contrast, we develop new algorithms that achieve optimal (constant-factor) performance guarantees in each model. Our mathematical analysis includes tighter linear programming (LP) relaxations that leverage distribution knowledge, and a new lossless randomized LP rounding scheme for \Indep.  We test our new LP relaxations and rounding scheme in simulations on real and synthetic data, and find that they {consistently} outperform well-established {matching algorithms}, especially on  real data sequences {that exhibit greater demand variance}.
}

\HISTORY{An earlier version of this paper appeared in the proceedings of EC 2023.  This version is forthcoming in Management Science.}

\maketitle

\section{Introduction} \label{sec:intro}

In online platforms and supply chain operations, the allocation of scarce resources to customers in real-time is often represented by an online matching problem. In this setting, customer queries, which can be of $m$ distinct types, arrive over a finite time horizon and must be irrevocably served with up to one of $n$ inventory-constrained resources.  A known reward of $r_{i,j}\ge0$ is collected each time a query of any type $j=1,\ldots,m$ is served by any resource $i=1,\ldots,n$.  Rewards can be arbitrary, with $r_{i,j}=0$ indicating imcompatibility between $i$ and $j$. The objective is to maximize the total reward collected over the time horizon.

To inform these sequential decisions,  decision-makers often rely on prior information about future demand. Intensive research  on online \textit{stochastic} matching has focused on designing matching algorithms under distributional assumptions regarding the sequence of arriving queries. The standard modeling framework assumes {\em serial independence} (SI) over a fixed horizon: each query type $j$ at time $t= 1,\ldots, T$ is drawn from a known distribution $(p_{t,j})_{j\in [m]}$, independent of the history.  This process naturally arises as a discrete-time approximation of independent Poisson processes with continuous-time arrival rates; the latter \textit{Poisson model} is also common and typically assumes a constant arrival rate $\lambda_j$ for each type $j$ over time.\footnote{If the continuous-time horizon is $[0,T]$, then the discrete-time approximation of the Poisson processes is obtained by setting $p_{t,j} =\lambda_j $ for all $t\in [T]$ and $j\in [m]$.} We refer to these models collectively as \textit{SI models}, as they impose the restrictive SI assumption, which we propose to reexamine in this work.

Consider first a critical implication of the serial independence assumption.  From the decision-maker's perspective, the total
number of queries of any type $j\in [m]$, 
which we call the \textit{demand} $D_j$ for type $j$, necessarily has its variance upper-bounded by its mean, i.e., $\var(D_j) \le\bE[D_j]$ for all $j \in [m]$.
This is because $D_j$ is the sum of $T$ independent Bernoulli outcomes. In the continuous-time version, where the demands follow Poisson distributions, the variance is exactly equal to the mean. 

Strikingly, this low-variance consequence is not consistent with demand models typically used in operations management. For example, when modeling unknown demand in supply chains, textbook examples use a Normal distribution with an arbitrary mean and standard deviation \citep{simchi2005logic}.
As another example, when managing revenue from different customer classes $j$, standard models allow the demands $D_j$ to be drawn from arbitrary distributions \citep[Chap.~2.2]{talluri2004theory}, which are independent across $j$ but do not need to satisfy $\var(D_j) \le\bE[D_j]$.
In fact, \citet[Chap.~2.5]{talluri2004theory} identify the same challenge when modeling the customer arrival process:
\begin{quote}
"\textit{Dynamic models} [e.g.\ online stochastic matching]
\textit{allow for an
arbitrary order of arrival, with the possibility of interspersed arrivals of
several classes [\ldots] the dynamic models require the assumption of Markovian
(such as Poisson)} [i.e.\ serial independence] \textit{arrivals to make them tractable. This puts restrictions
on modeling different levels of variability in demand.}"
\end{quote}

More troubling still, we find that the constraint $\var(D_j) \le\bE[D_j]$ is severely violated in real-world demand data that we analyze.
We analyze two separate data sets describing different online retail settings, the first one being released by the e-commerce platform JD.com \citep{shen2020jd}, and the second one being a proprietary data set from a large fashion retail platform. Both platforms must decide in real time from which warehouse to dispatch the items ordered by their customers. We analyze the demand at the warehouse-SKU level, counting the number of units of an SKU ordered by individuals local to a warehouse, over a duration approximately representing a replenishment cycle.
We find that the variance in the demand per warehouse-SKU exceeds the mean demand in the majority of cases, sometimes by an order of magnitude, even after controlling for several factors.
Details of this analysis are presented and further discussed in \textbf{Appendix~\ref{sec:empVal}}. 

In order to model high-variance stochastic demand, characterized by $\var(D_j) > \bE[D_j]$, we depart from the traditional SI model. High variance and serial correlations are inextricably linked. As a running example, we consider the Poisson setup with rates $\{\lambda_j\}_{j\in [m]}$. To capture higher levels of uncertainty, it is common to generalize this setup through a {\em mixture} of Poisson models: Suppose that the rate $\Lambda_j$ is now a random variable from the decision-maker's perspective (e.g., due to uncertainty around point estimates of the arrival rates). This approach  can capture high variance:
$$\var(D_j) = \bE[ \var(D_j| \Lambda_j)] + \var(\bE[D_j|\Lambda_j]) = \bE[\Lambda_j] + \var(\Lambda_j) = \bE[D_j] + \var(\Lambda_j) \ ,$$
whenever $\var(\Lambda_j)>0$---that is, when $\Lambda_j$ is not deterministic. At the same time, treating $\Lambda_j$ as random (with a probabilistic prior) induces serial correlation, violating the SI assumption. Indeed, a high realization of $\Lambda_j$ indicates ``high popularity'' of demand type $j$, effectively introducing serial correlation whereby the types of early and later queries are positively correlated. This model extension has consequences for the design of online matching algorithms, since higher than expected popularity could be inferred from observing early query types, making adaptivity more desirable.

Yet mixed Poisson processes of this sort are scarcely studied in the online matching and resource allocation literature. Can rigorous approximation algorithms be devised in such settings? Do these algorithms possess a simple structure? Furthermore, the mixed Poisson model is itself restrictive. One might consider other families of stochastic processes with parametrically specified correlations over time. Such parametric approaches, however, require committing to a particular correlation structure. While adversarial models capture arbitrary demand patterns, they are often overly pessimistic. Our main contribution is to introduce a nonparametric framework that captures serial correlation implicitly while retaining simple and tractable algorithms.

\subsection{Modeling framework}

\paragraph{Framework.} We propose a general framework---which may be of independent interest---and then focus on two  models that fall under this framework and generalize the SI model through different perspectives on serial correlations. 
Our framework abstracts away the specifics of the arrival process, instead focusing on modeling the {\em demand vector} $\bD=(D_1,\ldots,D_m)$, which counts for each  $j \in [m]$, the total number of queries $D_j$ from type $j$ that arrive in the sequence. We assume that $\bD$ is random and drawn from an arbitrary distribution known to the decision-maker.  This allows the total demand $D = \sum_{j=1}^m D_j$ to have high-variance---whereas $D$ is tied to the (deterministic) horizon length in the discrete SI model or has low-variance $\var(D) = \ex{D}$ in the Poisson model.

Once the realization of $\bD$ is drawn, we still need to specify the queries' arrival order. For this, we consider two assumptions from the online (non-stochastic) matching literature:
\begin{hitemize}
    \item {\em Adversarial order} (\Adv): the sequence is an arbitrary adversarially-chosen permutation of   $(\underbrace{1,\ldots,1}_{D_1}, \ldots,\underbrace{m,\ldots,m}_{D_m})$, where each type $j$ is repeated with multiplicity $D_j$;
     \item {\em Random order} (\Rand): the sequence of queries is a random permutation of the same vector, chosen uniformly at random.
\end{hitemize} 
Although the arrival order might affect the performance of online algorithms, it does not modify the {\em offline optimum}, which  computes the maximum total reward obtainable from matching queries to resources in hindsight, with full knowledge of the sequence. This benchmark, which is often used measure the performance of online algorithms, only depends on the realization of $\bD$---aligning with the focus of our framework.

Since this framework is very general, we impose  restrictions on the distribution of the demand $\bD$ which capture two different forms of serial correlations and high-variance demand.

\paragraph{ Model 1: \Indep-endent demand across types.} Our first model, called \Indep, allows the marginal distributions of $\bD$ to be arbitrary, but requires the demands per type to be mutually independent. That is, the demand coordinate $D_j$ of each type $j\in [m]$ is drawn independently from any nonparametric distribution, which may well differ across types. Our running example corresponds to $D_j\sim \Pois(\Lambda_j)$, where $\Pois(\lambda_j)$ denotes a Poisson distribution with mean $\lambda_j$; the \Indep model just requires that the random rates $\Lambda_j$ are mutually independent across types $j\in [m]$.  \Indep effectively captures any desired serial correlation pattern, but it requires cross-sectional independence, i.e,  the demand per type are mutually independent and serial correlations are ``confined'' within each type. This assumption  is motivated by spatial markets, where query types correspond to distinct geographic regions and the demand shocks are idiosyncratic to each region.\footnote{An analogous model is proposed by \citet[Chap.~2]{talluri2004theory}, and in particular, \Indep enables us to flexibly quantify the level of uncertainty per type, with ${\rm Var}(D_j)$ possibly larger than $\expar{D_j}$.} \Indep subsumes as special cases the SI and Poisson models (see \textbf{Appendix~\ref{app:generalizations}}).

\paragraph{Model 2: \Correl-ated demand across types.} Our second model, called \Correl, allows the distribution of the total demand $D = \sum_{j=1}^m D_j$ to be arbitrary, but conditional on the realization of $D$, requires that the {demand $D_j$ per type $j\in[m]$} to essentially follow an SI model with horizon $D$.  That is, conditional on $D$, each query $t\in [D]$ independently draws a type from $\{1,\ldots,m\}$ according to a known probability vector $\bp=(p_{1},\ldots,p_{m})$, which represents the average proportions of types, with $\sum_{j=1}^m p_{j}=1$.  {Our recurring example fits the $\Correl$ model by specifying $D\sim \Pois(\Lambda)$ for some random variable $\Lambda$, which amounts to having $D_j \sim \Pois(\Lambda \cdot p_j)$ for each type $j$. Under this specification, the query types exhibit linearly dependent random rates $\Lambda_j = \Lambda \cdot p_j$, where a common shock $\Lambda$ induces serial correlations across arrivals. }

Unlike the $\Indep$ model,  \Correl can capture positive cross-sectional correlations across $j$'s:  a high realization of $D$ leads to every demand $D_j$ being larger.
External common shocks may be due to unobserved calendar effects or competitor actions that can simultaneously affect the demands of all types.  Similarly to the \Indep model, $\Correl$ subsumes as a special case the SI and Poisson models
(see \textbf{Appendix~\ref{app:generalizations}}).

\subsection{Preview of our results}
The standard approach for online stochastic matching is as follows: solve a {\em linear program} (LP), which often represents the arrival process in fluid scale, and then match queries sequentially using a discrete randomized {\em rounding} of the LP solution. This fluid LP, also known as the certainty-equivalent LP, replaces the random demand with its expectation $\ex{D_j}$. For our nonparametric models, the central challenge is that the fluid approximation is too coarse when the SI assumption violated. We observe a "flaw of averages" in high-variance environments: the fluid LP's optimal value can exceed the best-achievable reward by an arbitrary factor. Our work essentially overcomes the failure of standard LP relaxations and devises algorithms for $\Indep$ and $\Correl$ with optimal performance guarantees. We also identify generalizable design principles for online matching algorithms in the face of high-variance, correlated demand.

\paragraph{Competitive and approximation algorithms.} Our main contribution is a $1/2$-{\em competitive} polynomial-time algorithm for the $\Indep$ model under adversarial arrival patterns (\Adv); we use the standard terminology that an $\alpha$-competitive algorithm achieves a fraction $\alpha$ of the offline optimum. This result is significant because $1/2$-competitiveness is the best-possible performance guarantee for the SI model---thus our more general model incurs no loss in worst-case performance. Our approach introduces a tighter LP relaxation, dubbed the {\em truncated LP}, that incorporates higher-order moment information about the distribution of the demand $D_j$. The crux of our algorithm is a {\em lossless rounding} scheme for this truncated LP that sequentially routes queries to resources at rates {\em exactly} matching our fractional LP solution---a property we formalize in \textbf{\Cref{sec:indep}}.

We study \Correl with $\Rand$ arrival patterns in \textbf{\Cref{sec:correl}}. While competitive algorithms against the offline optimum are not possible in this setting, we obtain a $1/2$-{\em approximate} algorithm relative to the optimum online algorithm (a weaker benchmark without foresight of future demand). Our approach combines somewhat similar ingredients: a tighter LP relaxation---dubbed here the {\em conditional LP}---and an online rounding scheme. In this setting, however, our matching algorithm is more explicitly adaptive to the realized demand.

We consider several extensions: sampling-based algorithms ({\bf \Cref{app:demandError}}) and  network-revenue management ({\bf\Cref{app:stochastic_gen}}), which suggest our theoretical findings are generalizable to some degree and leave other open questions for future research.

\paragraph{Simulations.} We use synthetic and real-world demand data to experiment with our algorithms for the $\Indep$ and $\Correl$ models in \textbf{\Cref{sec:simulations}}. Specifically we compare algorithms of the form ``LP relaxation'' plus ``online rounding.'' In high-variance environments, matching policies calibrated using our tighter LPs  achieve significant gains compared to those using the fluid LP. Moreover, in the case of \Indep, our lossless rounding scheme can outperform standard online rounding methods, or yield further improvements when combined with them. We find that the relative impact of lossless rounding is greater on real customer query sequences.

\paragraph{Design principles.} We construct simple examples that illustrate the inner workings of our algorithms and illuminate the mechanisms explaining why they achieve better performance than standard ones in high-variance, correlated demand environments (\textbf{Appendix~\ref{sec:toyExamples}}). Such examples suggest that our algorithms make improved matching decisions by  better anticipating the demand-supply mismatch or by preserving inventory to adapt to different demand scenarios later.

\section{Further Related Work} \label{sec:relatedWork}
Our work connects with the literature on online matching in computer science, tracing back to the work of~\cite{karp1990optimal}. Without any prior information on $\bs{D}$, competitive algorithms are not possible unless under additional structure is considered, such as binary or bounded rewards~\citep{karp1990optimal,aggarwal2011online,ma2020algorithms}, or flexibility for the matchmaker to amend their decisions~\citep{feldman2009online,ekbatani2022online}. While such adversarial models can capture high levels of demand uncertainty, their theoretical tractability requires  assumptions about rewards and decision processes that may be restrictive in our applications of interest (though we note that they have been used in practice for order fulfillment in retail; see~\cite{andrews2019primal}). Our work takes a fundamentally different approach:  our model integrates stochastic information that preserves the ability to model high-variance demand while allowing arbitrary edge weights.

Closer to our work there is an extensive line of research on online matching in stochastic environments. The core set-up is the SI model and its variants, which have connection to the prophet inequality. \citet{alaei2012online} establish the best-possible competitive ratio of 1/2 when the distribution over types $(p_{t,j})_{j\in [m]}$ is time-varying. The typical algorithm therein combines LP-based randomized routing of queries to resources with a prophet inequality-like  accept/reject decision; the key challenge in our setting is that a naive randomized routing introduces correlations between queries. Much of the literature has focused on obtaining improved competitive ratios in specialized settings like random arrivals~\citep[e.g.,][]{ehsani2018prophet} or IID stationary arrivals~\citep[e.g.,][]{feldman2009online,manshadi2012online,brubach2016new,huang2021online}. A recent line of work also considers a tighter form of analysis using the online optimum as a benchmark~\citep{papadimitriou2021online}, enabling approximation ratios better than 1/2. However, these works are all subject to the SI modeling assumptions, and consequently, assume low-variance demand.

Several works in online resource allocation capture more flexible uncertainty structures than the models discussed above. In particular, the \Correl model is closely connected to sequential decision problems with stochastic horizons~\citep{walczak2006modeling}. \citet{besbes2014dynamic} study a dynamic pricing problem with demand shift after a random time and propose a relaxation similar in spirit to our conditional LP. In a concurrent and independent work, \citet{bai2022fluid} develop an analogous LP relaxation for network revenue management. In both settings, however, the authors focus on asymptotic analysis. Perishable inventory and customer abandonment can be viewed as other forms of stochastic horizons, which have received attention in the literature~\citep{cygan2013catch,alijani2020predict}; an extreme representation of this idea is the dynamic matching problem where both demand/supply join or depart the market according to a stochastic process~\citep{aouad2022dynamic}. While these settings have a connection to correlated prophet inequalities~\citep{truong2019prophet}, the demand/supply uncertainty structures they capture are very different from our framework as they often focus on Poisson-like models. Moving beyond the SI model in online resource allocation connects our work to related streams of literature, including random-order arrival models~\citep{devanur2009adwords} and hybrid frameworks that combine stochastic and adversarial components~\citep{esfandiari2015online, hwang2021online}.

More broadly, online matching has a spectrum of applications beyond e-commerce fulfillment, that may  exhibit correlated arrival patterns, e.g., gig economy platforms \citep{dickerson2021allocation}, blood donation \citep{mcelfresh2020matching}, volunteer matching \citep{manshadi2022online}; we refer the reader to the recent monograph by~\citet{echenique2023online}. A limitation of our work is to focus on one-to-one matching---with the exception of the Network Revenue Management set-up of Appendix~\ref{app:generalizations}. Online resource allocation takes many other ``forms'' in operations management, such as multiple knapsack problems for advance scheduling~\citep{stein2020advance,keyvanshokooh2021online}. A relevant direction of future work is to integrate nonparametric demand models in such settings.

Finally, our lossless rounding adds to a sparse literature on {\em exact polynomial-time relaxations} for implementable allocation policies---in our context, the truncated LP is exact for a fixed query type with random demand. Results in this spirit include the LP formulation by~\citet{alaei2012bayesian} for the polytope of implementable interim allocation rules in multi-agent Bayesian auctions.\footnote{This is essentially a lifted version of an earlier LP by~\citet{border1991implementation}. The analogy with our work is that the approach uses a multi- to single-type reduction and the LP has exponentially many constraints. However, the Bayesian auction problem has a  different structure. It does not feature online decisions, and thus, the ex-post implementation algorithm has no clear connection with our lossless rounding. Additionally, the exponential constraints are parametrized by agent types, rather than resources.}   A well-known dependent rounding for the bipartite matching polytope is due to \citet{gandhi2006dependent}.  For the query-commit model of online bipartite matching, \cite{gamlath2019beating} propose an efficiently separable LP that describes the distribution of implementable query-then-match policies for any fixed vertex. To our knowledge, however, these results have no direct implication for our models.

\section{Technical outline} \label{sec:preliminaries}
We briefly review our model set-up and  performance measures, then state our main results.

\textbf{Model.}  For a positive integer $n$, we let $[n]$ denote the set $\{1,\ldots,n\}$.
An instance $\cI$ of our online stochastic matching model consists of the following parameters: the number of resource types $n$, the number of query types $m$, the corresponding reward values $(r_{i,j})_{i\in [n],j\in [m]}$, the starting inventories $(k_i)_{i\in [n]}$, the distribution of the demand random variable $\bD$, and a categorical variable for its arrival pattern which could be "adversarial " or "random". Each resource $i$ can be matched at most $k_i$ times.
Meanwhile, $\bD=(D_j)_{j\in[m]}$ is a random vector denoting for each type $j\in [m]$ the total number of queries $D_j$ of that type to arrive. The total demand, defined as $D=D_1+\cdots+D_m$, is the length of the sequence of all queries.

We consider two classes of distributions, \Indep and \Correl, for the demand random variable $\bD$.
Under the first class, entries $D_j$ of $\bD$ are drawn from arbitrary distributions, independently across $j$. Each type $j \in [m]$ may have a different distribution for its demand $D_j$. Meanwhile, the second class allows for entries $D_j$ of $\bD$ to be correlated in the following way: first the total demand $D$ is drawn from an arbitrary distribution, and then the types of these $D$ queries are specified by independent and identically distributed outcomes. Each type $j \in [m]$ is drawn with probability $p_j$, where we assume that $\sum_{j=1}^m p_j=1$.  
We use the notation $\cI\in\Indep$ or $\cI\in\Correl$ to indicate that the demand distribution for instance $\cI$ falls under each of the above classes, respectively.

Finally, the arrival pattern will generate a sequence of query types $\sigma$ representing the arrival order of the $D=D_1+\cdots+D_m$ queries.
In $\sigma$, each type $j$ has multiplicity $D_j$, i.e.\ appears exactly $D_j$ times. 
We denote by ${\cal S}(\bD)$ the set of all such sequences, noting that $|{\cal S}(\bD)|=\frac{D!}{D_1!\cdots D_m!}$.

\textbf{Algorithms and performance.}
An online algorithm provides a (randomized) policy for how to match queries on-the-fly, knowing only the instance $\cI$ ahead of time. Specifically, the algorithm has access to the full distribution of the demand and the arrival pattern, but it does not know the specific realization of $\bD$ and $\sigma$ until those are revealed by the sequence of queries. We let  $\ALG_\cI(\bD,\sigma)$ denote the total reward collected by the algorithm in expectation (over any randomness in the algorithm) when the demand vector realizes to $\bD$ and the arrival order is $\sigma$.

Recall that $\sigma$ is determined by an arrival pattern associated with instance $\cI$, which is either adversarial or random, in which case we write $\cI\in\Adv$ or $\cI\in\Rand$ respectively.
Under adversarial
order, $\sigma$ is chosen to minimize $\ALG_\cI(\bD,\sigma)$ knowing both $\bD$ and the algorithm\footnote{We assume that the adversary does not know the realization of the algorithm's random bits, for simplicity.  Our result for $\Indep\cap\Adv$ extends to the {\em almighty} adversary, if we take a prophet inequality designed to work against an almighty adversary \citep[e.g.][]{chawla2010multi} and modify it to compare to the LP.} being used.
Therefore, if $\cI\in\Adv$ then we define the algorithm's performance to be $\ALG(\cI):=\bE_\bD[\inf_{\sigma \in {\cal S}(\bD)} \ALG_{\cal I}(\bD,\sigma)]$.
Meanwhile, under random order, conditional on the realization of $\bD$, the sequence $\sigma$ is equally likely to be any element in ${\cal S}(\bD)$.
Therefore, if $\cI\in\Rand$ then we define the algorithm's performance to be $\ALG(\cI):=\bE_{\bD,\sigma}[\ALG_\cI(\bD,\sigma)]$.
We note that algorithmic performance can always be made  better for instances $\cI\in\Rand$ than for the corresponding instances in \Adv.

\textbf{Benchmarks.} We formalize standard benchmarks against which the performance of online algorithms is measured. We define $\OFF_\cI(\bD)$ as the maximum-weight \textit{offline} matching that could have been made knowing the demand realization $\bD$ in advance. Clearly, the offline matching does not depend on the arrival order $\sigma$.
Consequently, the offline optimum, or prophet optimum, is the quantity $\OFF(\cI):=\bE_{\bD}[\OFF_\cI(\bD)]$.
For instances $\cI\in\Rand$, we also consider the online optimum $\OPT(\cI)$, corresponding to the performance of an optimal online algorithm without any restriction on computational time. This algorithm is defined as an exponential-sized dynamic program that makes decisions in each stage to maximize total expected reward, based on the prior on $\bD$ and $\sigma$ and the information revealed thus far; see Appendix~\ref{sec:formalDP} for a formal definition.

\textbf{Competitive and approximation ratios.}
For every $c\in[0,1]$, we say that an algorithm is \textit{$c$-competitive} for a family of instances if $\ALG(\cI)/\OFF(\cI)\ge c$ for all such instances $\cI$. The maximum constant $c$ for which this holds, i.e.\ the quantity $\inf_\cI\ALG(\cI)/\OFF(\cI)$ with $\cI$ restricted to that family, is sometimes referred to as the \textit{competitive ratio} (for the family and algorithm in question).
Our upper bounds (negative results) will generally hold for the competitive ratio of \textit{any} algorithm.
In this paper, we consider competitive ratios for families of instances constructed by specifying
$\cI\in\Indep\cap\Adv$, or by specifying $\cI\in\Correl\cap\Rand$, 
with otherwise no restrictions on $n$, $m$, the reward values $r_{i,j}$, or the distributions.
We note that our lower bounds (positive results) from $\Indep\cap\Adv$ also apply to $\Indep\cap\Rand$ (or any other way of generating $\sigma$ under $\Indep$), because as explained earlier, $\Adv$ minimizes the algorithmic performance.

When restricting attention to instances $\cI\in\Rand$, we also define a notion of \textit{approximation ratio}, where the algorithm's performance is compared to the online optimum $\OPT(\cI)$. For every  $\alpha\in[0,1]$, we say that an algorithm is \textit{$\alpha$-approximate} for a family of \Rand instances if $\ALG(\cI)/\OPT(\cI)\ge \alpha$ on all instances $\cI$ lying in this family.

\paragraph{\bf Main results.} Our results essentially provide algorithms with best-possible performance guarantees for the $\Indep$ and $\Correl$ models.

Our central result is a generalization of the celebrated prophet inequality  to $\Indep$.
\begin{theorem}[Informal] \label{thm:indepInformal}
For the \Indep model with adversarial arrival patterns,
there is a polynomial-time online algorithm that achieves a $1/2$ competitive ratio.
\end{theorem}
The remainder of the paper focuses on proving this result. After observing the failure of the standard fluid LP, we propose a tighter LP relaxation and devise a new rounding scheme, which may be of independent interest. We also provide a data-driven version  of \Cref{thm:indepInformal} in \textbf{\Cref{app:demandError}}, where the distribution of the $\bD$ is unknown but the decision-maker has access to polynomially many samples. Extending this result to more general allocation settings, such as network revenue management, requires a combinatorial extension of our new LP relaxation, which we leave as an open question for future research; see our discussion in \textbf{\Cref{app:stochastic_gen}}.

For the $\Correl$ model,  competitive algorithms that garner a fixed fraction of the offline optimum are not possible as in \Cref{thm:indepInformal}. This issue has been noticed in related research that considers resources that perish after an unknown random time horizon. Nonetheless, we devise efficient algorithms with optimal performance guarantees against relaxations of the online optimum benchmark.

\begin{theorem}[Informal] \label{thm:correlInformal}
For the $\Correl$ model with random arrival patterns, there is a polynomial-time online algorithm that achieves a $1/2$ approximation ratio.
\end{theorem}
To keep the paper concise, we defer the formal statement and proof of~\Cref{thm:correlInformal} to \textbf{\Cref{sec:correl}}. A byproduct of our algorithm is  a near-optimal performance guarantee in the so-called large-inventory regime, i.e., we obtain an approximation ratio of $(1-O(1/\sqrt{k}))$ as $k:= \min_{i\in [n]} k_i$ grows large. We extend our approach to network revenue management in \textbf{\Cref{app:stochastic_gen}}, but leave the sample-based $\Correl$ model to future work. Since $\Correl$ requires learning only the distribution of aggregate demand $D = \sum_{i=1}^n D_i$ and type proportions $(p_j)_{j\in [m]}$---rather than $m$ separate distributions $D_j$ as in $\Indep$---we expect the data-driven version to be more tractable.

\section{Competitive Algorithm for the \textnormal{\textsc{Indep}} Model} \label{sec:indep}

In this \namecref{sec:indep}, we present our algorithmic results for the \Indep model. In
Section~\ref{subsec:truncLP}, we justify the need for a new LP relaxation, called the "truncated" LP. In Section~\ref{subsec:algIndep}, we develop our algorithm for the \Indep model, assuming the existence of a lossless rounding scheme for the truncated LP, and derive its competitive ratio. Finally, in Section~\ref{sec:lossless}, we devise the lossless rounding scheme.

\subsection{Failure of the fluid LP and formulation of the truncated LP} \label{subsec:truncLP}

We first show that a standard approach for establishing competitive ratios, based on a fluid LP, fails.
For any instance $\cI$, let $\LP(\cI)$ denote the optimal objective value of the following LP:
\begin{align}
\max  \ \ \ \ \ & \sum_{i=1}^n\sum_{j=1}^m r_{i,j}x_{i,j} \label{lp:start} \\
\text{s.t. } \ \ \ \ & \sum_{j=1}^mx_{i,j} \leq k_i &\forall i\in[n] \label{ineq:fluidResource}
\\ &\sum_{i=1}^n x_{i,j} \leq \bE[D_j] &\forall j \in [m] \label{ineq:fluidDemand}
\\ & x_{i,j} \ge 0 &\forall i\in [n],\forall j\in[m] \label{lp:end}\ .
\end{align}
The LP defined in~\eqref{lp:start}--\eqref{lp:end} amounts to a simplified problem formulation where the stochastic demands of query types are replaced by deterministic quantities---their expectations. This relaxation has been the starting point for the design of constant-factor competitive algorithms in a rich literature on online stochastic matching problems.
Similarly, this LP serves as a standard upper bound for asymptotic performance analysis in the literature on revenue management.
Indeed, it is well-known that $\OPT(\cI)\leq\OFF(\cI)\leq\LP(\cI)$ for all instances $\cI$, and hence establishing that $\ALG(\cI)/\LP(\cI)\geq c$ for some constant $c>0$
over a class of instances $\cI$
implies that $\ALG(\cI)/\OFF(\cI)\geq c$ and $\ALG(\cI)/\OPT(\cI)\ge c$ as well for all such $\cI$.

We show that for instances $\cI\in\Indep$, the optimal objective of the fluid relaxation $\LP(\cI)$ can be arbitrarily larger than the offline performance $\OFF(\cI)$. Thus, in sharp contrast with existing models, this LP does not provide an appropriate yardstick for algorithm design.
\begin{proposition} \label{prop:loose1}
Under \Indep, the fluid relaxation LP can be arbitrarily larger than the offline optimum, i.e., $\displaystyle\inf_{\cI \in \Indep}\frac{\OFF(\cI)}{\LP(\cI)}=0$. This holds even for the following restrictions of \Indep: (i) $n=m=1$, or (ii) $n>1$, $m=1$, and $\prpar{D_1\le\sum_{i=1}^n k_i} =1$.
\end{proposition}

\Cref{prop:loose1} is proved in \Cref{pf:prop:loose1} by constructing a family of instances where the ratio between $\OFF({\cal I})$ and $\LP({\cal I})$ converges to zero.  Construction (ii) shows that a simple fix to the fluid LP still does not suffice to obtain a benchmark comparable to the offline optimum. 
In light of this, we introduce a tighter LP.
\begin{definition} \label{def:lpprime}
For any instance $\cI$, we define $\LPtrunc(\cI)$ as the optimal objective value of the following truncated LP:
\begin{align}
\max\ \ \ \ \ & \sum_{i=1}^n\sum_{j=1}^m r_{i,j}x_{i,j} \label{lpTrunc:start}
\\ \text{s.t. }\ \ \ \ & \sum_{j=1}^m x_{i,j} \leq k_i &\forall i\in[n] \label{ineq:truncResource}
\\ & \sum_{i\in S} x_{i,j} \leq \bE\left[ \min\left\{ D_j, \sum_{i\in S} k_i \right\} \right] &\forall S\subseteq [n], \forall j \in [m] \label{ineq:truncDemand}
\\ & x_{i,j} \ge 0 &\forall i\in[n],\forall j\in[m]\ . \label{lpTrunc:end}
\end{align}
\end{definition}
It is straightforward to see that $\LPtrunc(\cI)$ is a tightening of the fluid relaxation $\LP({\cal I})$, i.e., $\LPtrunc(\cI)\le\LP(\cI)$ for all $\cI$, through a comparison of constraints~\eqref{ineq:truncDemand} and~\eqref{ineq:fluidDemand}, while noticing that all other ingredients of the formulation are unchanged. Indeed, by specifying $S = [n]$ in constraint~\eqref{ineq:truncDemand}, we obtain that any feasible solution of $\LPtrunc(\cI)$ satisfies 
\[
\sum_{i=1}^n x_{i,j} \leq \ex{\min\left\{ D_j, \sum_{i=1}^n k_i \right\}} \leq \ex{D_j} \ ,
\]
which is precisely what is required to meet constraint~\eqref{ineq:fluidDemand} in $\LP({\cal I})$.
For each set of resources $S\subseteq [n]$, constraint~\eqref{ineq:truncDemand} expresses the fact that the maximum cardinality of a matching between resource units in $S$ and queries of type $j$ never exceeds $\min\left\{ D_j, \sum_{i\in S} k_i \right\}$ on each realization of $D_j$. Hence, our new constraints~\eqref{ineq:truncDemand} place exponentially many cuts for every $S\subseteq [n]$ by leveraging the full knowledge of the distribution of demand entries $D_j$, compared to merely using their expectation as in the fluid relaxation.

We show in the next lemma that $\LPtrunc(\cI)$  is indeed a valid benchmark for competitive analysis; see \Cref{pf:lem:comparison1} for proof.

\begin{lemma} \label{lem:comparison1}
For any instance $\cI$, we have $\OFF(\cI)\le\LPtrunc(\cI)$.
\end{lemma}

Notice that $\LPtrunc({\cal I})$ has a connection with the Natural LP introduced by~\cite{huang2021online}; we incorporate tightening constraints on the demand side rather than the resource side. Further, \cite{huang2022power} proposed a hierarchy of LP relaxations for Poisson arrivals; its highest level is tighter than $\LPtrunc$, although \cite{huang2022power} use the second level of this hierarchy in their analysis. 

\paragraph{Solving $\LPtrunc(\cI)$.}
At face value, $\LPtrunc(\cI)$ has exponentially many constraints, and thus, it is unclear if it can be solved efficiently from a computational standpoint. However, the reader may notice that constraints~\eqref{ineq:truncDemand} form a polymatroid,  and thus, they are separable in polynomial time~\citep{huang2022power}. Moreover, we can exploit the constraints' structure to devise a fast separation oracle.
To that end, we  rewrite constraints~\eqref{ineq:truncDemand} as
\begin{align} \label{eqn:rewriteTrunc}
\sum_{i\in S} {x_{i,j}} &\leq
\sum_{\ell=1}^{\sum_{i\in S}k_i}\Pr[D_j\ge \ell] &\forall S\subseteq[n],\forall j\in[m]\ .
\end{align}
\begin{claim} \label{clm:sep}
Fix $j\in[m]$ and sort indices $i\in[n]$ so that $\frac{x_{1,j}}{k_1}\ge\cdots\ge\frac{x_{n,j}}{k_n}$.  Then a non-negative vector $(x_{i,j})_{i\in[n]}$ satisfies~\eqref{eqn:rewriteTrunc} (equivalently,~\eqref{ineq:truncDemand} from $\LPtrunc$)
if and only if 
$\sum_{l=1}^i x_{l,j} \le \sum_{\ell=1}^{k_1+\cdots+k_i} \prpar{D_j\ge \ell}$ for all $i\in[n]$.
\end{claim}

Claim~\ref{clm:sep} is proven in \Cref{sec:pfOfNewClaim}, and shows that checking at most $n$ inequalities suffices either to certify the feasibility of a given fractional solution or to identify a violated constraint for the exponential family~\eqref{eqn:rewriteTrunc}. Our numerical experiments in Section~\ref{sec:simulations} demonstrate that this separation oracle solves $\mathcal{L}_p^{\text{trunc}}(\mathcal{I})$ very quickly.

\subsection{Algorithm and results based on the truncated LP} \label{subsec:algIndep}

In this section, we describe our algorithm and performance analysis for the $\Indep$ model under any adversarially chosen arrival order.

\paragraph{Main result.} We assume unit inventories $k_1=\cdots=k_n=1$ throughout the rest of this \namecref{sec:indep}; this is without loss by creating $k_i$ copies of each original resource $i\in[n]$.  Unlike standard online matching problems, we cannot exploit large inventories $k_i$ to improve the competitive ratio under the \Indep model. However, we can still solve the original version $\LPtrunc(\mathcal{I})$ (with non-binary inventories) as it is more compact than the extended version with $k_i$ copies of each resource $i\in [n]$. Indeed, the proof of Claim~\ref{clm:sep} shows that taking the values of $x_{i,j}$ from the original $\LPtrunc(\mathcal{I})$ and dividing by $k_i$ yields a feasible and hence optimal solution to the extended $\LPtrunc(\mathcal{I})$.

\begin{algorithm}
\caption{Algorithm for \Indep model} \label{alg:indep}
\begin{algorithmic}
\State Solve $\LPtrunc(\cI)$, letting $(x_{i,j})_{ i\in [n], j\in [m]}$ denote an optimal solution
\For{$j=1,\ldots,m$}
\State Input $(x_{i,j})_{ i\in [n]}$ into \Cref{alg:losslessRounding} and let $\pi_j$ be the (independently random) function it outputs
\State $\mathsf{num[j]}=0$ \Comment{counts how many arrivals of type $j$ there have been}
\EndFor
\For{$i=1,\ldots,n$}
\State $\tau_i=\sum_{j=1}^m r_{i,j} x_{i,j}/2$ \Comment{reward threshold for resource $i$}
\State $\inv[i]=1$ \Comment{starting inventory of $i$, where we have assumed that $k_i=1$}
\EndFor
\For{$\sum_{j=1}^m D_j$ queries arriving in any order} \Comment{realizations $(D_j)_{j\in[m]}$ unknown to algorithm}
\State Let $j$ denote the type of the current query
\State $\mathsf{num}[j] = \mathsf{num}[j] + 1$
\State $i=\pi_j(\mathsf{num}[j])$ \Comment{$\pi_j$ routes the $\mathsf{num}[j]$-th arrival of type $j$ to resource $i$}
\If{$i\neq\perp$ and $r_{i,j}\ge\tau_i$ and $\inv[i]=1$}
\State Match the current query to resource $i$, collecting reward $r_{i,j}$
\State $\inv[i]=\inv[i]-1$
\EndIf
\EndFor \Comment{$\mathsf{num}[j]$ will equal $D_j$ for all $j$ at the end}
\end{algorithmic}
\end{algorithm}

\Cref{alg:indep} starts by solving the truncated LP relaxation. Next, it converts the LP optimal solution $(x_{i,j})_{i\in [n], j\in [m]}$ into randomized functions $(\pi_j)_{j\in [m]}$. This crucial step is achieved using another subroutine, \Cref{alg:losslessRounding}, which we present later in Section~\ref{sec:lossless}. For each  type $j \in [m]$, \Cref{alg:losslessRounding}'s output $\pi_j: {\bb N} \rightarrow [n] \cup \{\perp\}$ indicates how to route the arriving queries of type $j$ to resources: for each $\ell=1,2,\ldots$, the $\ell$-th arrival of type $j$ is {\em routed} either to a resource $\pi_j(\ell) \in [n]$, or to a "null resource" $\pi_j(\ell)=\perp$, which is our convention for discarding that query. Routing decisions in \Cref{alg:indep} can be interpreted as "match promises" which we may or may not fulfill; once the $\ell$-th query of type $j$ is routed to $\pi_j(\ell)$, it cannot be matched with any other resource.  While $\pi_j$ does not route more than one query of type $j$ to the same resource, $\pi_j$ may discard many different type-$j$ queries by routing them to the null resource $\perp$.

As we prove below, these preliminary routing decisions of \Cref{alg:indep} reduce our original problem to the standard prophet inequality problem. Indeed, the second step of \Cref{alg:indep} matches routed queries to resources based on static thresholds, eventually "fulfilling" some matching promises. Specifically, each resource $i\in [n]$ gets matched to its first routed query that generates a reward greater or equal to the threshold $\tau_i=\frac{1}{2}\cdot(\sum_{j=1}^m r_{i,j} x_{i,j})$. By formalizing the reduction to the standard prophet inequality problem, we derive the following competitive ratio for our algorithm.

\begin{theorem} \label{thm:main1}
For the \Indep model,
\Cref{alg:indep} is a polynomial-time online algorithm that is $\frac{1}{2}$-competitive, satisfying $\displaystyle\inf_{\cI\in \Indep \cap \Adv}\frac{\ALG(\cI)}{\OFF(\cI)}\ge\inf_{\cI\in \Indep \cap \Adv}\frac{\ALG(\cI)}{\LPtrunc(\cI)}\ge\frac{1}{2}$.
\end{theorem}

A competitive ratio of $1/2$ is essentially optimal for any online algorithm, as shown next. 

\begin{proposition} \label{prop:main1tight}
For any online algorithm, $\displaystyle\inf_{\cI\in \Indep \cap \Adv}\frac{\ALG(\cI)}{\OFF(\cI)}\le\frac{1}{2}$, even for a single resource and an arbitrarily large starting inventory and expected demand.
\end{proposition}

\Cref{prop:main1tight} is proved in \Cref{pf:prop:main1tight}. That $1/2$ remains unbeatable even for arbitrarily large inventories $k_i$ stands in stark contrast to traditional online stochastic matching models, where competitive ratios approaching $1$ are achievable as $\min_i k_i\to\infty$. This difference stems from the \Indep model: even when demands $D_j$ have large means, their coefficients of variation need not diminish and converge to zero.

\paragraph{Proof of Theorem~\ref{thm:main1}.}  To establish $1/2$-competitiveness, we leverage the fact that \Cref{alg:losslessRounding} is a \emph{lossless rounding scheme} for $\LPtrunc$ for each demand type $j$---a notion that captures the following properties:
\begin{henumerate}
\item The routing by $\pi_j$ does not create contention:  if $\pi_j(\ell)\in[n]$ then $\pi_j(\ell')\neq\pi_j(\ell)$ for all $\ell'\neq\ell$.  
\item The probability of each resource $i$ being routed a query of type $j$ is \textit{exactly} $x_{i,j}$, i.e., $\prpar{\sum_{\ell =1}^{D_j} {\bb I}[\pi_j(\ell) = i]}$.  Here, the expectation is over the randomness in the function $\pi_j$ outputted by \Cref{alg:losslessRounding} in conjunction with the randomness demand $D_j$ of type $j$.
\end{henumerate}

Now, assuming that \Cref{alg:losslessRounding} is a lossless rounding scheme, \Cref{thm:main1} directly follows from known results about prophet inequalities. To explain this reduction, we take the perspective of an individual resource $i$.  By property 1, it is routed a stream of queries of {\em different} types, and it can only spend its single inventory unit on one of them because $k_i=1$. Not all types will get routed, and the ones that do can arrive in any arbitrary order. Our objective is essentially to spend resource $i$ on a type $j$ to maximize its expected reward $r_{i,j}$, knowing beforehand that:
\begin{hitemize}
\item Each type $j \in [m]$ gets routed to $i$ at most once, with probability $x_{i,j}$. This follows from properties~1-2 of lossless rounding.
\item Whether each type $j$ gets routed to $i$ is independent across types $j\in [m]$. Indeed, the random functions $\pi_j$ are constructed independently in \Cref{alg:indep}, and the demands $D_j$ are also independent by the assumption of the \Indep model.
\end{hitemize}
The above description coincides with the standard prophet inequality problem (see \citet{chawla2010multi}), with the only difference being that we compare the expected reward of our policy to the fractional value $\sum_{j=1}^m r_{i,j}x_{i,j}$ appearing in the LP objective, rather than to the usual prophet (hindsight optimal) benchmark. Fortunately, this variant has been also examined : as long as $\sum_j x_{i,j}\le 1$---as implied by our LP constraint~\eqref{ineq:truncResource} with $k_i=1$---accepting the first routed type $j$ with $r_{i,j}$ exceeding the threshold
$\tau_i=\frac{1}{2}(\sum_{j=1}^m r_{i,j}x_{i,j})$ collects reward at least $\frac{1}{2}(\sum_{j=1}^m r_{i,j}x_{i,j})$ in expectation; see \citet[Thm.~2]{ma2021dynamic}. We  include in Appendix~\ref{app:generalizations} a self-contained proof that even allows for negative correlation.
By the linearity of expectation over resources $i \in [n]$, we obtain
\begin{align} \label{eqn:proofThm1}
\ALG(\cI)\ge\sum_{i=1}^n\sum_{j=1}^m r_{i,j} x_{i,j}/2=\LPtrunc(\cI)/2\ .    
\end{align}

This concludes the proof of \Cref{thm:main1} assuming that \Cref{alg:losslessRounding} is a lossless rounding. 

\paragraph{Necessity of lossless rounding.} We devote the next section to a description and analysis of \Cref{alg:losslessRounding}. Meanwhile, let us briefly highlight the difficulty resolved by our lossless rounding---and in particular explain why naive rounding methods no longer suffice.

{If demand $D_j$ were deterministic for all types $j$, then a $1/2$-competitive algorithm would follow from existing literature, and our lossless rounding would not be needed. Indeed, one could \textit{independently} round each arrival of type $j$. Even if a resource $i$ receives multiple queries of type $j$, these routing decisions would be governed by \textit{independent} Bernoulli random variables, and a $1/2$-competitive algorithm would follow from applying a threshold-based prophet inequality for each resource $i$.

However, when $D_j$ is random, independent rounding can induce positive correlation among these Bernoulli random variables. For example, if $D_j \in \{1, 3\}$, then the second and third arrivals of type $j$ being routed to resource $i$ are both contingent on $D_j = 3$, and thus positively correlated. Prophet inequalities fail under such positive correlations. To circumvent this issue, our lossless rounding scheme ensures that \textit{at most one} query of type $j$ gets routed to resource $i$---occurring with probability exactly $x_{i,j}$.}

\section{Lossless rounding} \label{sec:lossless}

In this section, we present \Cref{alg:losslessRounding} and establish the lossless rounding properties claimed in Section~\ref{subsec:algIndep}.  The query type $j$ is fixed, as \Cref{alg:losslessRounding} is applied independently to each type; we therefore omit the subscript $j$. We still assume each resource $i$ has a single unit, i.e., $k_i=1$.

Let us first recap the underlying randomized rounding problem.
There are $n$ single-unit resources and a random number of queries $D$ whose distribution is known.
We are given probabilities $(x_i)_{i\in[n]}$ that satisfy the following inequalities
\begin{align} \label{eqn:expConstrSimple}
\sum_{i\in S}x_i &\le\bE[\min\{D,|S|\} &\forall S\subseteq[n] \ ,
\end{align}
which correspond to our exponential family of constraints~\eqref{ineq:truncDemand} when $k_1=\cdots=k_n=1$. Upon arrival, each query must be routed to up to one resource $i$, in which case $i$ is depleted and no further queries can be routed to $i$.
The process abruptly "stops" after the $D$-th query arrives, where $D$ is unknown a priori. The goal is to route the arriving queries randomly such that that each resource $i \in [n]$ receives a query with probability exactly $x_i$.

We note that among all subsets $S\subseteq[n]$ of the same cardinality $\ell\in[n]$, constraints~\eqref{eqn:expConstrSimple} are tightest if $S$ takes the $\ell$ indices $i$ with highest $x_i$ values.  Therefore, once the LP has been solved and the $x_i$ values are fixed, we can sort and re-index the resources such that $x_1\ge\cdots\ge x_n$. Consequently, the constraints~\eqref{eqn:expConstrSimple} are equivalent to
\begin{align} \label{eqn:expConstrLinear}
\sum_{i=1}^\ell x_i &\le\bE[\min\{D,\ell\}] &\forall \ell\in[n]\ .
\end{align}
Of course, we do not know the sorting of indices before solving the LP, and thus, we cannot ex-ante eliminate the exponential family of constraints.
Using the simplified constraints~\eqref{eqn:expConstrLinear}, the following examples give intuition on the randomized rounding problem and illustrate its challenges.

\begin{example}[Examples of Lossless Rounding] \label{ex:lossless}
Let $n=3$ and consider the following distribution for $D$: $\prpar{D = 1} = \frac{1}{2}$, $\prpar{D = 2} = \frac{1}{4}$, and $\prpar{D = 3} = \frac{1}{4}$.
We first illustrate that some constraints on $\bx=(x_1,x_2,x_3)$ are clearly necessary in order for lossless rounding to be possible.
For example, because the expected number of queries is $\frac12\cdot1+\frac14\cdot2+\frac14\cdot3=7/4$, the expected number of resources that receive queries, which equals $x_1+x_2+x_3$, must satisfy $x_1+x_2+x_3\le7/4$.
Additional constraints are needed---as a bad example, if $x_1+x_2>3/2$, then a lossless rounding is impossible because the expected number of queries routed to one of two resources cannot exceed $\bE[\min\{D,2\}]$, which equals $\frac12\cdot1+\frac12\cdot2=3/2$.
This is precisely the motivation behind our exponential family of constraints.
Because $x_1\ge x_2\ge x_3$, constraints~\eqref{eqn:expConstrSimple} are equivalent to
\begin{align} \label{eqn:typeRoundEg}
x_1\le1; \qquad x_1+x_2\le3/2; \qquad x_1+x_2+x_3\le 7/4\ .
\end{align}

Surprisingly, enforcing the inequalities in~\eqref{eqn:typeRoundEg} is sufficient for lossless rounding.
At first, sufficiency may seem trivial---if all of the inequalities in~\eqref{eqn:typeRoundEg} are satisfied as equality, then we can just route the first query to resource 1, the second query (if it arrives) to 2, etc. It is not difficult to see that this approach satisfies the required properties of the rounding.
However, without equality, this naive method does not work. To see why, take the example where $\bx=(3/4,2/3,1/3)$. If we always satisfy resource 1 using the first query (i.e., route the first query to resource 1 with probability~$x_1=3/4$), then the probability of being able to route any query to resource 2 is at most
\begin{align} \label{eqn:7891}
\frac{1}{4}+\Pr[D\ge 2]\cdot\frac{3}{4}=\frac{5}{8}\ .
\end{align}
To explain~\eqref{eqn:7891}, we route the first query to resource 2 with the residual probability ${1}/{4}$, and then, we can route a query to resource 2 in the future only when: (i) $D\geq 2$, and (ii) the first query was not routed to resource 2.  Event (ii) occurs independently with probability $1-{1}/{4}={3}/{4}$, resulting in the upper bound of 5/8 which is too small for the desired probability of $x_2=2/3$.

Nonetheless, there is a solution to this example $\bx=(3/4,2/3,1/3)$, using our lossless rounding scheme.  It processes the resources $i =1,2, 3$ iteratively and routes to each $i$ a randomly chosen query to the pair of ``latest arriving'' queries that are still sufficient to satisfy the probability requirement $x_i$. For example, resource $i=1$ must receive a query with probability $x_1 = 3/4$. Because $\prpar{D\geq 1} =1 \geq x_1 \geq \prpar{D\geq 2} = \frac{1}{2}$, the latest arrivals are either the first or second query. If we only routed the second query, we would not meet the required probability $x_1$ since $x_1 > \prpar{D\geq 2}$, and thus, the routing decision needs to be randomized between the first two queries. Specifically, our scheme routes the first query to resource 1 with probability 1/2, and otherwise routes the second query to resource 1 (if it arrives). In total, a query is routed to resource 1 with probability
\[
\frac{1}{2}+\frac{1}{2}\cdot\Pr[D\ge 2]~=~\frac{3}{4}~=~x_1 \ ,
\]
as desired.  After satisfying resource 1, one of the first two arriving queries is still available with sufficiently high probability to satisfy resource 2.
In particular, we can flip an independent coin for resource 2, and with probability 5/6, route to it whichever of the first two queries that did \textit{not} get routed to resource 1; otherwise, with probability 1/6, route to it the third query.  Resource 2 would get routed a query with probability
\[
\frac{5}{6}\left(\frac{1}{2}+\frac{1}{2}\cdot\Pr[D\ge 2]\right)+\frac{1}{6}\cdot\Pr[D\ge 3]~=~\frac{5}{6}\left(\frac{3}{4}\right)+\frac{1}{6}\cdot\frac14~ = ~ \frac{2}{3} ~ = ~x_2 \ ,
\]
as desired. Finally,
we route to resource 3 the first remaining query that did not get routed to either resource 1 or 2, with probability exactly $\frac{1}{3}\cdot (\frac{3}{4}\cdot \frac{1}{6} +\frac{1}{4} \cdot \frac{5}{6})^{-1} = 1$.  In the end, our lossless rounding scheme returns a random routing function $\pi$, which, in this simple example, can be interpreted as a permutation of resources to which to route the first, second, and third queries.  The permutation is 
$(1,2,3)$ with probability~5/12, $(2,1,3)$ with probability~5/12, $(1,3,2)$ with probability~1/12, and $(3,1,2)$ with probability~1/12.
\Halmos\end{example}

\begin{figure}[htb]
\center{\includegraphics[width=0.9\textwidth]
{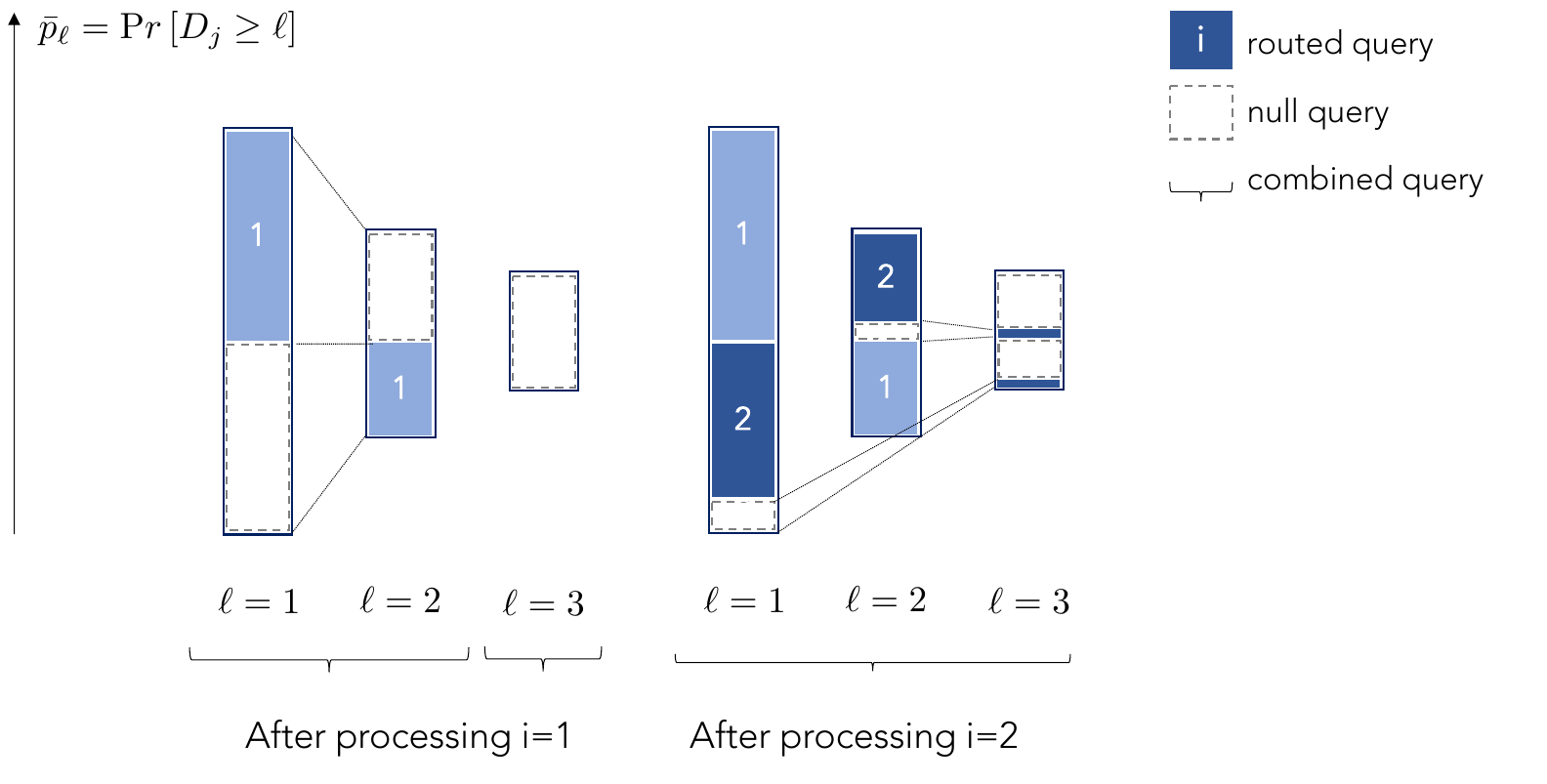}}
\caption{\label{fig:typescheme} 
Illustration of our lossless rounding for resources $\boldsymbol{i=1, 2}$ in Example~\ref{ex:lossless} with $\bx=\boldsymbol{(3/4,2/3,1/3)}$. To satisfy resource $\boldsymbol{i=1}$, we toss an unbiased coin and choose which of the first two queries is routed to $1$ (light blue shading). To satisfy resource $\boldsymbol{i=2}$, we independently toss a biased coin: with probability $\boldsymbol{5/6}$, we route the first idle query to $\boldsymbol{2}$, otherwise, we route the last query (dark blue shading). In each step, we merge two consecutive into a singly combined query, as  represented visually by the lines delimiting complementary probabilistic events.
}
\end{figure}

Our lossless rounding scheme is devised by formalizing the scheme in \Cref{ex:lossless}.   We inductively argue over $i = 1,\ldots, n$ that the {\em pair of latest arriving queries}, sufficient to satisfy the routing probability $x_i$ for resource $i$, always exists and can be computed in polynomial time. Hence, we toss a biased coin to determine which of these two queries is routed to resource $i$---the coin toss probability is calibrated to ensure $i$ receives one query with probability exactly $x_i$. The important observation is that one of the two queries is still available for the resources $i+1,i+2,\ldots$ that remain to be processed---hence, we can "merge" the two queries into a single {\em combined query} and update its arrival probability. This merging operation is visualized in Figure~\ref{fig:typescheme} for \Cref{ex:lossless}. For convenience, we assume that $x_i>0$ for all $i \in [n]$, ignoring resources that never need to receive a routed query.
We let $T$ denote an upper bound on the realization of $D$.
Hereafter, we refer to the $\ell$-th arriving query as {\em query $\ell$}, noting that it may not exist if $\ell>D$.  

\begin{algorithm}
\caption{Lossless Rounding Scheme} \label{alg:losslessRounding}
\begin{algorithmic}
\State Initialize empty list $\cL$
\For{$\ell=1,\ldots,T$} \Comment{combined queries in $\cL$ will have decreasing $\prob$'s}
\State Append to $\cL$ the combined query $\cq$ with $\cq.\start=\cq.\ennd=\cq.\free=\ell$, $\cq.\prob=\Pr[D\ge\ell]$
\EndFor
\For{$\ell=T+1,\ldots,T+n$} \Comment{add dummies for convenience}
\State Append to $\cL$ the combined query $\cq$ with $\cq.\start=\cq.\ennd=\cq.\free=\ell$, $\cq.\prob=0$
\EndFor
\State Initialize null routing $\pi(\ell)=\perp$ for all $\ell=1,\ldots,T+n$
\For{$i=1,\ldots,n$}
\State Find the \textit{last} combined query $\cq\in \cL$ for which $\cq.\prob\ge x_i$ \Comment{  exists by proof of Lemma~\ref{lem:lossless} }
\State Let $\cq'$ denote the successor of $\cq$ \Comment{$\cq.\prob\ge x_i>\cq'.\prob$}
\State Draw an independent random bit $B_i$ that is 1 with probability \Comment{lies in (0,1]}
\begin{align*}
\frac{x_i-\cq'.\prob}{\cq.\prob-\cq'.\prob}
\end{align*}
\State Update $\pi$ to route to the current resource $i$:
$
\begin{cases}
\pi(\cq.\free)=i &\text{ if $B_i=1$} \\
\pi(\cq'.\free)=i &\text{ if $B_i=0$}
\end{cases}
$
\State Replace $\mathsf{cq}$ and $\mathsf{cq}'$ with a single combined query $\mathsf{cq}"$, for which:
\begin{align*}
\cq".\start &=\cq.\start
\\ \cq".\ennd &=\cq'.\ennd
\\ \cq".\prob &=\cq.\prob+\cq'.\prob-x_i
\\ \cq".\free &=
\begin{cases}
\cq'.\free &\text{ if $B_i=1$} \\
\cq.\free &\text{ if $B_i=0$}
\end{cases}
\end{align*}
\EndFor
\end{algorithmic}
\end{algorithm}

\Cref{alg:losslessRounding} is the commented pseudocode of our lossless rounding scheme. In every iteration, \Cref{alg:losslessRounding} maintains the list $\cL$ of combined queries $\cq$, each representing an interval of merged queries $\{\cq.\start,\cq.\start+1,\ldots,\cq.\ennd\}$.
Precisely one of these queries, indicated by $\cq.\free$, has not already been routed and hence is still available, i.e., \ $\pi(\cq.\free)=\perp$, where $\perp$ is a null resource indicating availability. Note that $\cq.\free$ is random from our scheme's previous routing assignments.
The probability that query $\cq.\free$ exists is maintained by the variable $\cq.\prob = \prpar{\cq.\free\le D}$ (here, the probability is  over the randomness in both $D$ and $\cq.\free$). We say that $\cq'\in\cL$ is the \textit{successor} of $\cq\in\cL$ if $\cq'$ comes immediately after $\cq$ in list $\cL$; in each iteration, our rounding scheme merges one combined query with its successor.
Initially, the combined queries are just singletons corresponding to the original queries. Thus, we start with a list ${\cal L}$ of length $T+n$, containing each query $\cq.\start =\cq.\ennd = \ell$ with probability $\cq.\prob = \prpar{D\geq \ell}$. (Note that we initialize ${\cal L}$ with $T+n$ combined queries, although there are at most $T$ queries, to ensure that the last combined query has a successor in each step of the iterative rounding). Because in each iteration $i=1,\ldots,n$, a pair of successive combined queries are merged into a single combined query, our list ends with length $T$.

\begin{lemma} \label{lem:lossless}
Suppose $(x_i)_{i\in [n]}\in[0,1]^n$ satisfies $\sum_{i\in S}x_i\le\bE[\min\{D,|S|\}]$ for all $S\subseteq[n]$.
Then \Cref{alg:losslessRounding} is well-defined and the randomized routing function $\pi$ it returns satisfies
\begin{align*}
\Pr_{B_1,\ldots,B_n,D}\left[\sum_{\ell=1}^D\bI(\pi(\ell)=i)\right] &=x_i & \forall i\in[n]\ .
\end{align*}
\end{lemma}

\proof{Proof of \Cref{lem:lossless}.}
Denote by ${\cal L}^i$ and $\pi^i$ the (randomized) list and routing constructed by the end of iteration $i = 0,\ldots, n$. Based on the pseudocode \Cref{alg:losslessRounding}, we observe that the fields of each combined query $\cq \in \cL^i$, except $\cq.\free$, are all deterministic and independent of the realizations of the random bits $B_1,\ldots,B_i$ drawn thus far. We first establish that the following properties are preserved by induction over $i=0,\ldots,n$:
\begin{enumerate}
\item List $\cL^i$ has length $T+n-i$.  If $\cq'$ is the successor of $\cq$ in $\cL^i$, then $\cq'.\start=\cq.\ennd+1$.  The first $\cq$ in $\cL^i$ has $\cq.\start=1$.  The last $\cq$ in $\cL^i$ has $\cq.\ennd=T+n$.
\item For all $\cq\in\cL^i$, we have $\pi^i(\cq.\free)=\perp$ and the sets $I^i(\cq):=\big\{\pi^i(\ell):\ell\in[\cq.\start,\cq.\ennd]\setminus\{\cq.\free\}\big\}$
form a partition of $\{1,\ldots,i\}$, i.e., $\bigcup_{\cq\in\cL} I^i(\cq) = [i]$ and $I^i(\cq) \cap I^i(\cq') = \emptyset$ for all $\cq \neq \cq'$.
\item If $\cq'$ is the successor of $\cq$ in $\cL^i$, then $\cq'.\prob\le\cq.\prob$.  Moreover, the last $n-i$ combined queries $\cq$ in $\cL^i$ all have $\cq.\prob=0$.
\item For all $\cq\in\cL^i$, the stored probability $\cq.\prob$ satisfies 
\[
\cq.\prob = \prtwo{B_1,\ldots,B_i,D}{\cq.\free\le D} =\sum_{\ell=\cq.\start}^{\cq.\ennd}\Pr[D\ge\ell] - \sum_{i' \in I^i(\cq)}x_{i'} \ ,
\]
where recall that $B_1,\ldots,B_i$ are the random bits drawn by \Cref{alg:losslessRounding} in preceding iterations. Note that the probability $\prpartwo{B_1,\ldots,B_i,D}{\cq.\free\le D}$ is calculated with respect to the {\em ex-ante} distribution of $B_1,\ldots,B_i$, although these random bits have already been drawn at end of iteration $i$.
\item The first combined query $\cq\in {\cal L}^i$ satisfies $\cq.\prob\ge x_{i'}$ for every $i' =i+1,\ldots,n$.

\end{enumerate}
We remark that the validity of \Cref{alg:losslessRounding} follows immediately from Property~5. Indeed, it implies that at the beginning of each iteration $i=1,\ldots, n$, there exists some combined query $\cq \in {\cal L}^{i-1}$ such that $\cq.\prob \geq x_i$, as required in the execution of the ``for $i=1,\ldots,n$'' loop.

We proceed by induction over $i\in [0,n]$.  The base case $i=0$ holds by construction---it is not difficulty to verify that before the for loop in \Cref{alg:losslessRounding}, properties~1--5 are all satisfied. In particular, the first combined query $\cq \in {\cal L}^0$ satisfies $\cq.\prob = \pr{D \geq 1} = \expar{\min\{D,1\}}\geq x_{i'} $ for every $i' \in [n]$ based on the constraint with $S = \{i'\}$ in the lemma's statement.

We now consider iteration $i\in[n]$, assuming the properties hold at the end of iteration $i-1$.
Property~1 holds for $i$ from the fact that we replace the successive queries $\cq$ and $\cq'$ with a single combined query $\cq''$, which is set as $\cq''.\start = \cq.\start$ and $\cq''.\ennd = \cq'.\ennd$.
Property~2 holds because $\pi^i(\ell)$ is set to $i$ for one of $\ell\in\{\cq.\free,\cq'.\free\}$, while $\cq''.\free$ is set to the other one.
The first part of property~3 holds because $\cq''.\prob=\cq.\prob+\cq'.\prob-x_i$ lies in $[\cq'.\prob,\cq.\prob)$, as implied by $\cq.\prob\ge x_i>\cq'.\prob$.
The second part of property~3 holds because at most one of the trailing queries $\cq$ with $\cq.\prob=0$ can get merged during iteration $i$ (as the successor $\cq'$).
Finally, to show the first part of property~4, we derive:
\begin{eqnarray*}
&&\prtwo{B_1,\ldots,B_i,D}{\cq''.\free\le D} \\
&&\quad =\pr{B_i=1}\cdot \prtwo{B_1,\ldots,B_{i-1},D}{\cq'.\free\le D}+(1-\pr{B_i=1}) \cdot\prtwo{B_1,\ldots,B_{i-1},D}{\cq.\free\le D}
\\ &&\quad =\frac{x_i-\cq'.\prob}{\cq.\prob-\cq'.\prob}\cq'.\prob+\frac{\cq.\prob-x_i}{\cq.\prob-\cq'.\prob}\cq.\prob
\\ &&\quad =\cq.\prob+\cq'.\prob-x_i
\\ &&\quad =\cq''.\prob \ ,
\end{eqnarray*}
where the first equality holds by independence of $B_i$,
the second equality holds by applying the induction hypothesis from iteration $i-1$,
and the last equality follows from the definition of $\cq''.\prob$ for the combined query $\cq''$ that replaces $\cq$ and $\cq'$.

To show the second part of property~4, we note that
\begin{align*}
\cq''.\prob
&=\sum_{\ell=\cq.\start}^{\cq.\ennd}\Pr[D\ge\ell] - \sum_{i' \in I^{i-1}(\cq)}x_{i'}
+\sum_{\ell=\cq'.\start}^{\cq'.\ennd}\Pr[D\ge\ell] - \sum_{i' \in I^{i-1}(\cq')}x_{i'} - x_i \\
&=\sum_{\ell=\cq.\start}^{\cq.\ennd}\Pr[D\ge\ell] - \sum_{\MyAtop{\ell\in [\cq.\start,\cq.\ennd]:}{\pi^{i-1}(\ell)\neq\perp}}x_{\pi^{i-1}(\ell)}
+\sum_{\ell=\cq'.\start}^{\cq'.\ennd}\Pr[D\ge\ell] - \sum_{\MyAtop{\ell\in [\cq'.\start,\cq'.\ennd]:}{\pi^{i-1}(\ell)\neq\perp}}x_{\pi^{i-1}(\ell)} - x_i
\\ &=\sum_{\ell=\cq''.\start}^{\cq''.\ennd}\Pr[D\ge\ell] - \sum_{\MyAtop{\ell\in [\cq''.\start,\cq''.\ennd]:}{\pi^i(\ell)\neq\perp}}x_{\pi^i(\ell)}
\\ &=\sum_{\ell=\cq''.\start}^{\cq''.\ennd}\Pr[D\ge\ell] - \sum_{i' \in I^i(\cq'')}x_{i'} \ ,
\end{align*}
where the first equality holds from property~4 of the induction hypothesis and the definition of $\cq''.\prob$. The second equality follows from property~2 of the induction hypothesis. The third equality holds because at the end of iteration $i$, we have $\pi(\ell)=i$ for one of $\ell\in\{\cq.\free,\cq'.\free\}$ and $\pi(\ell)=\perp$ for the other.

It remains to establish Property~5.
To this end, observe that for the first combined query $\cq\in\cL^i$,
\begin{align}
\cq.\prob
=\sum_{\ell=1}^{\cq.\ennd}\Pr[D\ge\ell] - \sum_{i' \in I^{i-1}(\cq)} x_{i'} 
=\bE[\min\{D,\cq.\ennd\}] - \sum_{i' \in I^{i-1}(\cq)} x_{i'}
\geq x_i \ ,\nonumber 
\end{align}
where the first equality holds because of properties~1 and~4 of the induction hypothesis, and
the second equality follows from re-writing $\bE[\min\{D,\cq.\ennd\}]=\sum_{\ell=1}^{\cq.\ennd}\Pr[D\ge\ell]$.
At this point, we invoke the exponential constraints assumed in the statement of the \namecref{lem:lossless} with $S=\{i'\in I^{i-1}(\cq)\cup\{i\}$, to infer that $\bE[\min\{D,|S|\}] \geq \sum_{i' \in I^{i-1}(\cq) \cup \{i\}} x_{i'}$, and the last inequality holds because $|S|=|I^{i-1}(\cq)|+ 1 = \cq.\ennd$.

To conclude the proof of \Cref{lem:lossless}, we argue that every resource $i$ is routed a query with probability $x_i$ after the last iteration.  Property~2 ensures that each resource $i$ receives exactly one query, during iteration $i$, among $\cq.\free$ and $\cq'.\free$, where $\cq \in {\cal L}^i$ is the combined query identified in iteration $i$ of the ``for'' loop.  Now, the probability that this query arrives, or equivalently $\Pr[\sum_{\ell=1}^D\bI(\pi(\ell)=i)]$, is equal to
\begin{align*}
&\Pr[B_i=1]\Pr_{B_1,\ldots,B_{i-1},D}[D\geq \cq.\free]+\left(1-\Pr[B_i=1]\right)\Pr_{B_1,\ldots,B_{i-1},D}[D \geq \cq'.\free]
\\ &=\frac{x_i-\cq'.\prob}{\cq.\prob-\cq'.\prob}\cq.\prob+\frac{\cq.\prob-x_i}{\cq.\prob-\cq'.\prob}\cq'.\prob
\\ &=x_i
\end{align*}
where we again use the independence of $B_i$ and Property~4.  
\Halmos\endproof

The lossless rounding established in \Cref{lem:lossless} implies our main result \Cref{thm:main1} for the \Indep model, as explained in Section~\ref{subsec:algIndep}.  We end with a remark about the lossless rounding.
\begin{remark} \label{rem:losslessInvarianceAndOrder}
As noted in the proof of \Cref{lem:lossless}, the sequence in which \Cref{alg:losslessRounding} combines queries is invariant to the realizations of the bits $B_i$.
Indeed, $B_i$ does not affect the stored fields $\start$, $\ennd$, and $\prob$, which is what determines the choice of $\cq$ and $\cq'$ in each iteration $i=1,\ldots,n$. Additionally, we note that our lossless rounding works regardless of how we permute the resource indices $i$. In particular, it does not require us to sort the resources by non-increasing $x_i$'s as in \Cref{ex:lossless}.  Different processing orders over resources in  \Cref{alg:losslessRounding} may result in different distributions for the random routing $\pi$.
\end{remark}

\section{Simulations} \label{sec:simulations}

We compare online matching algorithms on
synthetic arrival sequences generated according to $\Indep$ and $\Correl$
and real arrival sequences from e-commerce customer order data.
In particular, we test algorithms of the form ``LP relaxation'' plus ``online rounding scheme,'' whereby we compare our new LP relaxations to the naive fluid LP, and benchmark our lossless rounding against standard online randomized rounding schemes. 

\subsection{Synthetic instances}

\paragraph{Distributions.}
We generate instances with the following demand distributions:
\begin{itemize}
\item N($\mu,\sigma$): a Normal distribution with mean $\mu$ and standard deviation $\sigma$;
\item Pois($\mu$): a Poisson distribution with mean $\mu$;
\item Unif($0:2\mu$): a uniform distribution over integers $0,1,\ldots,2\mu$ (assuming $\mu$ is integer);
\item Exp($1/\mu$): an Exponential distribution with mean $\mu$;
\item $5\mu\cdot\mathrm{Ber}(1/5)$: a weighted Bernoulli distribution that is $5\mu$ with probability 1/5 and 0 otherwise.
\end{itemize}
All of these distributions have the same mean $ \mu$.\footnote{The Normal and Exponential distributions are truncated and discretized to take non-negative integer values, so the mean is not exactly $\mu$ but very close.} The Normal distribution is the only two-parameter model that allows us to disentangle mean from standard deviation; we consider four different variants with $\sigma\in\{0.1\mu,0.3\mu,0.5\mu,\mu\}$, resulting in 8 distinct distributions in total.

\paragraph{Generating sequences of queries.}
Under the \Indep model, we select one of the preceding distributions and then draw each demand $D_j$ independently from the same distribution.
Under our \Correl model, we select one of these distributions to draw the total demand $D$, and then draw a probability vector $\mathbf{p}=(p_j)_{j\in[m]}$ uniformly\footnote{To do this, we draw independent $\mathrm{Exp}_1$ random variable $E_j$ for each $j\in[m]$ and then set $p_j$ proportional to $E_j$.} from the simplex to generate $(D_j)_{j\in[m]}$ from a $\mathrm{Multinomial}(D,\mathbf{p})$ distribution.
Having drawn each demand vector $\bD$, we construct a sequence of queries by randomly shuffling the arrival order because it is unclear how to determine an adequate "adversarial" order.

\paragraph{Supply, demand, and rewards.}
We fix the number of demand types to be $m=10$.
We consider distributions with mean $\mu=10$ in our \Indep model, so that the total expected demand---equivalently, the average length of the sequence---is $m\cdot \mu=100$.
We consider distributions with mean $\mu=100$ in our \Correl model, again resulting in a total expected demand of 100.
We balance supply with demand: we evenly distribute 100 units of inventory across $n$ supply locations.
For \Indep we fix $n=10$, but for \Correl we investigate performance as $n$ varies in $\{1,2,5,10\}$.
We draw random reward values $r_{i,j}$ independently and uniformly from [0,1] for each $i\in[n]$ and $j\in[m]$.

\paragraph{Summary: Array of synthetic instances.}
For \Indep, we consider 8 demand distributions and generate 200 draws of the reward matrix for each distribution, resulting in a total of $8\times 200=1600$ instances.
For \Correl, we generate 200 draws of the reward matrix and probability vector $\mathbf{p}$ for each distribution and each $n\in\{1,2,5,10\}$, resulting in a total of $8\times 4\times 200=6400$ instances.

\subsection{Real-data instance generation}

\paragraph{Arrival sequences.}
We extract arrival sequences from a large fashion retail platform, as described in \Cref{sec:myntra}. For the $100$ most frequently sold SKUs, we consider the sequence of customer orders for that SKU from different locations, where locations are aggregated at the destination fulfillment center We treat each month as a separate arrival sequence, noting that this is a plausible approximate replenishment period for SKUs and hence represents one time horizon (see \Cref{sec:myntra}). Over the year, each SKU yields up to $12$ sequences---one per month in which it was sold---with average length in the thousands, an order of magnitude higher than the synthetic arrival sequences.

\paragraph{Supply, demand, and rewards.} 
Like in our synthetic setup, we vary $n\in\{1,2,5,10\}$ and draw random rewards $r_{i,j}$ uniformly from [0,1] for all $i\in[n]$ and $j\in[m]$.
We note that for a given SKU, the number of demand types $m$ could be less than 10, if that SKU is only demanded from a subset of the 10 locations.
We take the same approach of balancing supply with expected demand:
for each SKU, we compute its average total demand (totaling over $m$ locations, averaging over up to 12 months) and evenly distribute that amount of starting inventory across $n$ supply locations.

\paragraph{Summary: Array of real-data instances.}
We separately consider the online matching problem for each SKU, and an instance is specified by the (up to 12) arrival sequences for that SKU.
For each combination of the 100 SKUs and 4 values of $n$, we draw the reward matrix 20 times, resulting in a total of $100\times 4\times 20=8000$ instances.

\subsection{Algorithms}
Each algorithm we test combines a primal LP relaxation with a randomized rounding scheme.
The LP is solved to prescribe frequencies with which each (query type-location) pair should be matched,
while the rounding converts these frequencies into online matching decisions.
We provide short descriptions here, with additional justification for using these methods and implementation details found in \Cref{sec:simulationsSupplement}.

\paragraph{LP relaxations.}
We consider 4 LPs whose optimal solutions guide the design of online algorithms:
\begin{enumerate}
\item \textbf{Fluid LP}: We compute an optimal solution $(x_{i,j})_{i\in[n],j\in[m]}$ of the LP defined by \eqref{lp:start}--\eqref{lp:end}.
\item \textbf{Truncated LP}: We compute an optimal solution $(x_{i,j})_{i\in[n],j\in[m]}$ of the LP defined by \eqref{lpTrunc:start}--\eqref{lpTrunc:end}.
\item \textbf{Conditional LP}: We compute an optimal solution $(y^t_{i,j})_{i\in[n],j\in[m],t\in[T]}$ of the LP defined by \eqref{lp:condObj}--\eqref{lp:condEnd}, noting that $T$ is specified as an upper bound on the total demand (length of arrival sequences).
\item \textbf{Offline LP}: We use an offline LP that computes the optimal matching ex-post for each realized sequence, and then average the resulting matchings over many samples of $(D_j)_{j\in[m]}$.  This is the most common approach in existing literature to integrate knowledge of the entire demand distribution into the LP (see \Cref{sec:simulationsSupplement} for further details).
\end{enumerate}

\paragraph{Rounding schemes.} We consider 4 methods for rounding LP solutions into online decisions.
\begin{enumerate}
\item \textbf{Independent Rounding (IR)}: For each arriving query, the algorithm observes its type $j$ and then independently samples a resource $i$ (possibly $i=\perp$) according to a probability $q_{i,j}$ based on the LP solution for type $j$.
For the Fluid, Truncated, and Offline LP's, the probabilities $q_{i,j}$ are stationary over time; for the Conditional LP, they are not (see \Cref{sec:simulationsSupplement} for details).  The arriving query is matched to the sampled $i$ if $i\neq\perp$ and $i$ has remaining inventory.
\item \textbf{In-stock Independent Routing (In-stock IR)}: While independent rounding is theoretically convenient, empirical performance improves by choosing an "in-stock" resource whenever possible (i.e., we choose a resource $i\neq\perp$ with non-zero remaining inventory). To that end, we sample only from resources $i\neq\bot$ with positive inventory, with probability proportional to $q_{i,j}$. For instance, if $n=3$ with $q_{1,j}=q_{2,j}=q_{3,j}=1/4$ and resource $2$ is out of stock, the algorithm routes to resource $1$ or $3$ with equal probability.

\item \textbf{Lossless Rounding (LR)}: We initialize a random routing $\pi_j$ for each query type $j$, which requires splitting all resources $i$ into $k_i$ units indexed by $u\in[k_i]$. Notice that we only need to solve the LPs once before splitting, because for any  $j \in [m]$, if $(x_{i,j})_{i\in[n]}$ satisfies the exponential family of constraints~\eqref{ineq:truncDemand} in $\LPtrunc$ before splitting, then setting $x_{i,j,u}:=x_{i,j}/k_i$ for each unit $u\in[k_i]$ of resource $i\in[n]$ satisfies the exponential family of constraints post-splitting (see the proof of Claim~\ref{clm:sep}).  Therefore, we implement \Cref{alg:losslessRounding} as described on the input probabilities $(x_{i,j}/k_i)_{i\in[n],u\in[k_i]}$.  Despite splitting the units as desired for \Cref{alg:losslessRounding}, it is logical to pool them when we run the online routing process: we assume that the routing $\pi_j$ maps to a resource $i$ instead of a specific unit $u\in[k_i]$ of resource $i$, effectively pooling all inventory units corresponding to the same resource.  This pooling approach improves empirical performance. Note that the resources can be processed in any order in \Cref{alg:losslessRounding}, as discussed in \Cref{rem:losslessInvarianceAndOrder}---thus, we randomly shuffle the units of the different resources before processing our lossless rounding. 
Having initialized the routing function $\pi_j$ for each $j\in[m]$, the algorithm simply routes the $\ell$'th arrival of each type $j$ to resource $\pi_j(\ell)$, for $\ell=1,2,\ldots$.  Each arriving queries is matched to $\pi_j(\ell)$ if and only if $\pi_j(\ell)\neq\perp$ and at that point in time, resource $\pi_j(\ell)$ still has remaining inventory.

\item \textbf{In-stock Lossless Rounding (In-stock LR)}: We run the same lossless rounding algorithm LR, initializing the routing functions $(\pi_j)_{j\in[m]}$ using \Cref{alg:losslessRounding}. However, we avoid rejecting incoming queries when the routing function points to no resource ($\pi_j(\ell) = \bot$) or when resource $\pi_j(\ell)$ has stocked out. In such cases, we increment $\ell$ by $1$ and attempt to match with $\pi_j(\ell+1)$. We repeat this process until we successfully match the arriving query to an "in-stock" resource or until $\ell$ exceeds the maximum number of units in the support of $\pi_j$.
\end{enumerate}

\paragraph{Summary: Algorithms.}
We implement IR and In-stock IR for each of the 4 LP relaxations described previously.
To implement LR and In-stock LR, constraints~\eqref{ineq:truncDemand} in $\LPtrunc$ must be satisfied, which is true only for the solutions of the Truncated and Offline LPs (see \Cref{sec:simulationsSupplement}). Therefore, in total, we consider 12 suitable combinations of LP relaxations and  online rounding schemes.

We note that our set-up does not account for sampling error---the algorithms, in particular the LP relaxations they use, have access to the demand distribution on synthetic instances and to the empirical demand distributions (formed by up to 12 sequences) on real-data instances.

\subsection{Results}

For each of the 1600 \Indep and 6400 \Correl synthetic instances, we draw 20 arrival sequences and use the same sequences to evaluate all algorithms, for variance reduction.
For each of the 8000 real-data instances, we use the (up to 12) actual order sequences to evaluate all algorithms.
Because the algorithms are randomized, we re-run each algorithm 20 times on each arrival sequence and take the average performance.
We then average such average performances over all arrival sequences associated with any given instance, and divide by the expected offline value $\OFF(\cI)$
to define the competitive ratio of an algorithm on that  instance $\cI$.

\begin{figure}
\centering
\includegraphics[width=\textwidth]{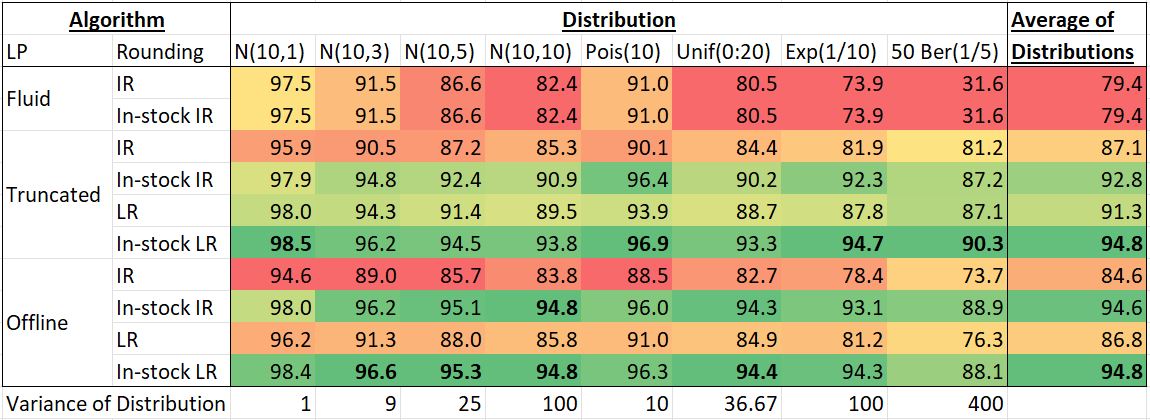}
\caption{INDEP instances: we plot the competitive ratio of algorithms (measured by \% of Offline) under different demand distributions.  A color gradient compares performance in each column (distribution), and the best number in each column is \textbf{bolded}.}
\label{table:simulations}
\end{figure}

\paragraph{Results on \Indep instances.}
We report the average competitive ratio over the 200 instances with each demand distribution, for 10 different algorithms, omitting the two based on the Conditional LP which is for the \Correl model.
We report our results in \Cref{table:simulations}.

In \Cref{table:simulations},
our Lossless Rounding improves over Independent Rounding almost uniformly, and
In-stock LR also improves over In-stock IR almost uniformly.
The best-performing algorithm for any distribution is In-stock LR using either the Truncated LP or Offline LP, and the best-performing algorithms on average are also these two.
We add that with our separation oracle\footnote{We implement this as a lazy callback in JuMP \citep{dunning2017jump}}, our Truncated LP despite being "exponentially-sized" solves an order of magnitude faster in practice than the Offline LP, taking an average running time of \textbf{0.027s} per instance instead of \textbf{0.347s} per instance.\footnote{This is assuming the Offline LP uses $M=100$ samples, which appears needed to model the true distributions.  To match the running time of the Truncated LP we would have to reduce $M$ to 5, but in that case its performance plummets.}
Finally, we note that higher distributional variance generally reduces the competitive ratio of all algorithms, but particularly brings out the disadvantages of Fluid LP. Another subtle ``disadvantage'' of the Fluid LP is that it always results in deterministic solutions, where there is no added benefit from In-stock IR relative to IR.\footnote{On all \Indep instances, $n=m$ and $k_i=\bE[D_j]=10$ for all $i\in[n]$ and $j\in[m]$, which implies that the Fluid LP always returns a solution where $x_{i,j}\in\{0,10\}$ for all $i,j$.
Consequently, the Fluid-based algorithm is deterministic, and In-stock IR is equivalent to IR.}

\begin{figure}
\centering
\includegraphics[width=\textwidth]{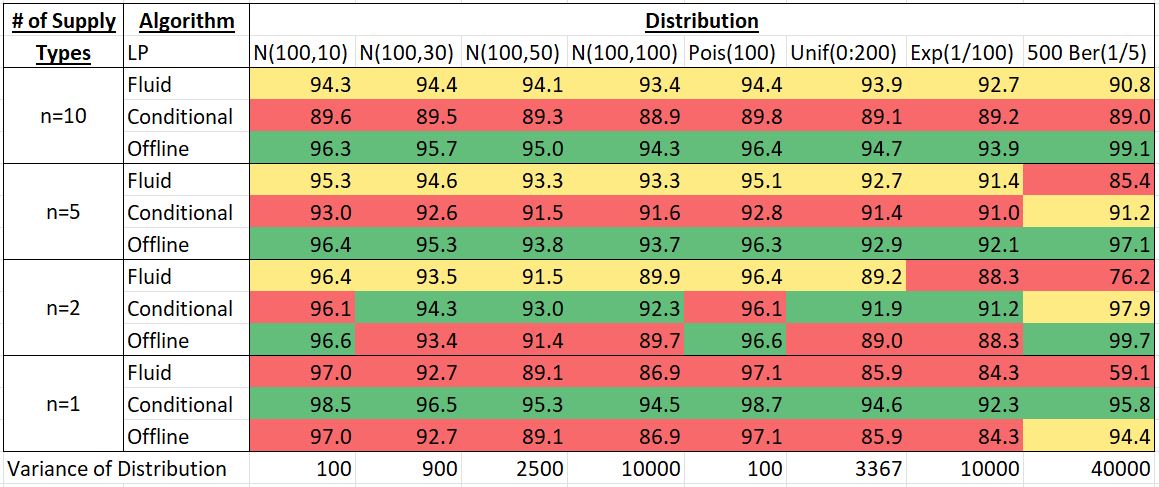}
\caption{CORREL instances: all algorithms use In-stock Independent Rounding.  We plot the ompetitive ratio of algorithms (measured by \% of Offline) under different demand distributions and different values of $n$.  A color gradient compares the performance of these three LP's, for each distribution and value of $n$.}
\label{table:simulations2}
\end{figure}

\paragraph{Results on \Correl instances.}
We report the average competitive ratio over the $200$ instances for each demand distribution and each value of $n\in\{1,2,5,10\}$. We focus on three algorithms: In-stock IR with the Fluid, Conditional, and Offline LPs. We omit the Truncated LP and Lossless Rounding because the corresponding algorithms are defined for the \Indep model. For readability, we also omit algorithms using standard IR, as their performance is uniformly worse than their In-stock IR counterparts. We report our results in \Cref{table:simulations2}.

In \Cref{table:simulations2}, our Conditional LP performs worse than the other LPs when $n=10$, but this reverses sharply as $n$ decreases to $1$. When $n=1$, the algorithm's decision is purely whether to accept or reject the incoming query---is the reward weight large enough to be worth accepting? By contrast, when $n=10$, the algorithm must decide among multiple matching options---which resource type to assign? The strong performance for $n=1$ aligns with the intuition from our \Cref{eg:dynamicPricing} in Appendix~\ref{sec:toyExamples}, showing that our Conditional LP excels at solving the accept/reject problem when the total number of arrivals is uncertain. However, our simulations suggest that this LP is less effective for matching decisions, even when the ground truth is \Correl. Nonetheless, the distributions with higher variance (N(100,50), N(100,100), Unif(0,200), and Exp(1/100)) tend to make our Conditional LP look better in comparison to other LP's.

\begin{figure}
\centering
\includegraphics[width=0.5\textwidth]{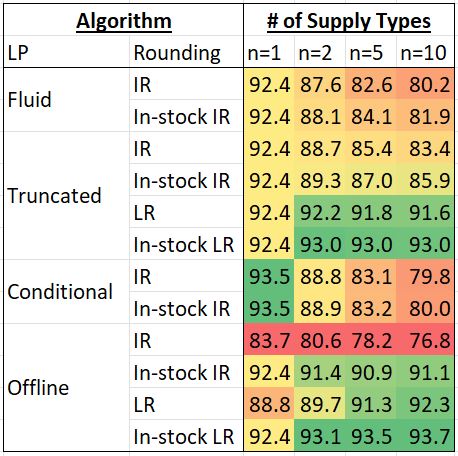}
\caption{Real-data instances: we plot the competitive ratio of algorithms (measured by \% of Offline) under different values of $n$.  A color gradient compares performance in each column.}
\label{table:simulations3}
\end{figure}

\paragraph{Results on real-data instances.}
We report the average competitive ratio of the 12 algorithms over the 2000 real-data instances (i.e., 20 instances for each of the 100 SKUs) for each value of $n\in\{1,2,5,10\}$, in \Cref{table:simulations3}.

On such real-data instances, Lossless Rounding really shines. Previously in \Cref{table:simulations}, LR of the Offline LP performed markedly worse than In-stock IR of the Offline LP; here in \Cref{table:simulations3}, LR often outperforms In-stock IR. We note that demand variance is extremely high in these real-world arrival sequences---for $99.8\%$ of (SKU, demand location) pairs, sample variance exceeds sample mean (see \Cref{sec:myntra}). Like in \Cref{table:simulations}, In-stock LR still performs best in \Cref{table:simulations3}, but the marginal impact of Lossless Rounding is now much greater. We also observe that the Conditional LP is best when $n=1$, consistent with our previous findings. When $n=1$, the Fluid, Truncated, and Offline LPs produce very similar solutions, and hence their performances are nearly identical.

\paragraph{Summary of numerical findings across instances.} Our algorithmic results provide two types of practical prescriptions for online matching systems: (1) to use Truncated/Conditional LP relaxations in place of Fluid/Offline LP relaxations, and (2) to use Lossless Rounding in place of Independent Rounding (with or without the "in-stock" improvement). To elaborate,
while the Conditional LP is beneficial in the special case $n=1$, our Truncated LP emerges as the most effective LP relaxation method overall: it achieves robust performance comparable to the Offline LP for all rounding schemes, while being faster to solve by an order of magnitude. 

Meanwhile, our Lossless Rounding (LR) scheme provides essentially a uniform improvement over the classical Independent Rounding scheme and runs equally fast, justifying its implementation despite the additional complexity. Also striking is that its improvement can be of a similar level as that of restricting to "in-stock" resources; the latter  requires real-time information about inventory levels (or at least stock-outs) and may be infrastructure-wise harder to implement.

It is noteworthy that our algorithmic contributions show the largest improvements on the real-data instances, not on instances generated under our \Indep and \Correl models. Therefore, they should be viewed as general data-driven tools for online matching, not algorithms tied to our specific models. We believe these benefits stem from two key phenomena. First, both our Truncated and Conditional LPs produce fractional solutions that account for high variance in the total arrivals of a demand type, unlike the Fluid LP. Second, our algorithms introduce limited adaptivity to the realized demand. For example, our Conditional LP's fractional solution changes over time as more arrivals occur---a nonstationary behavior absent even in the Offline LP. Similarly, even when the LP fractional solution (from the Truncated or Offline LP) is time-invariant, our Lossless Rounding converts it into routing decisions that evolve dynamically over time, generally prioritizing the largest fractional values first while avoiding repeated routing to the same resource. These benefits are illustrated by simple examples in \Cref{sec:toyExamples}.

\section{Summary and Future Directions}

We propose a new framework to capture correlated arrivals in online stochastic matching, that first models the distribution of the vector $\bD=(D_1,\ldots,D_m)$ containing the total arrivals of each query type, and then interleaves the $D_1+\ldots+D_m$  arrivals in some order (e.g., adversarial or random).  By contrast, existing frameworks
assume that the types of arriving queries are drawn independently over time and, therefore, fail to capture high variance in the types' demands.

We provide numerical evidence that high variance is present in e-commerce order fulfillment settings (\Cref{sec:empVal}) and then study two specific distribution classes for $\bD$, dubbed \Indep and \Correl, both of which allow for correlated types over time but differ in whether the random shocks are idiosyncratic or common across types (\Cref{sec:preliminaries}).
We show that directly applying the commonly used fluid relaxation leads to poor decisions, and then derive  tighter LP relaxations, the truncated LP and conditional LP, which better represent the matching decisions for the \Indep and \Correl models, respectively.
Next, we devise online matching algorithms based on these relaxations that yield strong worst-case performance guarantees for the \Indep model under the adversarial arrival order (\Cref{sec:indep}) and for the \Correl model under the uniformly random arrival order (\Cref{sec:correl}).
Our result for \Indep can also be extended to a sample-based setting (\Cref{app:demandError}), and our result for \Correl can be extended to Network Revenue Management  (\Cref{app:stochastic_gen}).
From a technical standpoint, our main contribution is a lossless rounding scheme for \Indep that, given any type $j$, routes its arriving customer queries to distinct resources while preserving marginal matching rates, without knowing the realized demand $D_j$ in advance (\Cref{sec:lossless}).  In additional to their theoretical properties, our LP's and rounding scheme exhibit excellent performance in simulations on synthetic and real data (\Cref{sec:simulations}).  We explain some of these empirical improvements using toy examples (\Cref{sec:toyExamples}).

The limitations of this work suggest several directions for future research.
First, our analysis of the e-commerce order fulfillment data sets only illustrates the limitations of standard online matching models, but it does not directly support the capability of the $\Indep$ or $\Correl$ models to  provide a better fit to demand data.
Doing so would require a more extensive statistical analysis that is beyond the scope of this work.
Second, our paper is restricted to online matching and resource allocation, but our general framework for capturing correlated arrival sequences can potentially be extended to other stochastic optimization problems.
Finally, a very concrete open problem from our paper is whether our 1/2-competitive guarantee for $\Indep\cap\Adv$ can be improved under $\Indep\cap\Rand$, where the arrival order is random.

\ACKNOWLEDGMENT{
The authors thank Rajan Udwani for sharing insights that helped articulate the benefit of lossless rounding over  naive rounding, Daniela Saban for providing comments on an early version of the paper that significantly improved the presentation, Kangning Wang for pointing us to the reference \citet{alijani2020predict}, and Jake Feldman for helping improve explanations in the paper.
}

\bibliographystyle{informs2014} %
\bibliography{bibliography} %

\clearpage

\begin{APPENDICES}
\crefalias{section}{appendix}
\crefalias{subsection}{appendix}

\section{Examples of Demand Data Sets}
In this section, we conduct simple analyses to understand the extent to which the assumption of serial independence (SI), standard in online stochastic matching, is valid on real demand data. We consider two distinct  data sets, a publicly available data set from the online retailer JD.com \citep{shen2020jd} and a private data set from a fashion e-commerce platform. In both settings, the platform needs to fulfil customer orders for various products by dispatching them from different warehouses to different locations. Because formally testing the serial independence (SI) assumption is not straightforward, we report various model-free and regression-based analyses. Recall that a consequence of the serial independence assumption is that $\var(D_j)\le\bE[D_j]$ for the demands of each type $j$. In the e-commerce context, each $j$ corresponds to a certain type of customer request based on the selected product and the closest warehouse to the customer location. We find that $\var(D_j)\le\bE[D_j]$ does not hold for a majority of types when comparing the sample mean and sample variance of the realized demand aggregated over a plausible replenishment horizon. In addition, we test for correlations in the panel data set formed by daily demand per type. We estimate fixed effects regressions and still observe that the sample variance of residual errors over query types often exceeds the corresponding sample mean demand.

\label{sec:empVal}
\subsection{JD.com data} \label{sec:empValJD}

\paragraph{Context and data description.} 
The online matching problem faced by JD.com consists of dynamically dispatching customer orders containing a stock-keeping unit (SKU) from various locations to different fulfillment centers (that still have inventory of that SKU) over time. Higher rewards are obtained when orders are dispatched to nearby distribution centers, resulting in an edge-weighted online matching problem. 

The JD.com data provides customer orders throughout China from a single product category in March 2018. We construct a dataset with daily demand level for each (SKU-location) combination. We focus our analysis on the 40 highest-selling SKUs and 40 largest locations, determined by the destination fulfillment center in China. When choosing the 40 highest-selling SKUs, we eliminate any SKU that had zero demand on any day because they may suggest an inventory stockout.  We otherwise assume that enough inventory was available so that the observed sales coincides with the true demand. Second, we remove the first day of sales, which we found to be three times higher than an average day due to promotions.
After this processing, we let $D^t_j$ denote the demand for a type $j$ on a day $t$, where a type $j$ refers to a (SKU-location) combination.

\paragraph{Model-free evidence.} We compare the sample mean and sample variance for the demand random variable $D_j$ across types. In stochastic matching models, the time horizon represents the duration between consecutive inventory replenishments. Thus, since our analysis focuses on highest-selling items, we consider this duration to be equal to be one week in JD.com context: a realization of the demand $D_j$ is the total amount that customers ordered in a week, with the type $j$ referring to a particular SKU and a particular location. Consequently, we aggregate our demand data set $D_j^t$ at the weekly level. Each week $w$ yields a sample of $D^w_j$ for every (SKU, location)-combination $j$.
We evaluate the sample mean and sample variance of $D^w_j$ for each type $j$. In the resultant scatter-plot \Cref{fig:high_varJ}, we visualize the log-ratio of sample variance to sample mean (y-axis) as a function of the sample mean (x-axis).

The property $\var(D_j)\le\bE[D_j]$ would imply that the dots in the scatter-plot \Cref{fig:high_varJ} are below the $y=0$ line with high probability. By contrast, we find that this is not the case and sample variance is greater than sample mean for a majority of types ($85\%$), often by orders of magnitude.  Only for a minority of (SKU, location)-combinations
is the empirical variance lower than the empirical mean.  This suggests that  standard online stochastic matching models might not accurately reflect the stochastic demand faced by JD.com. Counter-intuitively, the property $\var(D_j)\le\bE[D_j]$ appears to hold for SKUs $j$ with small mean demand rather than large mean demand. Considering the fact that the sample mean and sample variance are estimated from only 1 month of weekly sales data and therefore are noisy, we simulate this experiment on synthetic data where the demand is Poisson-distributed, and thus, $\var(D_j)\le\bE[D_j]$. We find that the probability that sample variance exceeds sample mean is of approximately $39\%$. Hence, the noise in our estimates cannot explain the fact a majority of types have sample variance above the $y=0$ line.

\begin{figure}[htb!]
    \centering
        \includegraphics[width=0.8\textwidth]{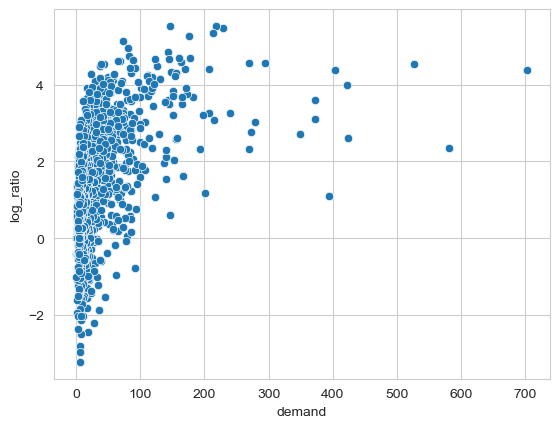}
    \hfill  
    \caption{Log-ratio mean-variance plots for the JD.com data set. Each dot represents a (SKU, location)-type $j$. The $x$-coordinate is the sample mean demand for that SKU from that location, and the $y$-coordinate is the natural logarithm of the ratio of sample variance to the sample mean.}\label{fig:high_varJ}
\end{figure}

Our analysis suggests that demand data violates the basic SI assumption. That said, the variance of SKU-level demands can be explained by several  factors, including ones that can be anticipated by the retailer in their forecasts. It is plausible that after controlling for contextual factors like marketing promotions and calendar events, the "unexplained" variance in demand might be lower than what is inferred in Figure~\ref{fig:high_varJ}. Another important limitation of our analysis is that we only have one month of sales data and one product category, which is a restriction of the JD.com data. Thus, we conduct a similar analysis in the next subsection using a data set with a longer time window and more detailed information.

\subsection{Fashion retail data} \label{sec:myntra}

\paragraph{Context and data description.} We replicate the analysis of Appendix~\ref{sec:empValJD} on a different data set, describing customer order fulfillment for a major fashion e-commerce platform in India. The online matching problem is similar to that of JD.com but the data and context are very different. While our analysis of~\Cref{sec:empValJD} was restricted to one product category over 1 month, the raw data describes customer orders from 11,800 location IDs over 12 months for 1,256,241 distinct SKUs across all categories. Products are dispatched from one of 14 warehouses.
We do not know which fulfillment center is the closest to the customer requests. That said, it is plausible that the platform prefers to dispatch from the closest warehouse, unless inventory is low or depleted, in which case it might choose a further warehouse. Similarly to Appendix~\ref{sec:empValJD}, we treat warehouse ID as our unit of geographic aggregation. That is, customer requests are classified into demand types, each corresponding to a SKU and the closest warehouse.

Similarly to~\Cref{sec:empValJD}, we construct a panel dataset with demand level $D_j^t$ for each (SKU-location) combination (or type) $j$ on each day $t$.  We drop all orders that have been subsequently cancelled, after which we focus our analysis on the 100,000 highest-selling types. SKUs in fashion retail may have short product cycles, hence we truncate the dataset to only consider any given SKU from the moment it has recorded at least one transaction until its last recorded transaction.   Finally, although customers might order multiple units of the same SKU, the demand $D_j^t$ is the count of unique customer orders on day $t$, rather than the total ordered quantity.
This assumption is conservative, as large order quantities mechanically inflate the variance-mean demand ratio. Our final data set $D_j^t$ comprises 16,046,034 demand observations for 100,000 distinct types $j$ with 77,938 SKUs and 12 warehouse locations.

\begin{figure}[!htb]
    \centering
    \begin{subfigure}{0.47\textwidth}
        \includegraphics[width=\textwidth]{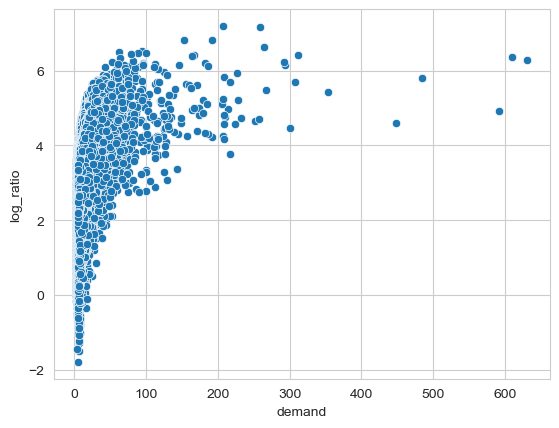}
        {\captionsetup{font=footnotesize} \caption{}\label{fig:high_varM}}
    \end{subfigure}
    \hfill  %
    \begin{subfigure}{0.47\textwidth}
        \includegraphics[width=\textwidth]{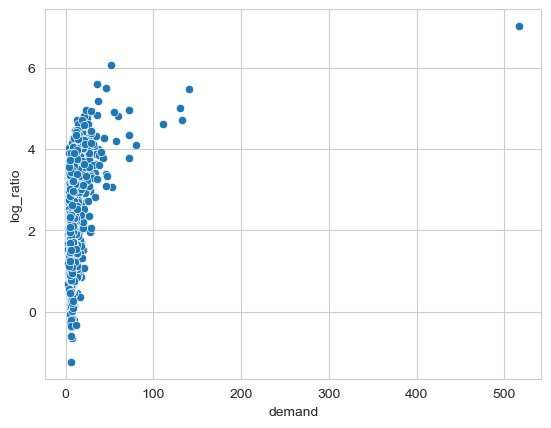}
        {\captionsetup{font=footnotesize}\caption{} \label{fig:high_resM}}
    \end{subfigure}
    \caption{Log-ratio mean-variance plots for the fashion e-commerce data set. Each dot represents a (SKU, location)-type $j$. The $x$-coordinate is the sample mean of the monthly aggregated demand for that SKU from that location. The $y$-coordinate is the natural logarithm of the ratio of sample variance to the sample mean. In (a), we compute the sample variance of monthly demands per type. In (b), we compute the sample variance of the linear model's residual errors per type (the linear model is estimated on a subsample of 1,500 types). In both cases, we restrict attention to SKUs offered for at least 6 months.}
\end{figure}

\paragraph{Model-free evidence.} We compare the sample mean and sample variance of the demand random variable $D_j$ across types. In fashion retail, the replenishment periods are less frequent, even for high-selling SKUs. Hence, we take the time horizon of the online matching problem to be equal to one month  rather than 1 week in the case of JD.com data. We aggregate our demand data set $D_j^t$ at that granularity. Each month $\tau$ yields a sample of $D^\tau_j$ for every (SKU, location)-type $j$.
We plot the log-ratio of sample variance and sample mean of $D^\tau_j$ for each type $j$ in Figure~\ref{fig:high_varM}. Similarly to our analysis of JD.com data, we find that the sample variance tends to be larger than the sample mean for nearly all types ($99.4\%$).

\paragraph{Controlling for variability.} In practice, the e-commerce platform can anticipate and forecast various sources of variability, reducing the uncertainty in the demand. We consider a simple two-way fixed effect model $D_j^\tau = \alpha_j + \beta_\tau + \gamma_j P_{j,\tau} +\eps_{j,\tau}$, where $\alpha_{j}$ is a type fixed effect, $\beta_\tau$ is the month fixed effect, $P_{j,\tau}$ is a per-unit sales-weighted price for type $j$ demand in month $\tau$, and $\eps_{j,\tau}$ is the residual error. This model controls for variability explained by price variation at the type-level and common seasonal patterns across types. Prices are inferred using as input the time-series of maximum retail price (MRP) for each item---for each SKU, we backfill missing prices chronologically, then normalize price as log-ratio between current price and the median non-zero price for that SKU.

We estimate the variance of (in-sample) residual errors $\eps_{j,\tau}$ for each type $j$ and month $\tau$ in our panel. Due to the scale of the data, we conduct our regression on random subsamples of 1,500 types (we repeat our analysis on multiple subsamples and the findings are unchanged).  The log-ratio mean-variance plot is shown in Figure~\ref{fig:high_resM}.  Despite better explaining variability in the demand, our linear model does not significantly reduce the level of uncertainty. The vast majority of (SKU-location) types still violate the SI assumption.

\paragraph{Limitations of our analysis.} Our analysis has several limitations.  Selecting the top 100,000 (SKU, region) pairs by total demand during the study period focuses our analysis on a
specific sample. High-volume types may systematically differ in variance properties from the full population. 

The linear model does not capture non-linear effects. Seasonal shifts in demand may be anticipated for subcategories of SKUs  (i.e., a refined model may use month-level fixed-effects per subcategory of SKUs instead of $\beta_t$-s). Our analysis overlooks other SKU-specific marketing interventions, known to the retailer, that affect the base demand. The MRP may not reflect true transaction price.  
Moreover, products in fashion retailing have a short product life cycle and strong seasonality, meaning that some of the variation in demand could be due to planned assortment rotations. Unfortunately, this information was not available to us in the dataset (i.e., we do not know when a SKU is offered and when it is discontinued). Note that we pre-processed the data to only consider "active" SKUs at any point in time using a simple heuristic, i.e., each (SKU-location) is included in the panel from the time it records at least one transaction until its last transaction. 

Lastly, we do not have access to inventory records, and thus, a zero demand may be due to stock-outs---something we do not control for.

\section{Connection to Standard Models} \label{app:generalizations}

We call \Poisson the online stochastic matching problem in continuous time where each type $j\in [m]$ arrives following an independent Poisson process of rate $\lambda_j$ over the time horizon $[0,1]$. We now explain why both our \Indep and \Correl models may capture \Poisson as a special case. To reveal the connection with \Poisson, suppose we operate under the random order $\Rand$, but we further assume that each arrival has a timestamp in the horizon $[0,1]$.\footnote{This can be achieved by assuming that each of the $D=D_1+\cdots+D_m$ total queries draws an independent arrival time uniformly from horizon [0,1]. This time-based process is combinatorially equivalent to $\Rand$. However, when time is an observable state variable, the online stochastic matching problem is different because a policy can set finer expectations for future demand based on which absolute time in the horizon $[0,1]$ it has reached.} Under the \Poisson model, we note that the demand $D_j$ of each type $j$ ends up having a distribution that is Poisson with expectation $\lambda_j$. Moreover, a basic property of stationary marked Poisson processes is that the density of arrivals is uniform over $[0,1]$ conditional on each realization of $\bD$. This implies that any instance of \Poisson with arrival rates $(\lambda_j)_{j\in [m]}$ can be captured within our framework by \Indep with random order, assuming that each demand entry $D_j$ follows a Poisson distribution with mean $\bE[D_j]=\lambda_j$. Similarly, the splitting property of Poisson processes straightforwardly implies that this \Poisson instance has an equivalent representation within \Correl, where $D$ is Poisson-distributed with mean $\bE[D]=\lambda_1+\cdots+\lambda_m$ and the type probabilities are $p_j=\lambda_j/\bE[D]$ for each $j\in [m]$.

We now explain why our results for \Indep and \Correl can capture the SI model beyond the limiting \Poisson case. In the case of $\Correl$, the reduction is straightforward. For a fixed horizon $T= D$ equal to the total demand, the type of the $t$-th query is drawn according to a distribution over types $(p_{t,j})_{j\in [m]}$, noting that our approximation ratio result for $\Correl \cap \Rand$ extends to such time-varying distributions; see Appendix~\ref{app:stochastic_gen}.
In the case of \Indep, the reduction is less straightforward but one possible approach may be to first link general SI to \Poisson, as has been done by~\cite{huang2021online}.  However, their link only applies to type distributions that IID over time.
For time-varying distributions, the demand $D_j$ of each type $j\in [m]$ follows a Poisson binomial distribution with $T$ trials and a success probabilities $(p_{t,j})_{t \in [T]}$, and the challenge is that the resulting demand vector $\bD$ does not satisfy the cross-sectional independence property required by $\Indep$. In fact, it is well known that the random variables  $D_j$ are {\em negatively associated}~\citep{dubhashi1998balls}, which means that $\expar{f(D_j,j\in J) \cdot g(D_j, j\in [m]\setminus J)} \leq \expar{f(D_j,j\in J)}\expar{g(D_j,j\in [m]\setminus J)}$ for every non-decreasing functions $f,g$ and every $J\subseteq [m]$. This strong negative dependence property, however, allows us to easily extend our analysis of the $1/2$-competitive algorithm for the $\Indep$ model (Algorithm~\ref{alg:indep} in Section~\ref{sec:indep}). In particular, we remark that the formulation of the truncated LP $\LPtrunc$ is unaffected by the dependence of the demand $D_j$ random variables. Thus, our lossless rounding scheme can be separately applied to each demand type to create a reduction to a prophet inequality setting, which is robust to weak notions of negative dependence for the sequence of random rewards~\citep{samuel1991prophet}. For completeness, we provide a proof below. Here, we use the weaker notion of Negative Right Orthant Dependence (NOD) property for a sequence of real-valued random variables $X_1,\ldots, X_m$, which requires that $\prpar{X_j \geq t_j, \ \forall j \in J\ | \ X_k \leq t_k, \ \forall k\in [m]\setminus J} \geq \prpar{X_j \geq t_j, \forall j \in J}$ for every vector $\boldsymbol{t} \in {\bb R}^m$ and $J\subseteq[m]$; see \citet[Prop.~3]{dubhashi1998balls}.

\begin{claim}
Suppose that $X_1,\ldots,X_n$ is a sequence of NOD two-outcome random variables with rewards  $0 = r_0 <r_1 <r_2< \ldots<r_n$ and probabilities $\prpar{X_i = r_i} = p_i$ and $\prpar{X_i = 0} = 1-p_i$ such that $\sum_{i=1}^n p_i \leq 1$. The static threshold  $\tau^* = \frac{1}{2}\sum_{i=1}^m p_i r_i$ yields  the prophet inequality $\expar{R(\tau^*)} \geq \frac{1}{2} (\sum_{i=1}^n r_ip_i)$, where $R(\tau^*)$ is equal to the first outcome above $\tau^*$, if any, and otherwise to  zero. 
\end{claim}

\proof{Proof.}
Let $\theta_i = \prpar{X_i \geq \tau^* | X_k < \tau^*, \ \forall k\leq i-1}$. By the NOD property, we clearly have that $\theta_i \geq \prpar{X_i \geq \tau^*} = p_i\cdot {\bb I}[r_i\geq \tau^*]$. Denoting by $\theta = \prpar{X_i < \tau^*\ , \forall i \in [n]}$, we can decompose the expected reward of the $\tau^*$-static threshold policy as
\begin{eqnarray*}
    \expar{R(\tau^*)} &=& \sum_{i=1}^n (r_i - \tau^*)\theta_i \pr{X_k < \tau^*,\ \forall k\leq i-1} + \sum_{i=1}^n  \tau^*\theta_i \\
    &\geq& \sum_{i=1}^n (r_i - \tau^*)p_i\cdot {\bb I}[r_i\geq \tau^*] \cdot \theta+ \tau^*(1-\theta) \\
      &\geq& \theta\cdot \left(\sum_{i=1}^n r_i p_i - \tau^*p_i\right)+ \tau^*(1-\theta) \\
      &\geq& \theta\cdot \frac{1}{2} \left(\sum_{i=1}^n r_ip_i\right) + \frac{1}{2} \left(1-\theta\right) \left(\sum_{i=1}^n r_ip_i\right) \\
        &=&  \frac{1}{2} \left(\sum_{i=1}^n r_ip_i\right)  \\
\end{eqnarray*}
where in the first inequality we use the fact that $\prpar{X_k < \tau^*,\ \forall k\leq i-1} \geq \theta$ for all $i\in [n]$. 
\Halmos\endproof

\section{Approximation Algorithm for the \textnormal{\textsc{\Correl}} Model} \label{sec:correl}

In this section, we focus on the $\Correl$ model. First, we observe in \Cref{subsec:stochasticHorizon} that instances in $\Correl \cap \Rand$ are connected to the online stochastic matching problem with a stochastic horizon length.
\Cref{subsec:condLP} introduces and justifies the "conditional" LP that we use as benchmark.
\Cref{subsec:algCorrel} defines our algorithm based on the conditional LP and establishes our results.

\subsection{Connection to the stochastic horizon setting} \label{subsec:stochasticHorizon}

As mentioned previously, the $\Correl$ model with random order $\Rand$ can be interpreted as a variant of the SI model with stochastic horizon. In the stochastic horizon setting, there is a maximum horizon length of $T$. We are given a vector $(p_{j})_{j\in[m]}$ satisfying $\sum_{j=1}^m p_{j}=1$ and denoting the probabilities that each arriving query is of type $j$. As per the SI model, the type of each new query in time $t=1,\ldots, T$ is drawn independently of the history. However, the true horizon length $D$ is unknown but randomly drawn from a known distribution with maximum value $T$. This means that the arrivals of queries suddenly "stop" after time $t=D$, i.e., the sequence is truncated at time $t=D$ and subsequent queries $t = D+1,\ldots, T$ never occur.

This stochastic horizon setting coincides with $\Correl\cap\Rand$.  To see why, recall that in $\Correl\cap\Rand$, the total demand $D$ is first drawn from a known distribution, after which $D$ types are drawn IID from probability vector $(p_j)_{j\in[m]}$ and arrive in a random order. However, the random permutation yields a sequence of the same distribution, because the $D$ draws are ex-ante identical. As such, it is equivalent to number the draws $t=1,\ldots,D$ and assume they arrive in that order, coinciding exactly with the aforementioned setting with the stochastic horizon $D$. 

This stochastic horizon setting has been previously studied by \citet{alijani2020predict}, in the context of a single item that perishes after an unknown random time $D$.
They show that the reward of the best online algorithm can be an arbitrary factor worse than that of a benchmark that knows the realization of $D$ in advance; i.e.,
\begin{align*}
\inf\limits_{\cI \in \Correl\cap\Rand}\frac{\OPT(\cI)}{\OFF(\cI)}=0\ .
\end{align*}
This result has also been discovered by \citet{brubach2023online,balseiro2023online} in different contexts.
Because $\LP(\cI)\geq \LPtrunc(\cI) \geq \OFF(\cI)$ for all instances ${\cal I}$, neither the commonly-used fluid relaxation nor the LP from Section~\ref{sec:indep} can provide a meaningful benchmark to analyze the performance of online algorithms.
We propose a second, tightened LP benchmark that allows for a constant-factor approximation ratio, comparing to the optimal online algorithm's reward $\OPT(\cI)$. 

\subsection{Formulation of the conditional LP} \label{subsec:condLP}

\begin{definition} \label{def:lpCond}
For any instance $\cI$, we define $\LPcond(\cI)$ as the optimal objective value of the following "conditional" LP:
\begin{align}
\max \ \ \ \ \ & \sum_{i=1}^n\sum_{j=1}^m \sum_{t=1}^{T}\prpar{D\geq t} \cdot r_{i,j} y^t_{i,j}  \label{lp:condObj} \\ 
\text{s.t. } \ \ \ \ &\sum_{j=1}^m \sum_{t=1}^T  y^t_{i,j} \leq k_i &\forall i\in[n] \label{ineq:condResource} \\ 
& \sum_{i =1}^n y^t_{i,j} \leq  p_{j} &\forall j \in [m], \forall t \in [T] \label{ineq:condDemand}  \\
&y^t_{i,j} \geq 0 &\forall i\in [n], \forall j \in [m], \forall t \in [T] \ . \label{lp:condEnd}
\end{align}
\end{definition}
Similar to Section~\ref{subsec:stochasticHorizon}, the parameter $T$ in $\LPcond(\cI)$ stands for the maximum value of $D$ that occurs with non-zero probability. Each decision variable $y^t_{i,j}$ represents the probability of matching a query of type $j$ to resource $i$ at time $t$, \textit{conditional} on $D\ge t$. This explains the objective function~\eqref{lp:condObj}, as well as  constraint~\eqref{ineq:condDemand}, in which $p_{j}$ is the probability that a type $j$ arrival occurs for time $t$ conditional on $D \geq t$. 
Finally, constraint~\eqref{ineq:condResource}, although appearing at first sight to be missing a coefficient $\Pr[D\ge t]$, is justified by the fact that each resource $i$ can be matched at most $k_i$ times conditional on $D=T$---the maximum value $D$ can ever take.
Because any online algorithm cannot foretell the realization of $D$, conditioning on $D=T$ should be equivalent to conditioning on $D\ge t$ for any time $t\in[T]$. This  informally justifies why $\LPcond(\cI)$ is a valid benchmark; the claim is formally proved in Appendix~\ref{pf:lem:comparison2}.

\begin{lemma} \label{lem:comparison2}
For any instance $\cI \in \Correl \cap \Rand$ and any online algorithm, we have $\ALG(\cI)\le\LPcond(\cI)$.  Therefore, $\OPT(\cI)\le\LPcond(\cI)$.
\end{lemma}

It is instructive to compare $\LPcond$ to the fluid LP in the same setting.  The fluid LP has an additional factor of $\prpar{D\geq t}$ on the left-hand side of~\eqref{ineq:condResource}, i.e., $\sum_{i=1}^n \prpar{D\geq t} \cdot y^t_{i,j} \leq p_j$. This is a relaxation of constraint~\eqref{ineq:condResource},  saying that the {\em expected consumption} of each resource $i$ cannot exceed $k_i$---a condition that holds for the offline optimum. Not knowing the realization of $D$ in advance is more constraining, and the online optimum needs to satisfy the tighter constraints~\eqref{ineq:condResource}. The two LPs are equivalent in the special case where $D$ deterministically equals $T$.

\subsection{Algorithm and results based on the conditional LP} \label{subsec:algCorrel}

Given $\LPcond(\cI)$ and the thought experiment of conditioning on $D=T$, our algorithm for $\Correl \cap \Rand$ is quite simple, and described in \Cref{alg:Correl}. 

\begin{algorithm}
\caption{Algorithm for $\Correl \cap \Rand$} \label{alg:Correl}
\begin{algorithmic}
\State Solve $\LPcond(\cI)$, letting $(y^t_{i,j})_{t\in[T], i\in [n], j\in [m]}$ denote an optimal solution
\State $\inv[i]=k_i$ for all $i\in[n]$ \Comment{starting inventory of resource $i$}
\For{$t=1,\ldots,D$} \Comment{realization $D$ unknown to algorithm}
\State Let $j$ denote type of query $t$ \Comment{drawn independently from probability vector $(p_{j})_{j\in[m]}$}
\State Choose $i$ independently from probability vector $\left(\frac{y^t_{i,j}}{p_{j}}\right)_{i\in[n]}$
\Comment{$\sum_{i=1}^n\frac{y^t_{i,j}}{p_{j}}\le1$ by~\eqref{ineq:condDemand}}

(set $i$ to $\perp$ with probability $1-\sum_{i=1}^n\frac{y^t_{i,j}}{p_{j}}$)
\If{$i\neq\perp$ and $\Acc_{i,t}(\inv[i])=1$} \Comment{$\Acc_{i,t}(\inv[i])$ is random bit for whether $i$ accepts}
\State Match the query to resource $i$, collecting reward $r_{i,j}$
\State $\inv[i]=\inv[i]-1$
\EndIf
\EndFor
\end{algorithmic}
\end{algorithm}

In \Cref{alg:Correl}, $\Acc_{i,t}(\inv[i])$ is a random bit returned by an Online Contention Resolution Scheme (OCRS) for resource $i$, indicating whether to accept a query at time $t$ if there are $\inv[i]$ units left.
It does not discriminate based on the type $j$ of the query, and the OCRS cannot accept if $\inv[i]=0$.
The purpose of an OCRS is to guarantee that the probability of accepting a query, at any time $t$, is always at least some constant $\gamma>0$.
Intuitively, an OCRS rejects early arrivals with some probability so that inventory is left for late arrivals; these probabilities are calibrated so that all arrivals have at least probability $\gamma$ of being accepted. In particular, our OCRS-based algorithm is adaptive to the unfolding of the sequence of queries. The largest achievable value of $\gamma$ then yields our desired approximation ratio.

The application of OCRS can be made precise by considering an equivalent, alternative world in which time $t$ always runs until the maximum of $T$, but only rewards collected in times $t\le D$ are counted.
From the perspective of any resource $i\in[n]$, it will be chosen by \Cref{alg:Correl} at each time $t\in[T]$ with probability
\begin{align} \label{eqn:active}
\sum_{j=1}^m p_{j}\frac{y^t_{i,j}}{p_{j}}=\sum_{j=1}^m y^t_{i,j},
\end{align}
in which case we say that $i$ is \textit{active} at time $t$.  Moreover, we have the following properties:
\begin{itemize}
\item[(Serial independence)] The events $\{i$ {\em is active in time} $t\}$ are mutually independent across $t \in [T]$, noting that the realization of $D$ does not induce any serial correlation in the alternative world where the horizon always runs until time $T$;
\item[(Feasibility in expectation)] For each resource $i\in [n]$, the expected number of active queries, equal to $\sum_{t=1}^T\sum_{j=1}^m y^t_{i,j}$ by~\eqref{eqn:active}, is at most $k_i$, as imposed by our LP constraint~\eqref{ineq:condResource}.
\end{itemize}

These are the two conditions required to utilize OCRSes devised in previous literature~\citep{jiang2022tight}.  Each resource $i$ runs a separate OCRS that prescribes random bits $\Acc_{i,t}(\inv[i])$, satisfying $\Acc_{i,t}(0)=0$ and guaranteeing that $\Pr[\Acc_{i,t}(\inv[i])=1]\ge\gamma^*_{k_i}$ for all $t$.  Here, $\gamma^*_{k_i}$ is a constant that depends on the starting inventory $k_i$ of resource $i$, and the probability is over the randomness in the inventory state $\inv[i]$ at time $t$.
Consequently, we leverage the OCRS guarantees for each combination of $(i,j,t)\in [n]\times [m]\times [T]$, noting that a reward is collected if: (i) at time $t$, query type $j$ is drawn and resource $i$ is chosen by \Cref{alg:Correl}; (ii) $D\ge t$; and (iii) $\Acc_{i,t}(\inv[i])=1$.  Because (i) and (ii) are independent from everything else, we obtain
\begin{eqnarray*} 
\ALG(\cI)
\ge\sum_{i=1}^n\sum_{j=1}^m \sum_{t=1}^{T}r_{i,j} y^t_{i,j} \Pr[D\ge t]  \gamma^*_{k_i}
\ge\gamma^*_{\min_i k_i} \cdot \left(\sum_{i=1}^n  \sum_{j=1}^m \sum_{t=1}^{T}r_{i,j} y^t_{i,j} \Pr[D\ge t] \right) 
=\gamma^*_{\min_i k_i} \cdot \LPcond(\cI)\ .
\end{eqnarray*}
To explain the second inequality, $\gamma^*_{k_i}$ is the best-possible OCRS guarantee for randomly rationing a resource with $k_i$ units.  This constant has been characterized in \citet{jiang2022tight}, and shown to be increasing in $k_i$.  Therefore, $\gamma^*_{k_i}\ge\gamma^*_{\min_i k_i}$ for every $i \in [m]$, resulting in the following \namecref{thm:main2}.

\begin{theorem} \label{thm:main2}
For the $\Correl\cap\Rand$ model, \Cref{alg:Correl} is a polynomial-time online algorithm that is $\gamma^*_k$-approximate, satisfying $\ALG(\cI)\ge\gamma^*_k \cdot \LPcond(\cI)\ge\gamma^*_k \cdot\OPT(\cI)$ for every instance $\cI$ in which all resources $i\in [m]$ have a starting inventory $k_i\ge k$.
\end{theorem}
We note that $\gamma^*_1=1/2$, and $\gamma^*_k$ is greater than an earlier lower bound of $1-1/\sqrt{k+3}$ proved in \citet{alaei2014bayesian} for all $k>1$.  This shows that \Cref{alg:Correl} is asymptotically optimal for $\Correl\cap \Rand$ as $k\to\infty$. Given the connection between $\Correl \cap \Rand$ and the stochastic horizon setting, Theorem~\ref{thm:main2} is also applicable in that setting. In fact, our conditional LP and OCRS reduction can be extended in two ways: (1) to a nonstationary version of the stochastic horizon setting, and (2) to Network Revenue Management where queries are matched to products that consume multiple resources.  These extensions are presented in Appendix~\ref{app:stochastic_gen}.

We conclude with two negative results, both only requiring a single unit of a single resource.

\begin{proposition} \label{prop:threshold}
For the $\Correl\cap\Rand$ model with a single unit of a single resource, any algorithm that accepts the first query above a fixed threshold has an approximation ratio of zero.
\end{proposition}

\Cref{prop:threshold} is proved in Appendix~\ref{pf:prop:threshold}.  Intuitively, static threshold policies perform poorly because they do not increase their thresholds in rare occurrences when the horizon "survives" past unlikely cutoffs. By contrast, our algorithm uses $\LPcond(\cI)$, which prescribes different matching probabilities $y^t_{i,j}$ and thus prescribes thresholds that are adapted to the time $t$.
\Cref{prop:threshold} also provides a noteworthy contrast to the single-item prophet inequality setting, in which a fixed threshold algorithm can be 1/2-approximate \citep{samuel1984comparison}.  This suggests that, despite the simplicity of our analysis above, the $\Correl\cap \Rand$ and SI models are definitely not equivalent and analyzing the former is crucially dependent on our conditional LP.

\begin{proposition} \label{prop:main2tight}
For the $\Correl\cap\Rand$ model, we have $\displaystyle\inf_{\cI\in\Correl\cap\Rand}\frac{\OPT(\cI)}{\LPcond(\cI)}\le\frac{1}{2}$.
\end{proposition}

Finally, \Cref{prop:main2tight} is proved in Appendix~\ref{pf:prop:main2tight}.  It shows that the approximation ratio of 1/2 achieved in \Cref{thm:main2} (where $\gamma^*_1=1$) is tight relative to $\LPcond(\cI)$, when $k=1$.  We note that this requires constructing new hard instances because we measure performance against a less powerful benchmark than the offline optimum, and allow for arbitrary online algorithms.

\subsection{Formal definition of $\OPT(\cI)$} \label{sec:formalDP}

We specify $\OPT(\cI)$, defined as the expected total rewards collected by an optimal policy found via computationally unconstrained dynamic programming, in the $\Correl\cap \Rand$ model of \Cref{sec:correl}. In that section of the paper, we compare our algorithm's performance against $\OPT(\cI)$ instead of $\OFF(\cI)$.  For every time $t\in[T]$ and remaining inventory levels $I_1,\ldots,I_n$ of each resource $i\in [n]$, let $J_t(I_1,\ldots,I_n)$ denote the value-to-go with remaining inventories $I_1,\ldots,I_n$ upon reaching the end of time $t-1$ (i.e., knowing that $D\ge t-1$ but not yet knowing whether $D\ge t$).  Here, the value function $J_{T+1}(I_1,\ldots,I_n)$ is understood to to be 0 for all vectors $(I_1,\ldots,I_n)$, and $J_t(0,\ldots,0)$ is understood to be 0 for all $t$.  The Bellman equation then tell us that
\begin{align*}
J_t(I_1,\ldots,I_n) &=\Pr[D\ge t|D\ge t-1]\sum_{j=1}^m p_{t,j}\max_{i:k_i>0}\big(r_{i,j}+J_{t+1}(I_1,\ldots,I_{i-1},I_i-1,I_{i+1},\ldots,I_n)\big)
&\forall t\in[T],(I_1,\ldots,I_n)
\end{align*}
because conditional on $D\geq t$ and type $j$ arriving next, we should match to the resource $i$ that maximizes the immediate reward $r_{i,j}$ plus the value-to-go $J_{t+1}$ with the $i$'th entry of the inventory vector being decremented by one unit.
With this definition at hand, $\OPT(\cI)$ simply corresponds to $J_1(k_1,\ldots,k_n)$, where $k_1,\ldots,k_n$ are the starting inventory levels of $\cI$.

\subsection{Proof of \Cref{lem:comparison2}} \label{pf:lem:comparison2}

Fix any online algorithm, and on a sample path of its execution, let $X^t_{i,j}\in\{0,1\}$ be the indicator random variable for a query of type $j$ being matched to resource $i$ in time $t$.  Let $Q^t_j\in\{0,1\}$ be the indicator random variable for a query at step $t$ having type $j$, where we note that $Q^t:=\sum_j Q^t_j\le 1$, with $Q^t=0$ whenever the total demand $D$ is less than $t$.

On any sample path, the reward collected by the algorithm is $\sum_{i,j,t}r_{i,j}X^t_{i,j}$.
Meanwhile, the constraint $\sum_{j,t}X^t_{i,j}\le k_i$ must be satisfied for every resource $i$, since $i$ can be matched at most $k_i$ times;
and the constraint $\sum_i X^t_{i,j}\le Q^t_j$ must be satisfied for every $j$ and $T$, since a query at step $t$ of type $j$ can be matched to at most one $i$, and only if $Q^t_j=1$.
Taking an expectation over \textit{only the sample paths where $Q_T=1$}, i.e., paths where $D$ realized to its maximum value of $T$, we have
\begin{align*}
&\sum_{j=1}^m \sum_{t=1}^T  \ex{\left. X^t_{i,j}\right|Q_T=1} \leq k_i &\forall i\in[n] \\ 
& \sum_{i =1}^n \ex{\left. X^t_{i,j}\right|Q_T=1} \leq p_{t,j} &\forall j \in [m], \forall t \in [T]
\end{align*}
noting that $\bE[Q^t_j]=p_{t,j}$.
Now, we derive for any $i$, $j$, and $t$ that
\begin{align}
\ex{\left. X^t_{i,j}\right|Q_T=1}
~=~\ex{\left. X^t_{i,j}\right|Q_t=1} 
~=~\frac{\bE[X^t_{i,j}Q_t]}{\bE[Q_t]} 
~=~\frac{\bE[X^t_{i,j}]}{\Pr[D\ge t]} \ , \label{eqn:2710}
\end{align}
where the first equality holds because an online algorithm's decision $X^t_{i,j}$ at time $t$ cannot distinguish between $Q_t=1$ and the stronger future event that $Q_T=1$,
and the final equality holds because $X^t_{i,j}=1$ implies $Q_t=1$ while $\bE[Q_t]=\Pr[D\ge t]$ by definition.
Applying~\eqref{eqn:2710}, the expected reward collected by the algorithm is $\bE[\sum_{i,j,t}r_{i,j}X^t_{i,j}]=\sum_{i,j,t}r_{i,j}\bE[X^t_{i,j}|Q_T=1]\cdot \Pr[D\ge t]$.

Letting $y^t_{i,j}=\bE[X^t_{i,j}|Q_T=1]$ for all $i$, $t$, and $j$, we see that this forms a feasible solution to $\LPcond(\cI)$ with objective value equal to the expected reward collected by the algorithm.
Since any online algorithm corresponds to such a feasible solution to $\LPcond(\cI)$, the optimal objective value of $\LPcond(\cI)$ can be no less than the expected reward of any online algorithm, thereby completing the proof of \Cref{lem:comparison2}.

\subsection{Proof of \Cref{prop:threshold}} \label{pf:prop:threshold}

Let $n=1$ with $k_1=1$.
Fix a large $T$ and a small $\eps>0$.
Let $D$ be distributed over $\{1,\ldots,T\}$ as follows: for any $t=1,\ldots,T-1$, the survival rate is $\Pr[D>t|D\ge t]=\eps$.
Types are defined so that conditional on any time $t\in[T]$ occurring, the reward for matching with the resource is $1/\eps^t$ with probability~$\eps$, and $1/\eps^{t-1}$ with probability~$1-1/\eps$.

We claim that $\OPT(\cI)\ge T-O(\eps)$ by considering the following online algorithm.
In any time $t$, it only accepts the query if the reward is the larger realization of $1/\eps^t$ (which occurs with probability~$\eps$).
Such an algorithm will accept a query in time $t$ if $D\ge t$, which occurs with probability~$\eps^{t-1}$; conditional on this, no query is accepted before time $t$, which occurs with probability~$(1-\eps)^{t-1}$.  Therefore,
\begin{align*}
\ALG(\cI)
&=\sum_{t=1}^T\eps^{t-1}(1-\eps)^{t-1}\eps\cdot\frac{1}{\eps^t}
\\ &\ge T \cdot (1-\eps)^T\ .
\end{align*}

Now consider any policy that sets a fixed threshold of $1/\eps^t$, for some $t\in\{0,\ldots,T\}$.
This policy can only collect reward in at most 2 times,  $t$ or $t+1$.
The maximum reward that can be collected in any time $t$ (assuming there is available inventory, and accepting either realization $1/\eps^t$ or $1/\eps^{t-1}$) is $$\Pr[D\ge t]\cdot \left(\eps\cdot\frac{1}{\eps^t}+(1-\eps)\frac{1}{\eps^{t-1}}\right)=\eps^{t-1}\left(\eps\cdot\frac{1}{\eps^t}+(1-\eps)\frac{1}{\eps^{t-1}}\right)\le 2\ .$$
Therefore, such a policy has expected reward upper-bounded by 4, which is an arbitrarily small fraction of $T\cdot (1-\eps)^T$ as  $T$ tends to $\infty$ and $\eps$ tends to $0$.

\subsection{Proof of \Cref{prop:main2tight}} \label{pf:prop:main2tight}

Let $n=1$ and $k_1=1$.  Fix a large integer $N$.  Let $m=2$, with the type rewards given by $r_{1,1}=1$ and $r_{1,2}=N^2$.
Let the distribution for the total demand $D$ be $D=1$ with probability~$1-1/N$, and $D=1+N^2$ with probability~$1/N$.
Each query draws an IID type that is 1 with probability~$p_1:=1-1/N^3$ and 2 with probability~$p_2:=1/N^3$.

The following is a feasible solution with respect to $\LPcond(\cI)$: set $y^1_{1,1}=1-(1+N^2)/N^3<p_1$ and $y^1_{1,2}=1/N^3=p_2$; set $y^t_{1,1}=0$ and $y^t_{1,2}=1/N^3$ for $t=2,\ldots,1+N^2$.
Intuitively, this fractional solution accepts both types of queries when $t=1$, and only accepts queries of type 2 when the time horizon $D$ "survives" past $t=1$.
Constraint~\eqref{ineq:condDemand} is satisfied by construction, and
constraint~\eqref{ineq:condResource} can be verified because its left-hand side equals $1-(1+N^2)/N^3+(1+N^2)\cdot 1/N^3=1$.
Now, the objective value is
\begin{align*}
1-\frac{1+N^2}{N^3}+N^2\left(\frac{1}{N^3}+\Pr[D\ge 2]\cdot N^2\cdot\frac{1}{N^3}\right)
&=1-\frac{1}{N^3}+N^2\left(\frac{1}{N}N^2\cdot\frac{1}{N^3}\right)
\\ &=2-O\left(\frac{1}{N^3}\right)\ .
\end{align*}

Meanwhile, consider any online algorithm.  If its plan is to accept type 1 in time 1, then its expected reward is $1-1/N^3+N^2\cdot 1/N^3=1+O(1/N)$.  On the other hand, if its plan is to reject type 1 in time 1, then its expected reward is at most $N^2\cdot(1+1/N\cdot N^2)\cdot 1/N^3$ which is also $1+O(1/N)$.
Taking $N\to\infty$ completes the proof.

\section{Extension to Sampling-Based Setting} \label{app:demandError}

Our result for \Indep extends to the setting where the decision-maker only has access to IID samples of the demand vector rather than its exact distribution. We do not analyze the \Correl model in this sampling-based setting, leaving the robustness of Theorem~\ref{thm:main2} to sampling error as an open question. We note that \Correl is more similar to the SI model, for which sample-based competitive algorithms are known. The key difference is the distribution of the stochastic horizon $D$, which intuitively should be easier to learn from data than the marginal distributions for each demand type in \Indep.

Here, we suppose that the distribution of $\bD$ is unknown but can be estimated from a sample of $N$ IID draws $\hat{\bD} = (\hat{\bD}_1,\ldots,\hat{\bD}_N)$ from the distribution of $\bD$. We present sample complexity results for \Indep, showing that for $N$ polynomially large, we can incur a small loss $O(\eps)$ in our performance guarantee. 

We can leverage existing results on the generalization error of hypothesis classes under product form distributions to obtain additive approximation errors. For every demand type $j\in [m]$, we define $\hat{E}_j$ as the empirical distribution over $\hat{D}_{1,j},\hat{D}_{2,j},\ldots, \hat{D}_{N,j}$. Here, we assume that the rewards are normalized such that $\max_{i\in [n],j\in [m]}r_{i,j}=1$ and $k_i = 1$ for every resource $i \in [n]$.
\begin{proposition} \label{prop:aditive}
There exists a universal constant $C>0$ such that for all $N \geq C \frac{n^3 m}{\eps^2}\log\frac{1}{\delta}$, running Algorithm~\ref{alg:indep} with respect to the product-form empirical distribution $\hat{\bE} = \hat{E}_1\times \cdots \times \hat{E}_m$ achieves $\inf\limits_{{\cal I} \in \Indep\cap \Adv} \ALG_{\hat{\bE}({\cal I})}({\cal I}) \geq \frac{1}{2} \cdot \LPtrunc({\cal I}) -O(\eps )$ with probability at least $1-\delta$.
\end{proposition}
This result follows from Theorem~5 of~\cite{guo2021generalizing}, noting that the demand for each type $j$ can be truncated to be at most $n$ without loss (i.e., $\prpar{D_j\leq n} = 1$ for all $j\in [m]$).
We maintain the assumption that $D_j\le n$ with probability 1 for all $j\in [m]$ throughout this section.

To elaborate on how to establish \Cref{prop:aditive}, let $\cI$ denote the instance with the true distribution, and $\hat{\cI}$ denote the instance with the same rewards $(r_{i,j})_{i\in[n],j\in[m]}$ and starting inventories $(k_i)_{i\in[n]}$ but using the empirical distribution instead.  Our algorithm trains an online policy (a fixed prescription of how to match queries on-the-fly) based on $\hat{\cI}$, which will satisfy $ \ALG_{\hat{\bE}({\cal I})}(\hat{\cI})\ge\frac12\cdot\LPtrunc(\hat{\cI})$. Here, we use $ \ALG_{\hat{\bE}({\cal I})}$ instead of $\ALG$ to emphasize that our actual policy is trained on the empirical distribution.  However, what the algorithm actually earns is $ \ALG_{\hat{\bE}({\cal I})}$, which we wish compare to $\LPtrunc(\cI)$.  Fortunately, the result of \citet{guo2021generalizing} says that with sufficiently many samples $N$, we can uniformly guarantee for any real-valued function of a distribution that its evaluation on the true vs.\ empirical distributions differ by at most a small $\eps>0$.  Because $ \ALG_{\hat{\bE}({\cal I})}(\cdot)$ and $\LPtrunc(\cdot)$ can both be viewed as real-valued functions that depend on the distribution in the instance,
we apply the result of \citet{guo2021generalizing} twice to get that
\begin{align*}
 \ALG_{\hat{\bE}({\cal I})}(\cI)
&\ge \ALG_{\hat{\bE}({\cal I})}(\hat{\cI})-\eps
\\ &\ge\frac12\cdot\LPtrunc(\hat{\cI})-\eps
\\ &\ge\frac12\cdot\LPtrunc(\cI)-2\eps.
\end{align*}
The required number of samples $N$ scales with the square of the range of the functions, which in this case is $[0,n]$, under the normalization of the rewards.  This explains why the sample complexity is $O(\frac{n^3 m}{\eps^2}\log\frac{1}{\delta})$ instead of $O(\frac{nm}{\eps^2}\log\frac{1}{\delta})$ as quoted in \citet[Thm~5]{guo2021generalizing}.

A drawback of Proposition~\ref{prop:aditive} is that the loss $O(\eps)$ might be large in the regime where $\LPtrunc({\cal I})$ is small. Thus, we propose a data-driven algorithm to achieve an $O(\eps)$-multiplicative error. The following result holds even if our algorithm does not observe any type $j_0\in [m]$ without any arrival in the sample  $\sum_{t=1}^N\hat{D}_{t,j_0}=0$, and it needs not know their corresponding rewards $(r_{i,j_0})_{i\in [n]}$. That said, our sample complexity bound implicitly assumes that we know an upper bound on the total number of customer types $m$.

\begin{theorem} \label{thm:multiplicative}
There  exist a polynomial-time algorithm and a universal constant $C>0$ such that for all $N \geq C \frac{nm}{\eps^3}\log\frac{nm}{\delta}$, $\inf\limits_{{\cal I} \in \Indep\cap \Adv} {\ALG_{\hat{\bD}~\sim \bD^N ({\cal I})}}({\cal I}) \geq (1-\eps)\cdot \frac{1}{2} \cdot \LPtrunc({\cal I})$ with probability at least $1-\delta$, where $\ALG_{\hat{\bD}~\sim \bD^N ({\cal I})}({\cal I})$ stands for the expected rewards in ${\cal I}$ for each sample instantiation $\hat{\bD}~\sim \bD^N ({\cal I})$.
\end{theorem}
The remainder of the section is devoted to proving \Cref{thm:multiplicative}. At a high level, the algorithm calls our original $\LPtrunc(\hat{\cal I})$ and lossless rounding for a specific input instance $\hat{\cal I}$, estimated using the sample $\hat{\bD}$, but it gives priority to matching infrequent types whose demand distribution cannot be accurately inferred from the sample. In what follows, we hide the dependence on ${\cal I}$ whenever it is clear which instance we refer to and only indicate which demand distribution is used to evaluate the outcomes.
We start by stating a few basic technical claims.
We denote by $ \bD \succeq \bs{E}$ the (first-order) stochastic dominance order $\prpar{D_j\ge\ell} \ge\prpar{E_j\ge\ell}$ for all $ j\in[m]$ and $\ell\geq 1$.

\begin{claim}[Dominated distribution] \label{clm:dominated}
For every product-form  distribution ${\cal D}$, having access to a sample of $N \geq \frac{3nm}{\eps^3}\log \frac{2nm}{\delta}$ independent draws from that distribution $(\hat{\bD}_1,\ldots,\hat{\bD}_N)  \sim \bD^N$, we can construct in polynomial time a product-form distribution $\bs{E}$ such that, with probability at least $1-\delta$, we have $ \bD \succeq \bs{E}$
and $\prpar{E_j\geq \ell} \geq (1-2\eps) \prpar{D_j \geq \ell}$ for all $j\in [m], \ell \in [n]$ for which $\prpar{D_j \geq \ell} \geq \frac{\eps}{m n}$.
\end{claim}
\proof{Proof.}
Fix $(j,\ell)\in [m]\times[n]$ for which $\prpar{D_j \geq \ell} \geq \frac{\eps}{m n}$. We invoke the multiplicative Chernoff bound, using the shorthand $p_{j,\ell} =\pr{D_j\geq \ell}$,
\begin{eqnarray} \label{ineq:chernoff}
    \prtwo{\hat{\bD}_1,\ldots,\hat{\bD}_N \sim \bD^N}{\left|\frac{1}{N}\cdot \sum_{t=1}^N {\bb I}\left[\hat{D}_{t,j}  \geq \ell\right] - p_{j,\ell}  \right| \geq \eps   p_{j,\ell}} \leq 2  e^{ -\frac{\eps^2 N p_{j,\ell} }{3} } \ .
\end{eqnarray}
It follows that, by defining $\hat{p}_{j,\ell} =   (1-\eps)\cdot \frac{1}{N}\cdot \sum_{t=1}^N {\bb I}[\hat{D}_{t,j}  \geq \ell]$, we obtain
\begin{eqnarray} \label{ineq:sw}
\prtwo{\hat{\bD}_1,\ldots,\hat{\bD}_N \sim \bD^N}{\hat{p}_{j,\ell} \notin ((1-\eps)^2 p_{j,\ell},(1+\eps)(1-\eps)p_{j,\ell})   }\le\frac{\delta}{nm} \nonumber
\\ \Longrightarrow \prtwo{\hat{\bD}_1,\ldots,\hat{\bD}_N \sim \bD^N}{ {\hat{p}_{j,\ell} \notin [(1-2\eps) p_{j,\ell},p_{j,\ell}]} } \leq \frac{\delta}{nm}
\end{eqnarray}
where we use inequality~\eqref{ineq:chernoff}, and the hypotheses $p_{j,\ell} =\prpar{D_j \geq \ell} \geq \frac{\eps}{m n}$ and $N \geq \frac{3nm}{\eps^3}\log \frac{2nm}{\delta}$. Consequently, we define $\prpar{E_j\geq \ell} = \min\{\hat{p}_{j,\ell}, \prpar{E_j\geq \ell-1}\}$ iteratively over $\ell \in[1,n]$ with $\prpar{E_j\geq 0} = 1$ and $\prpar{E_j\geq n+1} = 0$. The desired claim follows from the union bound over $(j,\ell)$ and inequality~\eqref{ineq:sw}.
\Halmos\endproof

\begin{claim}[LP Sensitivity] \label{clm:sensitivity}
For all product-form distributions $\bD \succeq \bs{E}$, and for all $\eps>0$,
\begin{itemize}
\item if $\prpar{ D_j \geq \ell} - \prpar{ E_j \geq \ell} \leq \eps$ for all $j\in [m]$, $\ell \in [n]$, then $\LPtrunc(\bs{E}) \geq \LPtrunc(\bD) - O(\eps nm)$.
\item  if $\prpar{ E_j \geq \ell} \geq (1-\eps) \cdot \prpar{ D_j \geq \ell} $ for all $j\in [m]$, $\ell \in [n]$, then $\LPtrunc(\bs{E}) \geq (1-\eps)\cdot \LPtrunc(\bD)$.
\end{itemize}
\end{claim}

\begin{claim}[Median prophet] \label{clm:prophet-var}
Suppose that $X_1,\ldots,X_n$ is a sequence of independent two-outcome random variables with rewards  $0 = r_0 <r_1 <r_2< \ldots<r_n$ and probabilities $\prpar{X_i = r_i} = p_i$ and $\prpar{X_i = 0} = 1-p_i$ such that $\sum_{i=1}^n p_i \leq 1$. The unique (randomized) perturbed threshold  $\tau_{\rm ptb}$ such that $\tau_{\rm ptb} \in \{r_i, r_{i+1}\}$ for some $i\in [0,n-1]$ and $\prpar{ \max_{i\in [n]} X_i \geq \tau_{\rm ptb} }=\frac{1}{2}$ yields  the prophet inequality $\expar{R(\tau_{\rm ptb})} \geq \frac{1}{2} (\sum_{i=1}^n r_ip_i)$, where $R(\tau_{\rm ptb})$ is equal to the first outcome above $\tau_{\rm ptb}$, if any, and otherwise to  zero. Moreover, when running this static threshold rule with noisy inputs $\tilde{p}_1,\ldots, \tilde{p}_n$ instead of $p_1,\ldots,p_n$, where $ \tilde{p_i} \geq (1-\eps)p_i$ and $\sum_{i=1}^n \tilde{p}_i \leq \sum_{i=1}^n (1-\eps)^{-1} p_i + \eps$ for all $i\in [n]$ with an $\eps \in (0,\frac{1}{2})$,  the resulting perturbed threshold $\tilde{\tau}_{\rm ptb}$ yields $\expar{R(\tilde{\tau}_{\rm ptb})} \geq \frac{(1-11\eps)}{2}\cdot\left( \sum_{i=1}^n r_ip_i \right)$ and $\prpar{ \max_{i\in [n]} {X}_i \geq \tilde{\tau}_{\rm ptb} } \leq \frac{(1+2\eps)}{2}$. 
\end{claim}

\proof{Proof.}
We prove the second claim, which implies the first one.  By construction, we have $ \prpar{\tilde{\tau}_{\rm ptb} \geq \max_{i\in [n]} \tilde{X}_i}=1/2$ for the sequence of independent modified random variables $\tilde{X}_i$ with $\prpar{\tilde{X}_i =r_i} = \tilde{p}_i$. In particular, there exists $i^* \in [0,n-1]$ such that $\prpar{\tilde{\tau}_{\rm ptb} = r_{i^*}} = \delta >0$ and $\prpar{\tilde{\tau}_{\rm ptb} = r_{i^*+1}} = 1-\delta$. Define $\delta_i = 0$ for $i\in[0,i^*-1]$, $\delta_i =1$ for $i\in[i^*+1,n]$, and $\delta_{i^*} = \delta$. The calibration of the threshold parameters $i^*, \delta$ such that $\prpar{\tilde{\tau}_{\rm ptb} \geq \max_{i\in [n]} \tilde{X}_i}=1/2$ is equivalent to the identity $\sum_{i=1}^n \tilde{p}_i \cdot \delta_i \cdot \prod_{i'=1}^{i-1} (1-\tilde{p}_{i'} \cdot \delta_{i'}) = \frac{1}{2}$. Defining $\eps_i = \tilde{p}_i-(1-\eps)^{-1}p_i$, it follows that 
\begin{eqnarray} 
\frac{1}{2} \ &\leq& \ \sum_{i=1}^n ((1-\eps)^{-1}{p}_i +\eps_i) \cdot \delta_i \cdot \left(\prod_{i'=1}^{i-1} \left(1-(1-\eps) {p}_{i'} \cdot \delta_{i'}\right) \right) \nonumber\\
&\leq& \ 3\eps + (1-\eps)^{-1}\sum_{i=1}^n {p}_i \cdot  \delta_i \cdot \left(\prod_{i'=1}^{i-1}  e^{ \log(1-{p}_{i'} \cdot \delta_{i'}) + 2\eps p_{i'} \delta_{i'}}  \right) \nonumber\\
\ &=& \  3\eps + (1-\eps)^{-1} e^{(\sum_{i=1}^{n} 2\eps p_i \delta_i)}\pr{\max_{i\in [n]} {X}_i \geq \tilde{\tau}_{\rm ptb} } \nonumber \\
\ &\leq& \ 3\eps + (1-5\eps)^{-1} \pr{\max_{i\in [n]} {X}_i \geq \tilde{\tau}_{\rm ptb} }  \label{ineq:sensitivity} \ 
\end{eqnarray}
where in the first inequality we use $ \tilde{p_i} \geq (1-\eps)p_i$ for all $i\in [n]$. In the second inequality, we use the fact that $\eps_i \geq -2\eps p_i$ and thus $\sum_{i=1}^n [\eps_i]^+ = \sum_{i=1}^n (\eps_i + 2\eps p_i) \leq 3\eps$ by hypothesis (because $\sum_{i=1}^n \eps_i \leq \sum_{i=1}^n\tilde{p}_i - \sum_{i=1}^n(1-\eps)^{-1} p_i  \leq\eps$), and notice that $\log(1 - (1-\eps)u) \leq \log(1 - u)  + \eps u \frac{1}{1-u}$ with $u = p_{i'}\delta_{i'} \leq \frac{1}{2}$. In the last inequality, we use the fact that $\sum_{i=1}^{n} 2\eps p_i \delta_i \leq 2\eps$ and $\eps \in (0,\frac{1}{2})$. Reciprocally, from $\tilde{p}_i \geq (1-\eps) p_i$ for all $i\in [n]$, we obtain
\begin{eqnarray} 
\frac{1}{2} \ & = & \ \sum_{i=1}^n \tilde{p}_i \cdot \delta_i \cdot \left(\prod_{i'=1}^{i-1} \left(1- \tilde{p}_{i'} \cdot \delta_{i'}\right) \right) \nonumber\\
\ &\geq& \ \sum_{i=1}^n (1-\eps) {p}_i \cdot \delta_i \cdot \left(\prod_{i'=1}^{i-1} \left(1-(1-\eps) {p}_{i'} \cdot \delta_{i'}\right) \right) \nonumber\\
\ &\geq& \ (1-\eps)  \sum_{i=1}^n {p}_i \cdot \delta_i \cdot \left(\prod_{i'=1}^{i-1} \left(1- {p}_{i'} \cdot \delta_{i'}\right) \right) \nonumber\\
\ &\geq& \ (1-\eps) \pr{\max_{i\in [n]} {X}_i \geq \tilde{\tau}_{\rm ptb} } \ , \label{ineq:sw2}
\end{eqnarray}
where the first inequality holds since $[ \bs{x}\mapsto \sum_{i=1}^n x_i \cdot \delta_i \cdot (\prod_{i'=1}^{i-1} (1- x_{i'} \cdot \delta_{i'}) )]$ is non-decreasing in $\bs{x}$.

Now, we use the standard decomposition
\begin{eqnarray*}
&&\ex{\min\left\{ X_i : i\in [n], {\bb I}[X_i \geq \tilde{\tau}_{\rm ptb}]\right\}}\\
&& \quad \geq \pr{\max_{i\in [n]} {X}_i \geq \tilde{\tau}_{\rm ptb} }\cdot r_{i^*}  + \left(1-\pr{\max_{i\in [n]} {X}_i \geq \tilde{\tau}_{\rm ptb} }\right)\sum_{i=1}^n \max\{0,r_i -r_{i^*}\} \cdot {p}_i \\
&& \quad \geq \frac{1-11\eps}{2}\cdot \left( \sum_{i=1}^n r_ip_i \right)\ ,
\end{eqnarray*}
where the first inequality holds since $r_i > r_{i^*}$ implies $\delta_i = 1$, and the last inequality follows from the fact that $\sum_{i=1}^n p_i \leq 1$ and inequalities~\eqref{ineq:sensitivity} and~\eqref{ineq:sw2}.
\Halmos\endproof

\paragraph{Data-driven algorithm.} Fixing $\bD$, we distinguish between frequent and infrequent queries with ${\cal T}_{\rm freq} = \{ (j,\ell)\in [m]\times [n]: \prpar{D_j \geq \ell} \geq \frac{\eps }{m n}\}$.   The first step of our algorithm is to use the sample $\hat{\bD}$ to construct the product-form distribution $\bs{E} \preceq \bs{D}$ achieving the properties stated in Claim~\ref{clm:dominated}. Next, we determine the subset of frequent types $\hat{\cal T}_{\rm freq} = \{ (j,\ell)\in [m]\times [n]: \prpar{E_j \geq \ell} \geq (1-2\eps)\cdot\frac{2\eps }{m n}\}$ and infrequent types ${\cal T}_{\rm infreq} =  [m]\times [n]\setminus \hat{\cal T}_{\rm freq}$. By Claim~\ref{clm:dominated}, with probability $1-\delta$, we have $\hat{\cal T}_{\rm freq} \subseteq {\cal T}_{\rm freq}$, $\prpar{D_j \geq \ell}\geq \prpar{E_j \geq \ell} \geq (1-2\eps)\prpar{D_j \geq \ell}$ for all $(j,\ell)\in \hat{\cal T}_{\rm freq}$, and $\prpar{D_j \geq \ell} \leq \frac{2\eps}{mn}$  for all $(j,\ell)\in \hat{\cal T}_{\rm infreq}$. Throughout the remainder of the analysis, we condition on the corresponding event and assume that these properties hold.

As a second step, we ignore infrequent arrivals and leverage \Cref{alg:indep} with respect to the residual demand distribution. That is, we consider the modified demand distribution $\bar{\bs{E}}$ after we eliminate every infrequent query $(j,\ell)\in \hat{\cal T}_{\rm infreq}$, i.e., $\prpar{\bar{E}_j \geq \ell }= \prpar{{E}_j \geq \ell }$ for every $(j,\ell)\in \hat{\cal T}_{\rm freq}$ and $\prpar{\bar{E}_j \geq \ell }= 0$ for every $(j,\ell)\notin \hat{\cal T}_{\rm freq}$ with renormalization $\prpar{\bar{E}_j = 0 }=1 - \sum_{\ell: (j,\ell)\in \hat{\cal T}_{\rm freq}} \prpar{\bar{E}_j \geq \ell }$. Consequently, let $\ALG_{\bar{\bs{E}}}$ denote the execution of \Cref{alg:indep} by specifying the modified distribution $\bar{\bs{E}}$ as input and the median threshold rule. That is, $\ALG_{\bar{\bs{E}}}$ solves $\LPtrunc(\bar{\bs{E}})$, then routes the arriving queries to resources using our lossless rounding~\Cref{alg:losslessRounding}, and finally it runs the threshold rule of Claim~\ref{clm:sensitivity} for each resource. The use of the threshold in Claim~\ref{clm:sensitivity} marks a slight difference with  \Cref{alg:indep}.

Lastly, as a third step, we consider the "augmented" algorithm $\ALG_{\bar{\bs{E}}}^{\uparrow}$ that deals with the arrivals of infrequent types, which are not expected to arrive as per $\bar{\bs{E}}$. Specifically, $\ALG_{\bar{\bs{E}}}^{\uparrow}$ takes the same matching decisions as $\ALG_{\bar{\bs{E}}}$ until an infrequent query $(j,\ell)\notin \hat{\cal T}_{\rm freq}$ arrives, if any. From that point onwards, it matches all subsequent arrivals of type $j$ greedily to the remaining resources and it no longer matches any query from other types. This completes the description of $\ALG_{\bar{\bs{E}}}^{\uparrow}$ as our  data-driven algorithm for \Cref{thm:multiplicative}.

\paragraph{Analysis.}The next claim relates the performance of $\ALG_{\bar{\bs{E}}}^{\uparrow}$ on the arrivals described by $\bD$ to  our LP benchmark, thereby completing the proof of \Cref{thm:multiplicative}.
\begin{lemma} \label{lem:final}
$\ALG_{\bar{\bs{E}}}^{\uparrow}(\bD) \geq \frac{(1-26\eps)}{2}\LPtrunc(\bD)$.
\end{lemma}
To prove Lemma~\ref{lem:final}, we first lower bound the contributions of infrequent queries to the expected total rewards. For the purpose of analysis, we create a monotone coupling $\omega$ between the arrivals in $\bar{\bs{E}}$ and those in $\bD$. Hence, we introduce $\zeta_{j,\ell,i}(\omega)$ as an indicator random variable such that $\zeta_{j,\ell,i}(\omega) = 1$ if the $\ell$-th arrival of type $j$ is routed to $i$ by $\ALG_{\bar{\bs{E}}}$ on the arrivals described by $\bar{\bs{E}}(\omega)$. Similarly, we define $\zeta_{j,\ell,i}^{\uparrow}(\omega)$ as an indicator random variable such that $\zeta_{j,\ell,i}^{\uparrow}(\omega) = 1$ if the $\ell$-th arrival of type $j$ is routed to $i$ by $\ALG_{\bar{\bs{E}}}^{\uparrow}$ on the arrivals described by $\bD(\omega)$. Call ${\cal E}^j_{\rm infreq}$ the event where at least one infrequent query of type $j$ arrives by the end of the process. Because $\omega$ is a monotone coupling between $\bar{\bs{E}}$ and $\bD$, and $\ALG_{\bar{\bs{E}}},\ALG_{\bar{\bs{E}}}^{\uparrow}$ make the same decisions until reaching an infrequent query, we infer that for all $(j,\ell) \in \hat{\cal T}_{\rm freq}$,
\begin{eqnarray}\label{ineq:routings}
\zeta_{j,\ell,i}^{\uparrow}(\omega) \geq \left(1-\sum_{j' \neq j} \bb{I}[{\cal E}^j_{\rm infreq}]\right)\cdot \zeta_{j,\ell,i}(\omega)\ ,
\end{eqnarray}
where we only sum over $j' \neq j$ because all infrequent queries of type $j$ arrive after the frequent ones, and therefore, they cannot affect the routing decisions of frequent queries within the same type $j$. Now, observe  
\begin{eqnarray} \label{ineq:eps}
\sum_{j=1}^m\pr{{\cal E}^j_{\rm infreq}} = \sum_{(j,\ell)\in \hat{\cal T}_{\rm infreq}} \pr{D_j \geq \ell} \leq \sum_{(j,\ell)\in {\cal T}_{\rm infreq}} \pr{D_j \geq \ell} \leq 2\eps \ ,
\end{eqnarray}
where the last inequality follows from the definition of ${\cal T}_{\rm infreq}$.
By combining inequalities~\eqref{ineq:routings} and~\eqref{ineq:eps}, we infer from the independence across types that $\expar{\zeta_{j,\ell,i}^{\uparrow}(\omega)} \geq (1-2\eps) \expar{\zeta_{j,\ell,i}(\omega)}$. Reciprocally, we observe that $\prpar{E_j \geq \ell} \geq (1-2\eps)\prpar{D_j\geq \ell}$ by construction for each $(j,\ell) \in \hat{\cal T}_{\rm freq}$, implying that, for any such frequent query,
\begin{eqnarray} \label{ineq:revineq}
\ex{\zeta_{j,\ell,i}(\omega)} \geq (1-2\eps)\ex{\zeta^{\uparrow}_{j,\ell,i}(\omega)} \ .
\end{eqnarray}
 Now, denote by $\xi_{j,\ell,i}^{\uparrow}(\omega)$ the indicator of whether $(j,\ell)$ is eventually matched to resource $i$ by  $\ALG_{\bar{\bs{E}}}^\uparrow$ using the threshold rule in Claim~\ref{clm:prophet-var}. We obtain
\begin{eqnarray*}
\ex{\sum_{i=1}^n\sum_{j=1}^m\sum_{\MyAtop{\ell \geq 1:}{(j,\ell)\in \hat{\cal T}_{\rm freq}}} r_{i,j}\xi_{j,\ell,i}^{\uparrow}(\omega)} &\geq& \frac{(1-22\eps)}{2}\left( \sum_{i=1}^n\sum_{j=1}^m\sum_{\ell \geq 1} r_{i,j}\ex{\zeta_{j,\ell,i}^{\uparrow}(\omega)}\right)\\
&\geq& \frac{(1-24\eps)}{2}\left( \sum_{i=1}^n\sum_{j=1}^m\sum_{\ell \geq 1} r_{i,j}\ex{\zeta_{j,\ell,i}(\omega)}\right) \\
&=& \frac{(1-24\eps)}{2}\LPtrunc(\bar{\bs{E}}) \\ 
&\geq& \frac{(1-26\eps)}{2}\LPtrunc(\bar{\bD}) \ ,
\end{eqnarray*}
where in the first inequality, we apply Claim~\ref{clm:prophet-var} to each resource, noting that we indeed route each query type $j$ with probability $p_{j} = \sum_{\ell \geq 1: (j,\ell)\in \hat{\cal T}_{\rm freq}}\expar{\zeta_{j,\ell,i}^{\uparrow}(\omega)}$, while our algorithm believes the true rate to be $\tilde{p}_{j} = \sum_{\ell \geq 1: (j,\ell)\in \hat{\cal T}_{\rm freq}}\expar{\zeta_{j,\ell,i}(\omega)}$, which satisfies $ \expar{\zeta_{j,\ell,i}^{\uparrow}(\omega)} \geq (1-2\eps)\expar{\zeta_{j,\ell,i}(\omega)}$ as previously noted, and $ \sum_{(j,\ell)}\expar{\zeta_{j,\ell,i}^{\uparrow}(\omega)} \leq \sum_{(j,\ell)\in \hat{\cal T}_{\rm freq}}\expar{\zeta^{\uparrow}_{j,\ell,i}(\omega)} +\prpar{\bigcup_{j=1}^m {\cal E}^j_{\rm infreq}}\leq (1-2\eps)^{-1}\tilde{p}_j +2\eps$, where the last inequality follows from~\eqref{ineq:eps} and~\eqref{ineq:revineq}.
Note that $\ALG_{\bar{\bs{E}}}^\uparrow$ does not alter the fact that the routed infrequent queries are independent from the perspective of each resource. The first equality follows from  Lemma~\ref{lem:lossless} using the fact that $\ALG_{\bar{\bs{E}}}$ implements our lossless rounding with respect to an optimal solution of $\LPtrunc(\bar{\bs{E}})$. The last inequality follows from Claim~\ref{clm:sensitivity}, where $\bar{\bD}$ is defined be to $\bD$ with infrequent types removed. 

The remainder of the proof looks at the contributions of infrequent queries to the expected rewards. Here, we first consider the natural coupling $\omega$ between lossless routing decisions for $\LPtrunc({\bD})$ and the routing decisions of our algorithm $\ALG_{\bar{\bs{E}}}^{\uparrow}$ on $\bD$. Specifically, we implement~\Cref{alg:losslessRounding} with respect to an optimal solution of $\LPtrunc(\bD)$ and denote by $\delta_{j,\ell,i}(\omega) \in \{0,1\}$ the corresponding routing decisions for every query $(j,\ell)$ and resource $i$. In parallel, we denote by $\zeta_{j,\ell,i}^{\uparrow}(\omega)$ our algorithm's routing decisions for frequent queries, as described above, on the same sequence of arrivals. Next, we ``realign'' the routing decisions to concentrate their differences on infrequent queries. To be more specific, define $\delta^{\rm swap}_{j,\ell,i}(\omega)$ as a new indicator variable that captures the following modified routing decision: at the end of the horizon, for each fixed $j \in [m]$, we consider the set of resources $i \in C^{\uparrow}_j(\omega)$ that have been routed a {\em frequent} arrival of type $j$ by our algorithm $\ALG_{\bar{\bs{E}}}^{\uparrow}$ and the subset of resources $i\in C_j(\omega) \subseteq C^{\uparrow}_j(\omega)$ that were {\em also} routed any arrival of type $j$ by the lossless rounding of $\LPtrunc(\bD)$, i.e., $i\in C_j^{\uparrow}(\omega) $ if and only if  $\sum_{\ell \geq 1: (j,\ell)\in \hat{\cal T}_{\rm freq}}  \zeta_{j,\ell,i}^{\uparrow}(\omega)=1$, and $i\in C_j(\omega) $ if and only if $\sum_{\ell \geq 1} \delta_{j,\ell,i}(\omega) = \sum_{\ell \geq 1: (j,\ell)\in \hat{\cal T}_{\rm freq}}  \zeta_{j,\ell,i}^{\uparrow}(\omega)=1$. Then, we construct an alternative routing strategy for $\LPtrunc(\bD)$, denoted by $\delta^{\rm swap}_{j,\ell,i}(\omega) = \delta_{j,\pi_j(\ell),i}(\omega)$, where $\pi_j$ is a permutation ensuring that all common resources in $C_j(\omega)$ are rerouted frequent queries but the set of resources with a routed query is unchanged, i.e., $\pi_j$ swaps each infrequent arrival $\ell$ such that $\delta_{j,\ell,i} = 1$ for some $i\in C_j(\omega)$ with an earlier $\ell'$ such that $(j,\ell') \in \hat{\cal T}_{\rm freq}$ but $\delta_{j,\ell',i'} = 1$ for some $i'\notin C_j(\omega)$. Because both processes are coupled by $\omega$ and thus there is the same realized demand per type, this swapping is always feasible. Importantly, the swapped routing might not be implementable as an online algorithm because defining $\pi_j$ requires knowing the entire sequence of arrivals, but this won't matter for analysis. The main properties of the alternative routing are stated next.
\begin{claim} \label{clm:swap}
The swapped routing of $\LP(\bD)$ on $\omega$ satisfies:
\begin{enumerate}
\item $\expar{\sum_{i=1}^n\sum_{j=1}^m \sum_{\ell\geq 1} r_{i,j} \delta^{\rm swap}_{j,\ell,i}(\omega) } = \LPtrunc(\bD)$.
\item Let $x_{i,j} = \expar{\sum_{{\ell \geq 1:}{(j,\ell)\in \hat{\cal T}_{\rm freq}}}\delta^{\rm swap}_{j,\ell,i}(\omega)}$. Then, $(x_{i,j})_{i\in [n], j\in [m]}$ is feasible in $\LPtrunc(\bar{\bD})$.
\end{enumerate}
\end{claim}
Property~1 immediately follows from the fact swapping queries of the same type does not alter the total number of routing decisions per resource and query type, and thus $\sum_{\ell \geq 1}\delta^{\rm swap}_{j,\ell,i}(\omega) = \sum_{\ell \geq 1}\delta_{j,\pi_j(\ell),i}(\omega)$ on every realization of $\omega$. Property~2 follows from the fact that the fractional flow $(x_{i,j})_{i\in [n], j\in [m]}$ is the expectation of feasible matchings in the random graph between frequent queries in $\bD$ and resources in $[n]$; in particular, it satisfies inequalities that, in expectation, correspond to the truncated LP's constraints in $\LPtrunc(\bar{\bD})$.

With this coupling at hand, we are ready to compare the expected reward contributions of infrequent types in $\ALG_{\bar{\bs{E}}}^{\uparrow}$ to those described by $\delta^{\rm swap}_{j,\ell,i}(\omega)$. Upon the first infrequent arrival $(j,\ell_j)$ of type $j$, let $i_1^j, i^2_j, \ldots, i_j^{U_j}$ be the random sequence of resources in $[n]\setminus C^{\uparrow}_j(\omega)$ with $U_j= n - \ell_j+1$ rearranged by decreasing order of revenue $r_{i_1^j} \geq r_{i^2_j}\geq \ldots \geq r_{i^U_j}$. Note that $\ell_j$ and $U_j$ are deterministic. Then, it is clear that
\begin{eqnarray} \label{ineq:infreq-ub}
    \sum_{i\in [n]} \sum_{\ell \geq \ell_j}r_{i,j}\delta^{\rm swap}_{j,\ell,i}(\omega) \leq \sum_{u=1}^{U_j} {\bb I}[D_j
    \geq u+ \ell_j - 1] \cdot r_{i_u^j} \ ,
\end{eqnarray}
where the inequality holds because, by construction, all resources in $C_j(\omega)$ were routed a frequent query by $\delta^{\rm swap}_{j,\ell,i}(\omega)$, and thus, they cannot be routed an infrequent query later. Moreover, all resources in $i\in C^{\uparrow}_j(\omega)\setminus C_j(\omega)$ satisfy $\sum_{\ell\geq \ell_j}\delta^{\rm swap}_{j,\ell,i}(\omega) \leq \sum_{\ell\geq 1}\delta_{j,\ell,i}(\omega) = 0$ by the definition of $C_j(\omega)$. Therefore, out of the remaining resources $i\in [n]\setminus C^{\uparrow}_j(\omega)$, the best we can do is to select them by decreasing order of their rewards. We now argue that the contributions of infrequent queries matched by $\ALG_{\bar{\bs{E}}}^{\uparrow}$ to the expected rewards generate a fraction at least $\frac{(1-2\eps)}{2}$ of the righthand side in inequality~\eqref{ineq:infreq-ub}. By extension of our definition, let $\xi_{j,\ell,i}^{\uparrow}(\omega)$ be the indicator for whether a query  $(j,\ell)$ gets {\em matched} to resource $i$ by $\ALG_{\bar{\bs{E}}}^{\uparrow}$. Recall that ${\cal E}^j_{\rm infreq}$ is the event where at least one infrequent query of type $j$ arrives. Denote by ${\cal F}^j_{\rm infreq} = {\cal E}^j_{\rm infreq} \cap (\bigcap_{j'\neq j} \bar{\cal E}^j_{\rm infreq}) $ the event where $j$ is the only infrequent type that arrives. On the event $\{C_j(\omega) = C\} \wedge {\cal F}^j_{\rm infreq}$, note that (i) when $(j,\ell_j)$ arrive, we have only routed and matched frequent queries thus far, (ii) all type $j$ queries have been routed to $C_j(\omega)=C$, and (iii) for all other query types $j'\neq j$ and $i\in [n]\setminus C$, we have
\begin{eqnarray} \label{ineq:conditioning}
\pr{\left. \sum_{\ell =1}^{\ell_{j'}-1} \zeta_{j',\ell,i} = 1 \right| C_j(\omega) = C,{\cal F}^j_{\rm infreq}} & = &\sum_{\ell =1}^{\ell_{j'}-1} \pr{\left. \zeta_{j',\ell,i} = 1 \right|  \bar{\cal E}^{j'}_{\rm infreq}} \nonumber\\
& = & \sum_{\ell =1}^{\ell_{j'}-1}\pr{\left. \zeta_{j',\ell,i} =1\right| D_{j'}\geq \ell}\pr{\left. D_{j'}\geq \ell \right|  D_{j'}\leq \ell_{j'}-1} \nonumber\\
& \leq & \sum_{\ell =1}^{\ell_{j'}-1}\pr{\left. \zeta_{j',\ell,i} =1\right| D_{j'}\geq \ell}\pr{D_{j'}\geq \ell } \nonumber\\
& = & \pr{\sum_{\ell =1}^{\ell_{j'}-1} \zeta_{j',\ell,i} =1}\ .
\end{eqnarray}
In particular, combining this inequality  with Claim~\ref{clm:prophet-var} for resource $i \in [n] \setminus C$ yields
\begin{eqnarray} \label{ineq:inter-still}
\pr{\left. \sum_{j'\neq j}\sum_{\ell =1}^{\ell_{j'}-1} \xi^\uparrow_{j',\ell,i} = 1 \right| C_j(\omega) = C,{\cal F}^j_{\rm infreq} } \leq \frac{1+2\eps}{2} \ ,
\end{eqnarray}
where we remark that inequality~\eqref{ineq:conditioning}, and the fact that routing decisions are independent across types $j'\neq j$, imply that the distribution of the set of queries of types $j'\neq j$ routed to resource $i$, conditional on $C^\uparrow_j(\omega) = C$ and ${\cal F}^j_{\rm infreq}$, is stochastically dominated by the unconditional distribution of the set of queries of types $j'\neq j$ routed to the same $i$. Because Claim~\ref{clm:prophet-var} considers a static threshold rule, this implies that the total match rate for resource $i$ over all types $j'\neq j$ is at most $\frac{1+2\eps}{2} $. Now, we observe that we can lower bound the rewards from infrequent items chosen by $\ALG_{\bar{\bs{E}}}^{\uparrow}$ as follows:
\begin{eqnarray*}
&&\ex{\sum_{i=1}^n\sum_{j=1}^m\sum_{\MyAtop{\ell \geq 1:}{(j,\ell)\in \hat{\cal T}_{\rm infreq}}} r_{i,j}\xi_{j,\ell,i}^{\uparrow}(\omega)} \\
& & \quad\geq \sum_{j\in [m]} \sum_{\ell\geq \ell_j} \ex{\sum_{C\subseteq[n]} {\bb I}\left[{\cal F}^j_{\rm infreq}\right] {\bb I}[C^\uparrow_j(\omega) = C] \cdot \left(\sum_{u=1}^{U_j} {\bb I}[D_j
    \geq u+ \ell_j - 1] \cdot {\bb I}\left[\sum_{j'\neq j}\sum_{\ell =1}^{\ell_{j'}-1} \xi^\uparrow_{j',\ell,i} = 0 \right] r_{i_u^j}\right) }\\
    & & \quad\geq \frac{(1-2\eps)}{2} \cdot   \ex{\sum_{j\in [m]}\sum_{C\subseteq[n]} \sum_{u=1}^{U_j} \pr{\left. D_j
    \geq u+ \ell_j - 1, {\cal F}^j_{\rm infreq} \right| C^\uparrow_j(\omega) = C} \cdot r_{i_u^j}}\\
    & & \quad\geq \frac{(1-4\eps)}{2} \cdot   \ex{\sum_{j\in [m]} \sum_{u=1}^{U_j}  {\bb I}\left[ D_j
    \geq u+ \ell_j - 1\right] \cdot r_{i_u^j}}\\
     & & \quad\geq \frac{(1-4\eps)}{2} \cdot  \ex{ \sum_{i\in [n]} \sum_{(j,\ell)\in \hat{\cal T}_{\rm infreq}}r_{i,j}\delta^{\rm swap}_{j,\ell,i}(\omega)}\\
     & & \quad\geq\frac{(1-4\eps)}{2}  \cdot \left(\LPtrunc(\bD) -  \LPtrunc(\bar{\bD})\right) \ ,
\end{eqnarray*}
where the first inequality proceeds from the fact that our algorithm on the event ${\cal F}^j_{\rm infreq}$ selects the resources $ i_u^j$ by decreasing rewards, if they were not previously matched to a frequent query of type $j'\neq j$. The second inequality follows from inequality~\eqref{ineq:inter-still} and the independence of demand across types. The third inequality holds because $\prpar{D_j \geq u+\ell_j - 1, {\cal F}^j_{\rm infreq} | C^\uparrow_j(\omega) = C}  \geq (1-2\eps) \prpar{D_j \geq u+\ell_j - 1| C^\uparrow_j(\omega) = C}$ from inequality~\eqref{ineq:eps} and the independence of demand across types. The third inequality follows from~\eqref{ineq:infreq-ub}. The last inequality is direct a consequence of Claim~\ref{clm:swap}, by combining properties 1 and 2.

\section{Extension to Network Revenue Management and Nonstationary Stochastic Horizon} \label{app:stochastic_gen}

Our result for \Correl extends to a Network Revenue Management setting where each matching option for an incoming query consumes multiple resources, and also allows a
nonstationary stochastic horizon where the type distribution may vary over time.

In the nonstationary stochastic horizon setting, we are given a probability vector $(p_{t,j})_{j\in [m]}$ for each time $t \in [T]$ satisfying $\sum_{j \in [m]} p_{t,j} =1$. The type of the new query in each time $t\in [T]$ is drawn independently of the history according to the probabilities $(p_{t,j})_{j\in [m]}$.
Like before, the horizon length $D$ is stochastic in that the sequence of queries may terminate after any time $t \in [T]$.

In the Network Revenue Management (NRM) extension, there are
\textit{products} $\rho\in\cP$, where product $\rho$ requires one unit each of a set of resources $A_\rho\subseteq[n]$ to produce.  Type $j$ yields a reward of $r_{\rho,j}$ whenever it is "matched" with product $\rho$, where $r_{\rho,j}=0$ indicates incompatibility between a type and product.
We note that this is a more general version of NRM that includes ``matching'' decisions: that is, each arrival type $j$ can be routed to a product $\rho$, generating a reward $r_{\rho,j}$. The basic version of NRM is captured by setting $\cP=\{1,\ldots,m\}$ and having $r_{\rho,j}=0$ unless $\rho=j$---in that simplified setting, there is a single reward per product/customer type.
On the other hand, the online matching problem (without NRM) is captured by setting $\cP=\{1,\ldots,n\}$ and $A_\rho=\{\rho\}$ for all products/resources $\rho\in[n]$.
(In terms of whether our \Indep result extends to NRM: we believe our Lossless Rounding naturally extends, but we are unsure how to efficiently write/solve the analogue of our $\LPtrunc$ when resource constraints are coupled, so we leave this as an open problem.)

\begin{definition} \label{def:lpCondExt}
For any instance $\cI$ of NRM with nonstationary stochastic horizons, we define $\LPcond(\cI)$ as the optimal objective value of the following LP:
\begin{align*}
\max \ \ \ \ \ & \sum_{j=1}^m \sum_{\rho\in\cP} r_{\rho,j} \sum_{t=1}^{T} \Pr[D\geq t] \cdot y^t_{\rho,j}  \\ 
\text{s.t. } \ \ \ \ & \sum_{j=1}^m \sum_{\rho\in\cP: i\in A_\rho} \sum_{t=1}^T y^t_{\rho,j} \leq k_i &\forall i\in[n] \\ 
& \sum_{\rho\in\cP} y^t_{\rho,j} \leq  p_{t,j} &\forall j \in [m], \forall t \in [T] \\
&y^t_{\rho,j} \geq 0 &\forall \rho\in\cP, \forall j \in [m], \forall t \in [T] \ .
\end{align*}
\end{definition}
Decision variable $y^t_{\rho,j}$ represents the probability of matching a query of type $j$ to product $\rho$ at time $t$, conditional on $D\ge t$.  Using the same proof as in \Cref{pf:lem:comparison2} that conditions on $D=T$, we can see that $\ALG(\cI)\le\LPcond(\cI)$ for all online algorithms and hence $\OPT(\cI)\le\LPcond(\cI)$.

Our approximate online algorithm is then given in \Cref{alg:Correl_gen}, where the comments highlight the changes from \Cref{alg:Correl}.

\begin{algorithm}
\caption{Algorithm for NRM matching with nonstationary stochastic horizon} \label{alg:Correl_gen}
\begin{algorithmic}
\State Solve $\LPcond(\cI)$ from \Cref{def:lpCondExt}, letting $(y^t_{\rho,j})_{t\in[T], \rho\in\cP, j\in [m]}$ denote an optimal solution
\State $\inv[i]=k_i$ for all $i\in[n]$
\For{$t=1,\ldots,D$}
\State Let $j$ denote type of query $t$ \Comment{nonstationary probability vector $(p_{t,j})_{j\in[m]}$}
\State Choose $\rho$ independently from probability vector $\left(\frac{y^t_{\rho,j}}{p_{t,j}}\right)_{\rho\in\cP}$
\Comment{$\rho$ is a product}

(set $\rho$ to $\perp$ with probability $1-\sum_{\rho\in\cP}\frac{y^t_{\rho,j}}{p_{t,j}}$)
\If{$\rho\neq\perp$ and $\Acc_{i,t}(\inv[i])=1$ for all $i\in A_\rho$} \Comment{every resource in $A_\rho$ must accept}
\State Match the query to product $\rho$, collecting reward $r_{\rho,j}$
\State $\inv[i]=\inv[i]-1$ for all $i\in A_\rho$
\EndIf
\EndFor
\end{algorithmic}
\end{algorithm}

We now require every resource in $A_\rho$ to accept in order to match a product $\rho$, where each resource $i$ runs its own OCRS with a starting inventory of $k_i$ units (ignoring the fact that the NRM problem ``couples'' the resources).  We recall that if $\Acc_{i,t}(\inv[i])=1$ then it is guaranteed for resource $i$ to have remaining inventory.
Similarly to our original analysis of Section~\ref{sec:correl}, we consider an alternative world where time $t$ always runs until the maximum $T$ but we only account for the rewards collected in times  $t\le D$:
\begin{align*}
\ALG(\cI)
&=\sum_{t=1}^T\sum_{j=1}^m p_{t,j} \sum_{\rho\in\cP} \frac{y^t_{\rho,j}}{p_{t,j}}\Pr\left[\bigcap_{i\in A_\rho}(\Acc_{i,t}(\inv[i])=1)\right] r_{\rho,j} \Pr[D\ge t]
\\ &=\sum_{t=1}^T\sum_{j=1}^m \sum_{\rho\in\cP} y^t_{\rho,j}\left(1-\Pr\left[\bigcup_{i\in A_\rho}(\Acc_{i,t}(\inv[i])=0)\right]\right) r_{\rho,j} \Pr[D\ge t]
\\ &\ge\sum_{t=1}^T\sum_{j=1}^m \sum_{\rho\in\cP} y^t_{\rho,j}\left(1-\sum_{i\in A_\rho}(1-\Pr[(\Acc_{i,t}(\inv[i])=1)])\right) r_{\rho,j} \Pr[D\ge t]
\\ &\ge\sum_{t=1}^T\sum_{j=1}^m \sum_{\rho\in\cP} y^t_{\rho,j}\left(1-\sum_{i\in A_\rho}(1-\gamma^*_{k_i})\right) r_{\rho,j} \Pr[D\ge t]
\\ &\ge(1-L(1-\gamma^*_k))\LPcond(\cI)
\end{align*}
where the first inequality applies the union bound,
the second inequality follows from the OCRS guarantee,
and the final guarantee assumes that $|A_\rho|\le L$ for all $\rho$, and $k_i\ge k$ for all $i$ (recall that the tight OCRS guarantee $\gamma^*_k$ is increasing in $k$).
This results in the following \namecref{thm:main2_gen}.
\begin{theorem} \label{thm:main2_gen}
For the nonstationary stochastic horizon and NRM extensions, \Cref{alg:Correl_gen} is a polynomial-time online algorithm that is $(1-L(1-\gamma^*_k))$-approximate for every instance in which all products $\rho\in {\cal P}$ require at most $|A_\rho|\leq L$ resources and all resources $i\in [m]$ start with inventory at least $k_i \geq k$.
\end{theorem}

We note that the preceding argument is based on the recent work of \citet{amil2022multi}, extending their performance guarantee to the  stochastic horizon setting.
\Cref{thm:main2_gen} can also be seen as improving the guarantee from \citet{bai2022fluid} in two ways: extending NRM to have ``matching'' decisions, and improving the convergence rate to 1 when $L$ is fixed and $k$ tends to $\infty$.  To see this, note that $\gamma^*_k\ge1-1/\sqrt{k+3}$ \citep{jiang2022tight,alaei2012bayesian}, and thus, our algorithm's approximation ratio is at least
\begin{align*}
1-\frac{L}{\sqrt{k+3}}=1-O_L\left(\sqrt{\frac1k}\right)
\end{align*}
whereas \citet{bai2022fluid} establish an approximation ratio that is $1-O_L(\sqrt{\frac{\log k}{k}})$.

\section{Concrete Examples of our Algorithmic Advancements} \label{sec:toyExamples}

We construct simple examples illustrating why our algorithms may outperform existing ones under the $\Indep$ and $\Correl$ models.
Although these are toy examples, they may reasonably depict practical tradeoffs in real-world scenarios.
In particular, \Cref{eg:resourceAllocation} represents a resource allocation setting where a central planner would like to maximize welfare, but the tradeoff is that the most valuable resources (which generate the largest rewards when allocated) are also the most flexible and perhaps should be saved for the most picky customer types.
\Cref{eg:Wgraph} is a typical service operations setting where two demand types each have a dedicated server and also share a common server.
Finally, \Cref{eg:dynamicPricing} is a revenue management setting where $k$ units of a single item can be dynamically priced, except that there could be a large common shock to the aggregate demand, i.e., the item is either a runaway hit or a total flop.

\begin{example}[Differences between $\LPtrunc$ and fluid LP] \label{eg:resourceAllocation}
Consider a simple example with two resources and two demand types.
Resource 1 is the superior resource, compatible with both demand types and yielding a high reward when matched.
Resource 2 is the inferior resource, compatible with only demand type 2 (the less picky type) and yielding a low reward when matched.
Suppose that both resources start with $k$ units and the expected demand of each type is exactly $k$.
The fluid LP, optimizing for a world with no demand variability, would suggest dedicating resource 1 to serving type 1 and resource 2 to serving type 2, perfectly matching supply and demand.
However, this dedicated service is undesirable in the true stochastic environment with demand variability---it could leave the most valuable resources unutilized with a large probability.
By contrast, our $\LPtrunc$ allocates a portion of resource 1 to demand type 2 to increase the probability that it is utilized; the portion size depends on the level of demand variance and the ratio of high to low reward.

To formalize this, let $n=2$, $m=2$, $r_{1,1}=r_{1,2}=r_H$, and $r_{2,2}=r_L,r_{2,1}=0$.  Here we let $r_H,r_L$ denote the High, Low rewards with $r_H\ge r_L$, and resource 2 is the inferior one,  which cannot serve the picky type 1.  The starting inventory is set as $k_1=k_2=k$ for some positive integer $k$, and let $D_1,D_2$ both be independent Poisson random variables with mean $k$, denoted by $\Pois(k)$.
Under the dedicated service policy described by the optimal fluid LP solution, total expected reward is
\begin{align} \label{lp:perf_actual}
r_H\ex{\min\{\Pois(k),k\}}+r_L\ex{\min\{\Pois(k),k\}}
&=(r_H+r_L)k \left(1-e^{-k}\frac{k^k}{k!}\right).
\end{align}
However, this is quite suboptimal when $r_H$ is much larger than $r_L$, because the expected fraction of the superior resource that goes untilized is $e^{-k}k^k/k! = \Theta((2\pi k)^{-\frac{1}{2}})$, which decays slowly as a function of $k$.  Strikingly, an "extreme policy" that always allocates the superior resource has total expected reward $r_H\bE[\min\{\Pois(2k),k\}]$, which can be higher than the expression in~\eqref{lp:perf_actual} when the ratio $r_H/r_L$ is sufficiently large.
In contrast $\LPtrunc$ balances between the two extremes of maximizing utilization, i.e., maximizing expected demand served via dedicated service versus maximizing the expected amount of resource 1 allocated.  Depending on the ratio $r_H/r_L$ and the exact demand distributions, our $\LPtrunc$ prescribes a solution that lies in-between these two extremes, partially using resource 1 to serve type 2.

It is not easy to quantify the exact benefit of $\LPtrunc$ in the Poisson setting as a closed-form expression. To exhibit concrete numbers, we consider a modification to the preceding example where $k=1$, $D_1$ is equally likely to be 0 or 2, and $D_2$ is deterministically 1.  The fluid LP still suggests dedicated service, with an expected total reward of $r_H/2+r_L$.   In contrast, if type $1$ queries arrive before the type $2$ query, our truncated LP yields the exact optimal solution $x_{1,1}=x_{1,2}=x_{2,2}=1/2,x_{2,1}=0$.  Now, if the type $2$ query arrives before type $1$ queries, the expected reward would still be $(3/4)r_H+r_L/2$, exceeding $r_H/2+r_L$ as long as $r_H/r_L > 2$.
\Halmos\end{example}

We remark that in \Cref{eg:resourceAllocation}, the fluid LP's solution was undesirable even though the Poisson demand distribution was "low variance" in that its variance is no greater than its mean.

\begin{example}[Lossless Rounding outperforms Independent Rounding] \label{eg:Wgraph}
First, we explain the shortcomings of Independent Rounding and its "In-stock" improvement that was tested in \Cref{sec:simulations}.
Suppose for a type $j$ that $D_j\in \{1,2\}$ with $\prpar{D_j=2} = \rho$ and $\prpar{D_j = 1}= 1-\rho$.
Consider a fractional solution where $x_{j,j}=1$ and $x_{0,j}=\rho$, with $j$ and 0 being different resources (note that this solution satisfies constraints~\eqref{ineq:truncDemand} in our $\LPtrunc$).
Independent Rounding routes each arrival of type $j$ independently to resources $j$ and $0$ with probabilities $1/(1+\rho)$ and $\rho/(1+\rho)$ respectively.
This preserves the marginals in that the expected numbers routed to resources $j$ and $0$ are $1$ and $\rho$, which equal $x_{j,j}$ and $x_{0,j}$, respectively; however, the downside is that multiple arrivals may be routed to the same resource and that resource may only start with one unit of inventory.
In-stock Independent Rounding routes the first arrival of type $j$ to resources $j$ and $0$ with probabilities $1/(1+\rho)$ and $\rho/(1+\rho)$ respectively, but differs by always routing the second arrival to the other resource.
This guarantees to route to different resources; however, marginals are not preserved---it can be checked that the expected numbers routed to resources $j$ and $0$ are $(1+\rho^2)/(1+\rho)$ and $2\rho/(1+\rho)$, which do not equal 1 and $\rho$ respectively unless\footnote{
The condition
$\rho\in\{0,1\}$ means that $D_j$ is deterministic.  Even in this case, it is possible to construct examples where In-stock Independent Rounding does not satisfy both criteria.  Indeed, suppose $\prpar{D_j=2}=1$ and there are 3 resources with $x_{1,j}=1,x_{2,j}=x_{3,j}=1/2$.  In-stock Independent Rounding routes to resource set $\{2,3\}$ with probability $\frac{1/2+1/2}{1+1/2+1/2}\cdot\frac{1/2}{1+1/2}=1/6$, which means that resource 1 only gets routed a query with probability 5/6.
} $\rho\in\{0,1\}$.
Our Lossless Rounding satisfies both the criteria of preserving the marginals and routing to different resources, which on this example means routing the first arrival of type $j$ to resource $j$, and the second to resource 0.

To see why this can improve the overall performance, consider an example where resources $i=0,1,2$ each have starting inventory $k_i = 1$.
There are $m=2$ demand types with $r_{1,1}=r_{2,2}=r_{0,1}=r_{0,2}=1$, and all other rewards are 0.
Demands $D_1$ and $D_2$ are each independently equal to $1$ or $2$ with probabilities $1-\rho$ and $\rho$, respectively.
An optimal solution to $\LPtrunc$ sets $x_{1,1}=x_{2,2}=1,x_{0,1}=x_{0,2}=\rho$, and all unspecified $x_{i,j}$'s are equal to 0.
Regardless of arrival order, our Lossless Rounding always matches resources 1,2 and matches resource 0 with probability $1-(1-\rho)^2$, for a total expected reward of
\begin{align} \label{eqn:lossless}
2+2\rho-\rho^2.    
\end{align}
Independent Rounding would match each resource $1,2$ with probability
$
1-\frac{\rho}{1+\rho}(1-\rho\frac1{1+\rho})=\frac{1+\rho+\rho^2}{(1+\rho)^2}
$
and resource 0 with probability
$
1-(\frac1{1+\rho})^2(1-\rho\frac\rho{1+\rho})^2=1-\frac{(1+\rho-\rho^2)^2}{(1+\rho)^4},
$
for a total expected reward of
\begin{align} \label{eqn:indepRound}
2\frac{1+\rho+\rho^2}{(1+\rho)^2}+1-\frac{(1+\rho-\rho^2)^2}{(1+\rho)^4}.
\end{align}
In-stock Independent Rounding has a more nuanced state trajectory that depends on the arrival order.  Whichever type $j$ arrives first, it will be routed to resource 0 with probability $\rho/(1+\rho)$, conditional on which the total expected reward is $2+\rho$ (resources 0 and $2-j$ get matched; resource $j$ gets matched if there is another arrival of type $j$).  On the other hand, conditional on the first arrival not being routed to resource 0, the best-case arrival order is that the same type arrives again.  If it does (happening with probability $\rho$), then the total reward is 3; otherwise, the conditional total expected reward is again $2+\rho$.  Putting it together, the unconditional total expected reward of In-stock Independent Rounding is
\begin{align}
\frac{\rho}{1+\rho}(2+\rho)+\frac1{1+\rho}\rho(3)+\frac1{1+\rho}(1-\rho)(2+\rho)
&=\frac{2+\rho}{1+\rho}+\frac{3\rho}{1+\rho} \nonumber
\\ &=2+2\rho-\frac{2\rho^2}{1+\rho}. \label{eqn:indepResamp}
\end{align}
We compare these expressions in the following plot, where the best performance (highest curve) is attained by Lossless Rounding (\cref{eqn:lossless}), the second-best performance is attained by In-stock Independent Rounding (\cref{eqn:indepResamp}), and the worst performance is attained by Independent Rounding (\cref{eqn:indepRound}).
\begin{center}
\includegraphics{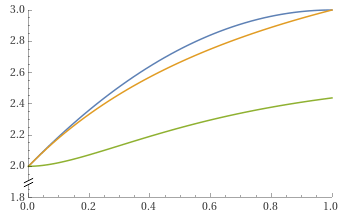}
\end{center}
Our Lossless Rounding outperforms In-stock Independent Rounding even under its best-case arrival order, and the difference in performance is most pronounced for medium values of $\rho$, which represent the highest variance of the demand distribution.
\Halmos\end{example}

\begin{example}[Differences between $\LPcond$ and fluid LP] \label{eg:dynamicPricing}
Support there is a single resource with $k$ units, where $k$ is a large even positive integer.
We will omit the resource index $i$.
Consider our $\Correl$ model where the total demand $D$ is equally likely to be $k/2$ or $3k/2$.
There are two demand types indexed by $j=L$ (Low) and $j=H$ (High), with $r_L<r_H$ and $p_L=1/3,p_H=2/3$.
Similar conclusions could be drawn for an arbitrary price curve, but this simple price curve is sufficiently illustrative.

We first explain our algorithm's decisions.  Our $\LPcond$ can be expressed as follows, after defining aggregate variables $\ypre_j:=\sum_{t=1}^{k/2}y^t_j$ and $\ypost_j:=\sum_{t=k/2+1}^{3k/2}y^t_j$ for $j\in\{L,H\}$:
\begin{align*}
\max\ &r_L\ypre_L+r_H\ypre_H+\frac{r_L}{2}\ypost_L+\frac{r_H}{2}\ypost_H
\\ \text{s.t. } &\ypre_L+\ypost_L+\ypre_H+\ypost_H \le k
\\ &0\le \ypre_L\le k/6
\\ &0\le \ypre_H\le k/3
\\ &0\le \ypost_L\le k/3
\\ &0\le \ypost_H\le 2k/3\ .
\end{align*}
The optimal solution always sets $\ypre_H$ to its upper bound of $k/3$, because this variable has the largest objective coefficient of $r_H$.
The optimal solution would then fill variables $\ypre_L,\ypost_H$ to their upper bounds, with the prioritization depending on whether $r_L\ge r_H/2$.
Indeed, if $r_L\ge r_H/2$, then it is optimal to set $\ypre_L=k/6,\ypost_H=k/2,\ypost_L=0$ (because $k/3+k/6+k/2=k$), in which case the algorithm would accept both the High and Low types at the beginning, and then only accept the High type conditional on surviving past time $k/2$ (i.e, conditional on $D$ realizing to $3k/2$).
On the other hand, if $r_L<r_H/2$, then it is optimal to set $\ypre_L=0,\ypost_H=2k/3,\ypost_L=0$, in which case the algorithm would only accept High types at the beginning, so that a maximum amount of resource is preserved in case $D$ realizes to $3k/2$.

The fluid LP, by contrast, sets $\ypre_L=k/6$ (i.e., accepting all Low types at the beginning) regardless of whether $r_L\ge r_H/2$, because the expected demand is $k$ which is no greater than the starting inventory.  If $r_L<r_H/2$, then the fluid LP foregoes $(r_H/2-r_L)k/6$ in expected rewards\footnote{As a rough approximation, we assume that a fraction exactly 1/3 or 2/3 of the arrivals has types $L$ and $H$, respectively.  There could be $O(\sqrt{k})$-deviations from this quantity, which are of second order compared to the fluid LP's loss of $(r_H/2-r_L)k/6$ when $k$ is large.} compared to using our $\LPcond$. It is worth noting that this logic holds even for an "intelligent re-solving" heuristic, which continuously reoptimizes the fluid LP based on the remaining inventory and conditional expected demand; in that context, only High types are accepted after the $k/2$-th arriving query.  The fluid LP foregoes revenue because it is agnostic to the variance of $D$; it fails to compare the ratio $r_H/r_L$ of high to low reward to the probability 1/2 that the demand is large $D=3k/2$. In fact, there is a newsvendor-like tradeoff in choosing the service levels in the first stage and second stage of the demand realization. Although the distribution of $D$ in this example is quite stylized to ease calculations, we believe that the intuition can be extrapolated to more general demand distributions.
\Halmos\end{example}

\section{Additional Proofs from Section~\ref{sec:indep} }

\subsection{Proof of Proposition~\ref{prop:loose1}} \label{pf:prop:loose1}
We first present construction (i).  Consider a family of instances parametrized by $\eps\in (0,1)$. Each instance comprises a single resource $i=1$ with starting inventory $k_i=1$ and a single query type $j=1$ with the normalized reward $r_{1,1} = 1$. The demand random variable $D_1$ takes two values $\{0,\lceil \frac{1}{\eps} \rceil\}$ with probabilities $\prpar{D_1 =0} = 1-\eps$ and $\prpar{D_1 = \frac{1}{\eps} \rceil} = \eps$. Because $\expar{D_1} =1$, it is clear that $\LP({\cal I}) = 1$. However, the single resource can be matched to at most one query, which only occurs with probability $\prpar{D_1 >0} = \prpar{D_1 = \frac{1}{\eps} \rceil} =\eps$. We have just shown that $\frac{\OFF({\cal I})}{\LP({\cal I})} = \eps$, which proves part (i) of \Cref{prop:loose1}.

We now present our construction for part (ii).  Fix a large $n$, and let $k_i=1$ for all $i\in[n]$.  Let $m=1$, and let $r_{i,1}=0$ unless $i=1$, in which case $r_{1,1}=1$.  Let $D_1$ take the value $n$ with probability~$\frac{1}{n}$, and take the value 0 with the residual probability $1-\frac{1}{n}$.  Note that $D_1$ is indeed no greater than $\sum_{i=1}^n k_i=n$ with probability~1.  It is feasible in the fluid relaxation LP to set $x_{1,1}=1$, yielding an objective value $1$.  Meanwhile, any algorithm can match query type 1 to resource type 1 with probability at most $\frac{1}{n}$ (when $D_1=n$).  Therefore, we have shown that $\frac{\OFF(\cI)}{\LP(\cI)}=\frac{1}{n}$, where $n$ can be  arbitrarily large, thereby completing the proof of \Cref{prop:loose1}.

\subsection{Proof of Lemma~\ref{lem:comparison1}} \label{pf:lem:comparison1}

To show that $\OFF({\cal I}) \leq \LPtrunc({\cal I})$, we represent the offline-optimum as the output of a certain exponentially sized linear program.  The offline benchmark solves a max-weight matching problem with respect to any specific realization $\bd$ of the demand $\bD$. Letting $\bar{\bd}= \{\bd : \prpar{\bD = \bd >0}\}$ be the support of $\bD$, the offline benchmark can be formulated as:
\begin{align}
\OFF({\cal I}) = \ \ \max\ \ &\sum_{\bd \in  \bar{\bd}} \pr{\bD = \bd} \cdot \left(\sum_{i=1}^n\sum_{j=1}^m r_{i,j}x^{\bd}_{i,j} \right) &
\nonumber\\ 
\text{s.t. }&\sum_{j=1}^m x^{\bd}_{i,j} \leq k_i &\forall i\in[n], \forall \bd \in \bar{\bd} \label{cons:a1}\\ 
&\sum_{i=1}^n x^{\bd}_{i,j} \leq \bd_j & \forall j \in [m], \forall \bd \in \bar{\bd} \label{cons:a2}
\\ 
&x^{\bd}_{i,j} \geq 0 &\forall i\in[n],\forall j\in[m], \forall \bd \in \bar{\bd} \nonumber\ .
\end{align}
Fix a feasible solution $(x^{\bd}_{i,j})_{i\in[n],j\in[m],\bd\in\bar{\bd}}$ of the above LP. For all $\bd \in \bar{\bd}$, $j\in [m]$, and $S\subseteq [n]$, we have $\sum_{i\in S} x^{\bd}_{i,j} \leq \sum_{i=1}^n x^{\bd}_{i,j} \leq \bd_j$ based on constraint~\eqref{cons:a2}. By summing  inequalities~\eqref{cons:a1} over all $i\in S$, we infer that $\sum_{i\in S} x^{\bd}_{i,j} \leq \sum_{i\in S} \sum_{j=1}^m x^{\bd}_{i,j} \leq \sum_{i\in S} k_i$. By combining these inequalities, we infer that any solution of the offline LP must satisfy $\sum_{i\in S} x^{\bd}_{i,j} \leq \min\{\bd_j,\sum_{i\in S} k_i\}$. Now, we consider the vector $(x_{i,j})_{i\in[n],j\in[m]}$ with $x_{i,j}  =\sum_{\bd \in \bar{\bd}} \prpar{\bD = \bd}\cdot x^{\bd}_{i,j}$, obtained as the weighted sum of the offline assignment variables. Based on the previous observation, for each $S\subseteq [n]$ and $j\in [m]$, we have
\[
\sum_{i\in S} x_{i,j} = \sum_{\bd \in \bar{\bd}} \prpar{\bD = \bd}\cdot\left( \sum_{i\in S} x^{\bd}_{i,j} \right) \leq \sum_{\bd \in \bar{\bd}} \prpar{\bD = \bd}\cdot \min\left\{\bd_j,\sum_{i\in S} k_i\right\} = \ex{\min\left\{D_j, \sum_{i\in S} k_i\right\}} \ ,
\]
which indicates that $(x_{i,j})_{i\in[n],j\in[m]}$ satisfies constraint~\eqref{ineq:truncDemand} of $\LPtrunc({\cal I})$. It is straightforward to verify that all other constraints of $\LPtrunc({\cal I})$ are also met by $(x_{i,j})_{i\in[n],j\in[m]}$. By exploiting this mapping, it follows that $\OFF({\cal I}) \leq \LPtrunc({\cal I})$.

\subsection{Proof of \Cref{clm:sep}} \label{sec:pfOfNewClaim}

We seek to establish that for any $j\in[m]$, if we re-index $i$ so that $\frac{x_{1,j}}{k_1}\ge\cdots\ge\frac{x_{n,j}}{k_n}$, then the exponential family of inequalities~\eqref{eqn:rewriteTrunc} for type $j$ is implied by the following linearly many inequalities:
    \begin{align} \label{eqn:sufficientTrunc}
x_{1,j}+\cdots+x_{i,j} &\le \sum_{\ell=1}^{k_1+\cdots+k_i}\Pr[D_j\ge \ell]  \ \ \ \ \  \ \ \ \ \ \ \forall i\in [n] \ .
\end{align}
In fact, we show that~\eqref{eqn:sufficientTrunc} implies a family of constraints which is even more general:
\begin{align} \label{eqn:granularTrunc}
\sum_{i=1}^n S_i\frac{x_{i,j}}{k_i} &\le \sum_{\ell=1}^{S_1+\cdots+S_n}\Pr[D_j\ge\ell] &\forall S_1\in[k_i]\cup\{0\},\ldots,S_n\in[k_n]\cup\{0\}\ .
\end{align}
To show that~\eqref{eqn:sufficientTrunc} implies~\eqref{eqn:granularTrunc}, take any inequality in~\eqref{eqn:granularTrunc} and let $\ell'$ denote $S_1+\cdots+S_n$, which we assume is non-zero.
We must show that the LHS of~\eqref{eqn:granularTrunc} is no greater than $\sum_{\ell=1}^{\ell'}\Pr[D_j\ge\ell]$.
Notice that the LHS of~\eqref{eqn:granularTrunc} can be interpreted as selecting $S_i$ units of every resource $i\in[n]$, and accumulating $x_{i,j}/k_i$ for each unit of $i$ selected.  From this interpretation, we can see that subject to the constraint $S_1+\cdots+S_n=\ell'$, the LHS of~\eqref{eqn:granularTrunc} is maximized by setting $S_i=k_i\ \forall i<i'$, $S_{i'}=\ell'-\sum_{i<i'} k_i$, $S_i=0\ \forall i>i'$, for some unique index $i'$ in $[n]$ such that this definition of $S_{i'}$ lies in $[k_{i'}]$ (this is because the indices $i$ are sorted in decreasing order of $x_{i,j}/k_i$).
Adding $1-S_{i'}/k_{i'}$ times constraint~\eqref{eqn:sufficientTrunc}, where we set $i=i'-1$ (with the understanding that this constraint is $0\le 0$ if $i'=1$), to $S_{i'}/k_{i'}$ times constraint~\eqref{eqn:sufficientTrunc}, where we set $i=i'$, we obtain
\begin{align*}
\sum_{i<i'} x_{i,j} + \frac{S_{i'}}{k_{i'}}x_{i',j}
&\le\sum_{\ell=1}^{k_1+\cdots+k_{i'-1}}\Pr[D_j\ge\ell]+\frac{S_{i'}}{k_{i'}}\sum_{\ell=k_1+\cdots+k_{i'-1}+1}^{k_1+\cdots+k_{i'}}\Pr[D_j\ge\ell]
\\ &\le\sum_{\ell=1}^{k_1+\cdots+k_{i'-1}}\Pr[D_j\ge\ell]+\sum_{\ell=k_1+\cdots+k_{i'-1}+1}^{k_1+\cdots+k_{i'-1}+S_{i'}}\Pr[D_j\ge\ell]
\end{align*}
where the second inequality holds because $\Pr[D_j\ge\ell]$ is decreasing in $\ell$.  This shows that even by maximizing the LHS of~\eqref{eqn:granularTrunc}, we cannot exceed its RHS for this value of $\ell'$, completing the proof that~\eqref{eqn:sufficientTrunc} implies~\eqref{eqn:granularTrunc} which in turn implies~\eqref{eqn:rewriteTrunc}.  

The other direction of \Cref{clm:sep} is immediate because~\eqref{eqn:rewriteTrunc} is a superset of the constraints in~\eqref{eqn:sufficientTrunc}.  This completes the proof of \Cref{clm:sep}.

\subsection{Proof of \Cref{prop:main1tight}} \label{pf:prop:main1tight}

Let $n=1$ and fix any starting inventory $k_1$, which can be arbitrarily large.  There are $m=2$ query types, with $r_{1,1}=1$ and $r_{1,2}=1/\eps$, for some small $\eps>0$.  The demand vector $\bD$ realizes to $(k_1,k_1)$ with probability $\eps$, and $(k_1,0)$ with the residual probability $1-\eps$.  The arrival order is such that all queries of type 2 arrive after all queries of type 1.

The offline optimum collects reward $\frac{k_1}{\eps}$ when $D_2=k_1$, and $k_1$ when $D_2=0$, which implies in expectation that 
\[
\OFF(\cI)=\eps\cdot \frac{k_1}{\eps}+(1-\eps)\cdot k_1=(2-\eps)k_1\ .
\]
Meanwhile, any online algorithm does not know whether $D_2$ will be $k_1$ or 0 during the first $k_1$ arrivals, corresponding to all queries of type 1.
Suppose it accepts $k$ queries of type 1, for some $k\le k_1$.  Then, it will collect expected rewards $k+\eps\cdot(k_1-k)\cdot \frac{1}{\eps}=k_1$.
Therefore, $\ALG(\cI)\le k_1=\frac{1}{2-\eps}\cdot \OPT(\cI)$, and taking the limit $\eps\to0$ completes the proof.

\section{Supplement to Simulation \Cref{sec:simulations}} \label{sec:simulationsSupplement}

\textbf{Offline LP}: we employ variables $(x_{i,j})_{i\in[n],j\in[m]}$ from an optimal solution to the following LP:
\begin{align*}
\max\ \ \ \ \ & \sum_{i=1}^n\sum_{j=1}^m r_{i,j}x_{i,j}
\\ \text{s.t. } \ \ \ \ & x_{i,j}=\frac1M\sum_{s=1}^M x^s_{i,j} &\forall i\in[n],\forall j\in[m]
\\ & \sum_{j=1}^m x^s_{i,j} \leq k_i &\forall i\in[n],\forall s\in[M]
\\ &\sum_{i=1}^n x^s_{i,j} \leq D^s_j &\forall j \in [m],\forall s\in[M]
\\ & x^s_{i,j} \ge 0 &\forall i\in [n],\forall j\in[m],\forall s\in[M] \ .
\end{align*}
In the offline LP, each $s$ indexes a demand scenario $(D^s_j)_{j\in[m]}$, and $(x^s_{i,j})_{i\in[n],j\in[m]}$ is an optimal offline matching under that scenario.  The employed variable $x_{i,j}$ is the average value of $x^s_{i,j}$ across scenarios $s$, and the optimal objective value coincides with the expected offline value that we earlier denoted using $\OFF(\cI)$ (assuming the $M$ scenarios fully capture the distribution of $(D_j)_{j\in[m]}$).
Although the Fluid LP is convenient for theoretical analysis, it has been observed that the Offline LP usually improves empirical performance \citep{talluri1999randomized,devalve2021understanding}.

The feasible region of the Offline LP lies within that of the Truncated LP.
Indeed, constraints~\eqref{ineq:truncDemand} in $\LPtrunc$ are satisfied if the $M$ demand scenarios fully capture the distribution of $(D_j)_{j\in[m]}$, because the Offline LP's solution satisfies $\sum_{i\in S}x^s_{i,j}\le\min\{D^s_j,\sum_{i\in S}k_i\}$ for every scenario $s$.  Therefore, we can use Lossless Rounding also on the Offline LP's solution.

\textbf{Defining the LP's for different instances}:
For \Indep synthetic instances, it is clear how to define the Fluid and Truncated LP's, and we omit the Conditional LP.
For \Correl synthetic instances, it is clear how to define the Fluid and Conditional LP's, and we omit the Truncated LP.
In the Offline LP, we always sample $M=100$ times for the synthetic distributions, which appears sufficient to capture the true distributions.

For real-data instances, to define the Fluid and Truncated LP's, we use the true distribution of (up to 12) monthly demands from a location $j$ (for the SKU) to evaluate the expectations involving $D_j$.
For the Conditional LP, we evaluate probabilities involving $D=\sum_{j=1}^m D_j$ in the same way, and set each $p_j$ to be the fraction of all orders across months (for the SKU) to come from location $j$.
For the Offline LP, each demand scenario $(D^s_j)_{j\in[m]}$ represents the actual demands over a month $s\in[M]$ (where $M\le12$).
We remark that in these ways of defining the LP's from real data, the Fluid and Truncated LP's are agnostic to the correlation of $D_j$ across $j$, the Conditional LP is agnostic to how a given $D_j$ may vary across months if it is not proportional to $D$, while the Offline LP is given the most information.

\textbf{Defining $q_i$ for the IR variants}: We have $q_i=x_{i,j}/\bE[D_j]$ for the Fluid LP, $q_i=x_{i,j}/\bE[\min\{D_j,\sum_{i'=1}^n k_{i'}\}]$ for the Truncated LP, $q_i=y^t_{i,j}/p_j$ for the Conditional LP, and $q_i=x_{i,j}/(\frac1M\sum_{s=1}^M D^s_j)$ for the Offline LP.  (Note that $q_i$ depends on the time $t$ for the Conditional LP.)

\textbf{Motivation for testing In-stock IR}: This rounding method is intuitive and is also known as "boosting" \citep{ma2023fairness} or "sampling without replacement" \citep{huang2024online}, with both papers demonstrating its strong empirical performance.

\section{De-randomization Methods}

\paragraph{\Indep model: De-randomizing Algorithms~\ref{alg:indep} and~\ref{alg:losslessRounding}.} De-randomizing Algorithm~1 is non-trivial because an adversary selects the arrival order, and hence simulation into the future is no longer possible.  Nonetheless, technically there should exist a deterministic algorithm that is 1/2-competitive because our performance guarantee holds even against an almighty adversary that knows the realizations of the random algorithm's bits.  Therefore, letting $\ALG$ denote the {\em realized deterministic} algorithm, we have
\begin{align*}
&& \bE_{\bD,\ALG}[\inf_\sigma\ALG_\cI(\bD,\sigma)]&\ge\bE_\bD[\OFF(\bD)]/2
\\ \Longrightarrow && \bE_{\ALG}[\bE_{\bD}[\inf_\sigma\ALG(\bD,\sigma)]]&\ge\bE_\bD[\OFF(\bD)]/2
\\ \exists\ALG\textrm{ (deterministic)  s.t.} && \bE_{\bD}[\inf_\sigma\ALG(\bD,\sigma)]&\ge\bE_\bD[\OFF(\bD)]/2\ .
\end{align*}
That said, finding the deterministic algorithm that is 1/2-competitive could be algorithmically intractable. Fortunately, the use of randomness in \Cref{alg:losslessRounding} is very explicit. The lossless rounding uses $n$ random bits per resource, so $nm$ random bits in total. Therefore, applying the method of conditional expectations \citep[Chap.~16]{alon2016probabilistic}---fixing the value of each random bit to 1 or 0 iteratively---can achieve a competitive ratio if $\frac{1}{2} - O(\eps)$. To do so, we must be able to simulate the expected total rewards for a {\em worst-case ordering} on each realization of the demand $\bD$ and each draw of the random bits in our lossless rounding.
Now, this is feasible because our prophet inequality algorithm and lossless rounding decouple how the adversary chooses ``permutations'' across resources: we let the almighty adversary choose a distinct permutation for each resource after we reveal which queries have been routed to it. Note that these permutations could be inconsistent across types{.  Regardless}, we  independently sample a pseudo-polynomial number of simulation draws to estimate the expected total rewards, while losing only $O(\frac{\eps}{nm})$ in the approximation factor in each of the $nm$ iterations.

\paragraph{\Correl model: De-randomizing Algorithm~\ref{alg:Correl}.} De-randomizing Algorithm~3 for $\Correl \cap \Rand$ is relatively easier since there is no adversarial element.  At each time $t$, both the resource $i$ that the query is routed to, and whether the resource accepts the query, can be determined by a random algorithmic decision.  In principle, de-randomization is possible---there are only $n+1$ ultimate decisions (matching to one of $n$ resources, or rejection), and the future from each one can be \textit{simulated} (still using the same \textit{randomized} algorithm for all future decisions).  The outcome with the highest simulated future rewards can then be deterministically chosen at time $t$, and this would not decrease the expected total reward of the algorithm.  In this way, we can sequentially de-randomize the decision at each time step $t$.  To formalize this argument, we must ensure we draw enough simulation samples to correctly choose the best decision at each time step, and in the end we will lose an $O(\epsilon)$ in the approximation factor. Please see \citet[Sec.~3]{ma2021dynamic} which formalizes this argument and achieves pseudo-polynomial running time.

\end{APPENDICES}

\end{document}